\documentclass[twocolumn,tighten,twocolappendix]{aastex631}
\usepackage{apjfonts}
\usepackage{amssymb}
\usepackage{amsbsy}
\usepackage{amsmath}
\usepackage{amsfonts}
\usepackage{mathrsfs}
\usepackage{color}
\usepackage{multirow}
\usepackage{bm}

\def\ihep{Key Laboratory for Particle Astrophysics, Institute of High Energy
Physics, Chinese Academy of Sciences, 19B Yuquan Road, Beijing 100049, China;
\href{mailto:dupu@ihep.ac.cn, wangjm@mail.ihep.ac.cn}{dupu@ihep.ac.cn,
wangjm@mail.ihep.ac.cn}}

\def\UCASastro{School of Astronomy and Space Sciences, University of Chinese
Academy of Sciences, Beijing 100049, China}

\def\UCASphys{School of Physics, University of Chinese Academy of Sciences,
Beijing 100049, China}

\def\NAOC{National Astronomical Observatories of China, Chinese Academy of
Sciences, 20A Datun Road, Beijing 100101, China}

\def\KNAO{Key Laboratory of Space Astronomy and Technology, National Astronomical Observatories, Chinese Academy of Sciences, Beijing 100101, China}

\def\KLOA{Key Laboratory of Optical Astronomy, National Astronomical Observatories, Chinese Academy of Sciences, Beijing 100101, China}

\def\DongGuan{Dongguan Neutron Science Center, 1 Zhongziyuan Road, Dongguan
523808, China}

\def\YNAO{Yunnan Observatories, Chinese Academy of Sciences, Kunming 650011,
China}

\def\KIAA{The Kavli Institute for Astronomy and Astrophysics, Peking University,
Beijing 100871, China} \def\PKUDoA{Department of Astronomy, School of Physics,
Peking University, Beijing 100871, China}

\def\CAHA{Centro Astronomico Hispano Alem\'an, Sierra de los filabres sn, E-04550
Gergal. Almer\'ia, Spain} 

\def\IAA{Instituto de Astrof\'isica de --Andaluc\'ia (CSIC),
Glorieta de la astronom\'ia sn, E-18008 Granada, Spain}

\def\QNU{College of Physics and Information Engineering, Quanzhou Normal University, Quanzhou, Fujian 362000, China}

\def\ANU{College of Physics and Electrical Engineering, Anyang Normal University, Anyang, Henan 455000, China}

\begin{document}

\title{\bf\large Supermassive Black Holes with High Accretion Rates in Active Galactic Nuclei. XV. \\
Reverberation Mapping of Mg {\sc ii} Emission Lines}

\author[0009-0009-5246-5719]{Hua-Rui Bai}
    \affiliation{\ihep}
   \affiliation{\UCASphys}
\author[0000-0002-5830-3544]{Pu Du}
   \affiliation{\ihep}
\author{Chen Hu}
   \affiliation{\ihep}
\author[0000-0003-4280-7673]{Yong-Jie Chen}
   \affiliation{\ihep}
   \affiliation{\DongGuan}
\author[0009-0000-1228-2373]{Zhu-Heng Yao}
   \affiliation{\NAOC}\affiliation{\KNAO}   
\author[0000-0001-5841-9179]{Yan-Rong Li}
   \affiliation{\ihep}
\author[0009-0004-5259-0900]{Yi-Xin Fu}
\author[0009-0009-9936-9871]{Yi-Lin Wang}
\author[0009-0006-7629-1459]{Yu Zhao}
\author[0000-0002-5595-0447]{Hao Zhang}
    \affiliation{\ihep}
   \affiliation{\UCASphys}
\author[0000-0003-3086-7804]{Jun-Rong Liu}
   \affiliation{\ihep}
\author{Sen Yang}
   \affiliation{\ANU}
\author[0009-0006-2592-7699]{Yue-Chang Peng}
   \affiliation{\ihep}
\author[0009-0009-9945-673X]{Feng-Na Fang}
   \affiliation{\ihep}
\author[0000-0003-4042-7191]{Yu-Yang Songsheng}
   \affiliation{\ihep}
   \affiliation{\DongGuan}
\author[0000-0001-5981-6440]{Ming Xiao}
   \affiliation{\ihep}
\author[0009-0005-4152-2088]{Shuo Zhai}
   \affiliation{\KLOA}
\author[0000-0003-3823-3419]{Sha-Sha Li}
   \affiliation{\YNAO}
\author[0000-0002-2310-0982]{Kai-Xing Lu}
   \affiliation{\YNAO}
\author{Zhi-Xiang Zhang}
   \affiliation{\QNU}
\author[0000-0003-2024-1648]{Dong-Wei Bao}
   \affiliation{\NAOC}   
\author[0000-0001-9457-0589]{Wei-Jian Guo}
   \affiliation{\KLOA}
\author{Jia-qi Feng}
   \affiliation{\ihep}
   \affiliation{\UCASphys} 
 \author{Yi-peng Zhao}
   \affiliation{\ihep}
   \affiliation{\UCASphys} 
\author{Jes\'us Aceituno}
    \affiliation{\CAHA}
    \affiliation{\IAA}
\author{Jin-Ming Bai}
   \affiliation{\YNAO}
\author[0000-0001-6947-5846]{Luis C. Ho}
   \affiliation{\KIAA}
   \affiliation{\PKUDoA}
\author[0000-0001-9449-9268]{Jian-Min Wang}
   \affiliation{\ihep}
   \affiliation{\UCASastro}
   \affiliation{\NAOC}
   
\collaboration{28}{(SEAMBH collaboration)}

\begin{abstract}
As the 15th paper in a series reporting on a large reverberation mapping (RM) campaign of super-Eddington accreting massive black holes (SEAMBHs) in active galactic nuclei (AGNs), we present the results of measurements of the Mg {\sc ii} lines in 18 SEAMBHs monitored spectroscopically from 2017 to 2024. Among these, the time lags of Mg {\sc ii} have been successfully determined for 8 of the 18 objects, thereby expanding the current Mg {\sc ii} RM sample, particularly at higher accretion rates. By incorporating measurements of the line widths, we determine the masses of their central supermassive black holes. Based on these new measurements, we update the relation between the Mg {\sc ii} radius and the monochromatic luminosity at 3000 \AA\ ($R_{\rm MgII}-L_{3000}$ relation), yielding a slope of $0.24 \pm 0.03$, which is slightly shallower than, yet still consistent with, previously reported values. Similar to the H$\beta$ lines, the Mg {\sc ii} time lags in SEAMBHs are shorter than those of AGNs with normal accretion rates at comparable luminosities. The deviation of AGNs from the best-fit $R_{\rm MgII}-L_{3000}$ relation shows a strong correlation with the accretion rate, while no significant correlation is found between the deviation and the flux ratio of UV iron to Mg {\sc ii}. 

\end{abstract}


\section{Introduction}

Accurate determination of the mass of supermassive black hole (SMBH) is crucial
for understanding the evolution of SMBHs over cosmic timescales and their
co-evolution with host galaxies \citep[e.g.,][]{FF2005, KH2013}. In the past
three decades, reverberation mapping (RM, e.g., \citealt{BM1982, Peterson1993})
has been demonstrated as an effective way to measure the masses of SMBHs
in the centers of active galactic nuclei (AGNs). The underlying physics of this
approach is based on the nearly virial kinematics of the gas in broad-line
regions (BLRs), which are primarily governed by the gravitational potential of
the central SMBHs \citep[e.g.,][]{PW1999, PW2000, OP2002}. RM determines the
sizes of BLRs through measuring the time delays between variations of broad
emission lines and those of the continuum fluxes. Then the SMBH mass can be
obtained by 
\begin{equation} \label{equation1} 
    M_{\bullet} = f \frac{\Delta V^2 R_{\rm BLR}}{G}, 
\end{equation}
where $R_{\rm BLR} = c \tau$ represents the typical size of the BLR, $c$ is the
speed of light, $\tau$ is the time lag, $G$ is the gravitational constant,
$\Delta V$ is the width of the emission line, and $f$ is a factor that accounts
for the geometry, kinematics, and inclination of the BLR.

However, RM remains a resource-intensive observational technique, requiring
extensive telescope time, which limits its practicality for large samples of
AGNs. For AGNs with only single-epoch spectroscopic data, an empirical
correlation between continuum luminosity and BLR size (called
``radius-luminosity relation'', see e.g., \citealt{Kaspi2000, Bentz2013}), as
established through RM campaigns in the past decades, enables mass estimation of
their SMBHs despite the absence of direct RM measurements. The single-epoch
virial mass estimation method, which combines the radius-luminosity relation with broad
emission-line width measurements, has been widely applied to large AGN samples
from the local universe to high redshifts and provided the foundation for our
current knowledge of AGN evolution \citep[e.g.,][]{Greene2005, Vestergaard2006,
Shen2011, YangJY2023}.

Most RM campaigns \citep[e.g.,][]{Peterson1998, Kaspi2000, Bentz2010, Grier2012,
Du2014, Barth2015, DeRosa2018, U2022, Shen2024, Woo2024} primarily focus on
H$\beta$ emission lines. 
This preference is due to the accessibility of H$\beta$ in the optical spectra obtained with ground-based telescopes, and older spectrographs and CCDs were mainly sensitive to blue wavelengths.
Additionally, the flux of
H$\beta$ can be calibrated more easily using the adjacent narrow [O {\sc
iii}]$\lambda$5007 emission line, which is commonly employed as a reliable
reference for flux calibration in most RM campaigns. Through these RM campaigns,
the radius-luminosity relation for H$\beta$ (hereafter referred to as the
$R_{\rm H\beta}-L_{5100}$ relation) has been relatively well established
\citep[e.g.,][]{Kaspi2000, Bentz2013} and can be applied as the single-epoch
mass estimator for low-redshift AGNs. 

For AGNs with redshifts $z\gtrsim1$, the H$\beta$ line falls outside the optical band. In such cases, the Mg\,\textsc{ii} $\lambda2800$ emission line serves as the primary diagnostic for AGNs in the range $1 \lesssim z \lesssim 2$, while C\,\textsc{iv} $\lambda1549$ becomes the dominant feature at $2 \lesssim z \lesssim 4$.
Accordingly, for AGNs at higher redshifts where H$\beta$ becomes observationally
inaccessible, ultraviolet emission lines such as Mg {\sc ii} (and in some cases
C {\sc iv}) must be employed as surrogate virial indicators.
Prior to the
development of robust Mg {\sc ii} radius-luminosity relations, currently limited
by the scarcity of high-quality RM campaigns for Mg {\sc ii} (and also for C
{\sc iv}), the single-epoch mass estimators calibrated either based on the SMBH
masses derived from H$\beta$ RM measurements \citep[e.g.,][]{MJ2002, Wang2009,
Bahk2019} or based on the established $R_{\rm H\beta}-L_{5100}$ relation
\citep[e.g.,][]{McGill2008, Woo2018, Le2020} were adopted. Nevertheless, the
validity of these Mg {\sc ii} calibrations (effectively extrapolations from the
H$\beta$ lines) requires rigorous verification through direct Mg {\sc ii} RM
measurements. Furthermore, implementing SMBH mass measurements derived
explicitly from Mg {\sc ii} lag measurements, rather than relying on
H$\beta$-based ``extrapolation'', could potentially yield significant
improvements in mass estimation accuracy.

Moreover, considering that the Mg {\sc ii} line may be dominated by collisional
excitation, unlike the recombination lines (e.g., H$\beta$ and H$\alpha$), a
question arises regarding the existence of a solid radius-luminosity relation
for the Mg {\sc ii} line. Theoretically, calculations based on CLOUDY models
\citep[e.g.,][]{KG00, Guo2020, Du2023} suggest that the Mg {\sc ii} line
exhibits a relatively weaker response to continuum variations compared to
H$\beta$, implying that RM results may be more challenging to obtain for Mg {\sc
ii}.  
This could be attributed to the larger optical depth for the Mg {\sc
ii} line. For a given continuum flux, compared to recombination lines, Mg {\sc
ii} line photons originating close to the continuum sources are less likely to escape. 
Consequently, flux variations tend to be dominated by gas clouds at larger radii, 
where variations in the continuum may be diluted \citep[e.g.,][]{KG04, Yang2020}.
From an observational perspective, prior to successful RM measurements,
variations in the Mg {\sc ii} lines were indeed found to respond to continuum
emission variations; however, these responses exhibited smaller amplitudes
compared to those of H$\beta$, as shown in multi-epoch observations
\citep[e.g.,][]{Kokubo2014, Sun2015, Zhu2017, Yang2020}.  In contrast, some
studies \citep[e.g.,][]{Goad1999, Trevese2007, Woo2008, Hryniewicz2014,
Cackett2015} found that the Mg {\sc ii} lines in certain AGNs did not show
significant variations despite substantial changes in the continuum,
highlighting the difficulties in obtaining RM results for Mg {\sc ii}.

With the investment of more observational resources, successful Mg {\sc ii} RM
observations gradually occur although the number of objects is still limited
\citep[e.g.,][]{MOP2006, Shen2016, Lira2018, Czerny2019, Homayouni2020,
Zajacek2020, Prince2022, Prince2023, Yu2021, Yu2023}. Specifically, two
relatively large samples are from Sloan Digital Sky Survey (SDSS) RM
\citep{Homayouni2020, Shen2024} and Australian Dark Energy Survey (OzDES) RM
\citep{Yu2023} projects. \cite{Homayouni2020} reported Mg {\sc ii} lags for 57
quasars, among which 24 were identified as ``gold sample'' with most reliable
measurements of lags. \cite{Shen2024} further expanded the sample by reporting lag measurements for 125 objects in the final results of the SDSS-RM project. 
Additionally, \cite{Yu2023} presented Mg {\sc ii} lags for 25
quasars. With the limited samples of successful and accurate Mg {\sc ii} measurements,
merely a preliminary radius-luminosity relation of Mg {\sc ii} (hereafter
$R_{\rm MgII}-L_{3000}$ relation) was built \cite[e.g.,][]{Homayouni2020,
Yu2023, Shen2024}. The slope of the derived $R_{\rm MgII}-L_{3000}$ relation is around
0.3, which is shallower than the typical slope ($\sim$0.5) of the $R_{\rm
H\beta}-L_{5100}$ in \cite{Bentz2013}. In order to refine $R_{\rm
MgII}-L_{3000}$ relation, more RM observations for Mg {\sc ii} lines are highly
required.

Moreover, recent studies have shown that super-Eddington AGNs exhibit shorter
H$\beta$ time lags compared to AGNs with normal accretion rates at the same
luminosities, suggesting that the accretion rate is a parameter that influences
the scatter of the $R_{\rm H\beta}-L_{5100}$ relation \citep[e.g.,][]{Du2015,
Du2016, Du2018, DW2019}. \cite{MA2020} found a similar phenomenon in the $R_{\rm
MgII}-L_{3000}$ relation for Mg {\sc ii}, indicating that its scatter may be
also primarily driven by the accretion rate. To further investigate the possible
dependency of the scatter of $R_{\rm MgII}-L_{3000}$ on the accretion rate, it
is essential to expand the RM sample of Mg {\sc ii} lines, particularly by
including AGNs with high accretion rates.

As the fifteenth paper of the series reporting the RM campaign of
super-Eddington accreting massive black holes (SEAMBHs) in AGNs, here we report
the results of Mg {\sc ii} lines in eighteen luminous quasars with high
accretion rates at intermediate redshift to advance the investigation of $R_{\rm
MgII}-L_{3000}$ relation. The time lags of eight among them are successfully
measured and the latest $R_{\rm MgII}-L_{3000}$ relation were established based on this
sample and the historical samples. The paper is organized as follows. In Section
\ref{sec_obs}, we describe the target selection, observation, and data
reduction. The data analyses and measurements are provided in Section
\ref{sec_analysis}, including the mean and root-mean-square (rms) spectra, the
light curves, the line widths, the time lags, and the SMBH masses. The $R_{\rm
MgII}-L_{3000}$ relation based on this new sample is provided in Section
\ref{sec_results}. Discussions regarding the comparison between the Mg {\sc ii} and H$\beta$ lags, as well as the implications of the shortened lags, are given in Section \ref{sec_dis}. Section \ref{sec_summary} offers a brief summary. Throughout the
paper, we use the standard $\Lambda$CDM cosmology with $H_0=67~{\rm
km~s^{-1}~Mpc^{-1}}$, $\Omega_{\rm M}=0.32$ and $\Omega_{\Lambda}=0.68$
\citep{Planck2020}.

\section{Observations and Data Reduction}
\label{sec_obs}
\subsection{Target Selection}

We selected the SEAMBH targets for Mg {\sc ii} measurements primarily based on
the dimensionless accretion rate derived from the standard accretion disk model
\citep[e.g.,][]{Du2014, Du2018}. The accretion rate, based on the monochromatic
luminosity $L_{3000}$ at 3000\AA, is given by
\begin{equation}
   \dot{\mathscr{M}}=7.0 \left(\frac{\ell_{44}}{\cos i}\right)^{3/2}m_7^{-2}
\end{equation}
where $\ell_{44} = L_{3000} / 10^{44}\ {\rm erg\ s^{-1}}$, and $m_7 = M_\bullet
/ 10^7 M_{\odot}$ is the SMBH mass. Here, $i$ denotes the inclination angle of
the disk relative to the line of sight, and we adopt $\cos i = 0.75$ as a
typical value throughout this paper (see further details in \citealt{Du2016}).

We sorted the quasars from SDSS Data Release 7 \citep{Shen2011} by their
accretion rates and selected targets with the highest values. Only those targets
with redshifts ($z=0.7-0.9$) for which the Mg {\sc ii} emission lines fall
within the wavelength range where the spectrographs (described in Section
\ref{sec_spec_phot}) exhibit the highest sensitivity, along with SDSS
r$^{\prime}$--band magnitudes brighter than 17.5, were selected, ensuring that
we could achieve optimal signal-to-noise ratios (S/N). Additionally, only
objects suitable for the observational seasons and the latitude of the
observatories (see Section \ref{sec_spec_phot}) with predicted time lags in the
observed frames of less than 450 days were considered. During the target
selection process for the present sample, there were very few successful Mg {\sc
ii} RM observations; therefore, we assumed that the Mg {\sc ii} and H$\beta$
lags are the same and simply estimated the Mg {\sc ii} lags by the $R_{\rm
H\beta}-L_{5100}$ relation, given that their line widths are not significantly
different \citep[e.g.,][]{MJ2002, Shen2008, Wang2009}. The SDSS names and
redshifts of the 18 selected targets in the present sample are listed in Table
\ref{Object_Observation}.

\subsection{Spectroscopy and Photometry} 
\label{sec_spec_phot}

The Mg {\sc ii} campaign started from the fall of 2017. Spectroscopic and
photometric observations of twelve targets were conducted using the Lijiang 2.4
m telescope at the Yunnan Observatories of the Chinese Academy of Sciences.
Additionally, observations of six other targets were carried out at the 2.2 m
telescope of the Calar Alto Astronomical Observatory, part of the Centro
Astronómico Hispano en Andalucía (CAHA) in Spain. The locations of the
observations and the detailed monitoring periods are listed in Table
\ref{Object_Observation}. The spectroscopic and photometric observations at the
Lijiang and CAHA are briefly described below.

\subsubsection{Spectroscopy}
\label{sec_spec}

Data were collected using the Yunnan Faint Object Spectrograph and Camera
(YFOSC) mounted on the Lijiang 2.4 m telescope, and the Calar Alto Faint Object
Spectrograph (CAFOS) on the CAHA 2.2 m telescope. Both instruments are capable
of performing spectroscopic and photometric observations. Due to the absence of
narrow emission lines in the vicinity of the Mg {\sc ii} lines, we employed a
comparison-star-based calibration method for our spectroscopic observations.
The field de-rotators of the Lijiang 2.4 m and CAHA 2.2 m telescopes allow the
slits to rotate, accommodating a nearby comparison star (listed in Table \ref{Object_Observation})
while simultaneously exposing the target, thereby enabling the star to serve as
the flux calibration standard. 

With the in-slit comparison star, we achieved accurate relative flux
measurements, even under relatively poor weather conditions
\citep[e.g.,][]{Maoz1990, Kaspi2000, Du2014}. The typical separation between
the target and the chosen comparison star was approximately 1-3 arcminutes.
More details regarding this comparison-star-based calibration method can be
found in, e.g., \cite{Du2014} and \cite{Du2018}. To minimize light loss, we
utilized long slits with relatively broad widths (5.05\,\arcsec\ at Lijiang and
3\,\arcsec\ at CAHA). We used Grism 3 at Lijiang, which provided a wavelength
coverage of 3400–9100 \AA\ and a dispersion of 2.9 \AA\ pixel\(^{-1}\), and
Grism G-200 at CAHA, which covered a wavelength range of 4000-8500 \AA\ with a
dispersion of 4.47 \AA\ pixel\(^{-1}\). The typical exposure time for each
observation was 1800 seconds. 

To mitigate the impact of cosmic rays, we conducted two consecutive exposures
for each epoch, which were combined to produce the individual-night spectra.
All spectra were extracted using a uniform aperture of 30 pixels (approximately
8.5 arcseconds), with background windows of [-26 pixels, -50 pixels] and [26
pixels, 50 pixels] on the left and right sides (both for the Lijiang and
CAHA data). Wavelength calibration was performed based on the standard neon
and helium lamps. The data were reduced using the standard IRAF v2.16.

Flux calibration was performed using a two-step procedure. Firstly, we
generated the fiducial spectra of the comparison stars using spectrophotometric
standards (e.g., G191B2B, Feige 56, and Feige 34) during photometric nights (when
conditions were good). Secondly, for each individual-night spectrum of the
target/comparison star pair, we compared the flux of the in-slit star with that
of the fiducial spectrum to generate a wavelength-dependent sensitivity
function, which was then applied to the target spectrum for flux calibration
(see also \citealt{Du2014} and \citealt{Du2018}).

\subsubsection{Photometry}\label{sec_phot}

As a double check on the accuracy of the flux calibration for the spectroscopic
observations and to ensure that the comparison stars did not exhibit
significant variability during the campaign, we also conducted photometric
observations of the targets and comparison stars. We obtained three broad-band
photometric images of the targets just prior to the spectral exposures each
night, with typical exposure times of 80 seconds for each image. The filters
used at Lijiang and CAHA are provided in Table \ref{Object_Observation}.

The data were reduced using standard IRAF procedures for bias and flat-field
correction. We performed differential photometry of the target and comparison
stars with respect to 3-6 nearby reference stars in the same field of view. The
fluxes of the targets and stars were extracted using circular apertures with
radii of 4.2\,\arcsec\ and 5\,\arcsec\ for the Lijiang and CAHA
observations, respectively. As shown in Appendix \ref{appD}, the comparison stars during the campaign are fairly stable.

To supplement and extend the coverage of the continuum light curves, we
utilized archival time-domain photometric data from the Zwicky Transient
Facility\footnote{\url{https://www.ztf.caltech.edu/}} (ZTF; see
\citealt{Masci2019}). ZTF provides photometric light curves in three bands
(ZTF-g, r, i). In our subsequent analysis, we adopted only the r band light
curves from ZTF because the g band overlaps with the Mg {\sc ii} lines in
wavelength, and the i band light curves provide less temporal coverage than
those in the r band.

\section{Analysis}
\label{sec_analysis}
\subsection{Mean and rms Spectra}

\begin{figure*}
\centering
\includegraphics[width=0.85\textwidth]{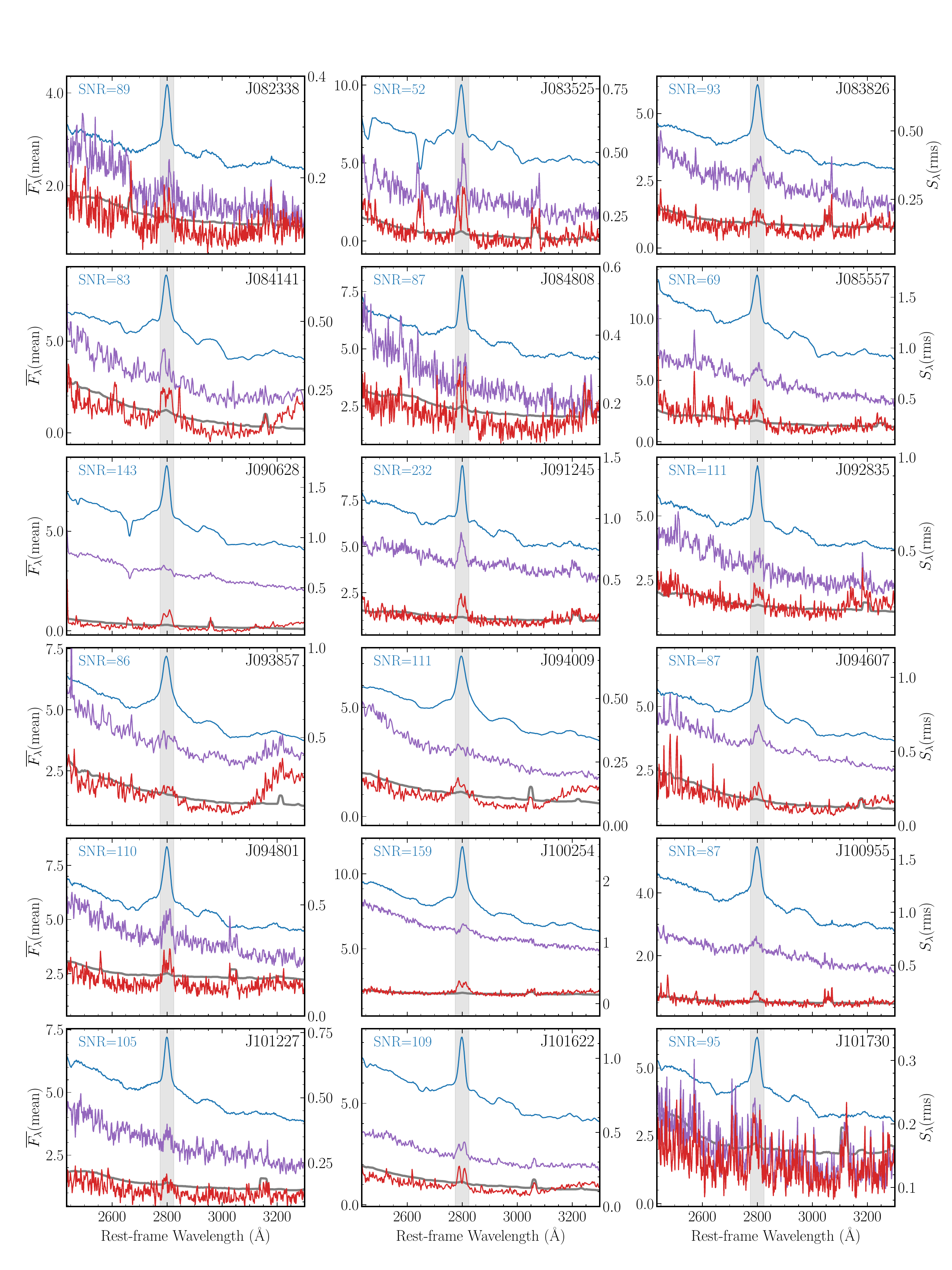}
\caption{Mean and rms spectra. The blue, purple, and red spectra in each panel represent the mean spectra, rms spectra, and rms spectra generated from the residuals after subtracting the power law and iron template (see more details in Section \ref{sec_lightcurve}), respectively. The spectra are plotted with rest-frame wavelength versus Galactic-extinction-corrected flux in the observed frame, consistent with the subsequent figures. The gray lines represent the error spectra for comparison. The gray shadow marks the 2775--2825 \AA\  integration window for measurements of Mg {\sc ii} light curves. The tick labels on the left and right y axes correspond to the mean and rms spectra, respectively. 
All spectra are in units of $10^{-16} \, {\rm erg} \, {\rm s}^{-1} \, {\rm cm}^{-2} \, \text{\AA}^{-1}$. The signal-to-noise ratio (SNR) at 3000{\AA} of the mean spectrum is listed at each panel. }
\label{fig_mean_rms}
\end{figure*}

The mean and root-mean-square (rms) spectra are always a useful way to quickly
check the profiles of the emission lines and to evaluate their variabilities
during the RM campaign. The mean and rms spectra can be derived from the
formula
\begin{equation}
    \overline{F}_\lambda=\frac{1}{N} \sum_{i=1}^{N}F^i_\lambda,
\end{equation}
and 
\begin{equation}
    S_\lambda=\left[ \frac{1}{N} \sum_{i=1}^{N} (F^i_\lambda - \overline{F}_\lambda)^2  \right]^{1/2},
\end{equation}
where $F^i_\lambda$ is the $i$-th spectrum of the object and $N$ is the number
of its spectra.  For comparison, we also present the rms spectra generated from the residual spectra obtained after subtracting the power law and iron template components from the spectral fitting detailed in following Section \ref{sec_lightcurve}. The mean spectra, rms spectra, and rms spectra generated from the residuals for all 18 targets throughout the campaign are shown in Figure \ref{fig_mean_rms}.

Additionally, the observational uncertainties of spectra may be responsible for some of the signatures showing on the rms spectra. We also present the error spectra defined as 
\begin{equation}
    E_\lambda=\left[ \frac{1}{N} \sum_{i=1}^{N} (e^i_\lambda)^2  \right]^{1/2}
\end{equation}
in Figure \ref{fig_mean_rms}, where $e^i_\lambda$ is the observational error of the $i$-th spectrum at wavelength $\lambda$, as obtained from the IRAF spectrum extracting procedure. The error spectra clearly indicate that some features observed around the 3000 \AA\ to 3200 \AA\ wavelength range in the rest frames are attributable to observational errors (specifically the Poisson noise caused by the strong 5577 \AA\ emission line from the atmosphere). Considering both the rms spectra and the error spectra, the Mg {\sc ii} variations in J082338, J084808, J092835, and J101730 are not significant.

\subsection{Light Curves}
\label{sec_lightcurve} 

The light curves of the H$\beta$ emission lines can be measured using either a
direct integration scheme \citep[e.g.,][]{Peterson1998, Kaspi2000, Bentz2010,
Du2014, Du2018mahaI} or a spectral fitting scheme \citep[e.g.,][]{Barth2013,
Hu2015, Hu2021, U2022, Woo2024}. The advantages and disadvantages of these two
schemes have been discussed in references such as \cite{Du2018mahaI} and
\cite{Hu2015}. For Mg {\sc ii} lines, strong iron emission features are often
present \citep[e.g.,][]{Wills1980, Verner1999}, which can influence the direct
integration scheme due to their contributions and variabilities. Therefore, the
fitting scheme is typically adopted for Mg {\sc ii} measurements
\citep[e.g.,][]{Prince2023, Yu2023}. 

Furthermore, for SEAMBHs, the relative strength of the iron emission in the UV
and optical, with respect to H$\beta$ \citep[e.g.,][]{boroson1992,
sulentic2000, marziani2001, shen2014, du2016FP, Panda2019MS} and Mg {\sc ii}
\citep[e.g.,][]{dong2011, shin2021, jiang2024, pan2025}, is generally stronger
than that in AGNs with normal accretion rates. Consequently, we measured the
flux of the Mg {\sc ii} lines in our sample using the fitting scheme. 

The spectral fitting was performed using the software
{\textsc{DASpec}}\footnote{\url{https://github.com/PuDu-Astro/DASpec}}
\citep{Du2024}, which is a multi-component spectral fitting code for AGN spectra
based on the Levenberg-Marquardt algorithm \citep{Press1992} and features a
user-friendly graphical interface. We adopted the following components in the
fitting: (1) a power law to model the AGN continuum, (2) a UV iron template 
compiled by \cite{Shen2019} convolved with a Gaussian kernel to model the iron
emission, and (3) two Gaussian functions to model the Mg {\sc ii} emission line. 
The template compiled by \cite{Shen2019} is a composite one: it incorporates 
the template from \cite{Salviander2007} for the wavelength range of \( 2200-3090\,\text{\AA} \), 
and the template from \cite{Tsuzuki2006} for the range of \( 3090-3500\,\text{\AA} \).
\cite{Yu2021} tested three different iron templates from \cite{VW2001},
\cite{Tsuzuki2006}, and \cite{Salviander2007} in their Mg {\sc ii} measurements
and found that the three templates produced generally consistent results.  
As a simple test, we also performed the analysis by replacing 
the \cite{Salviander2007} template by the \cite{Tsuzuki2006} template and obtained 
fully consistent results. More details are provided in Appendix~\ref{app_template}.
In addition, the contributions from the host galaxies in the
spectra were not considered in the fitting, as they are expected to be
negligible for the targets in our sample, given their high luminosities
\citep{Shen2011}.

\begin{figure*}[t]
\centering
\includegraphics[width=\columnwidth]{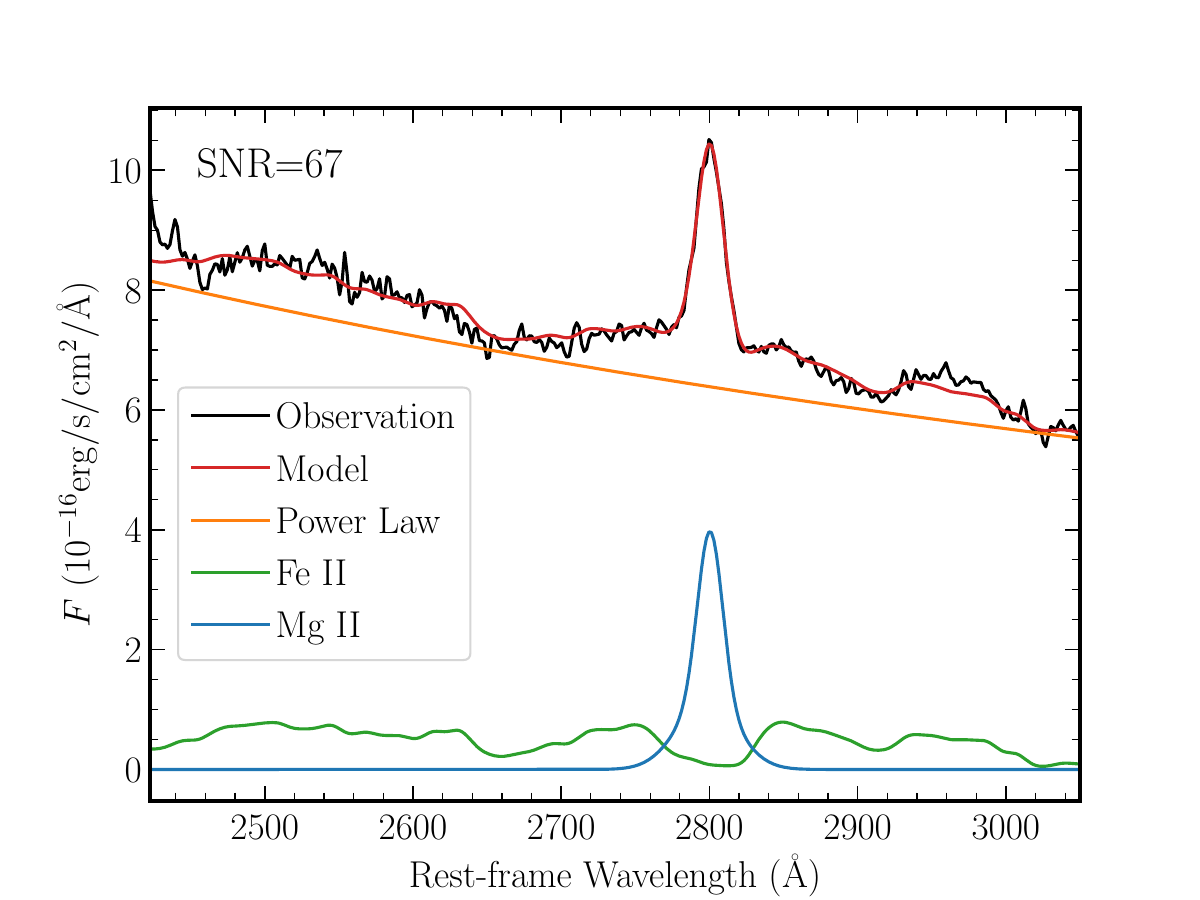}
\caption{An example of the spectral fitting results. The black line represents a single epoch spectrum of J091245 in the rest frame after correcting for Galactic extinction. The signal-to-noise ratio (SNR) of this spectrum is explicitly written.
The red line indicates the best-fit model. The orange, green, and blue lines correspond to the power law, iron template, and Mg {\sc ii} components, respectively (see Section \ref{sec_lightcurve}).}
\label{fig_fit}
\end{figure*}

We first corrected for Galactic extinction \citep{schlafly2011} and transformed the observed spectra to their rest frames, after which we performed the fitting in the wavelength range of 2260 – 3050\,\AA\ in those frames. For several objects (J084808, J085557, J091245, J092835 and J093857), a slightly smaller wavelength range (starts from $\sim$2300\text{\AA}) was used due to their redshifts. An example of the spectral fitting results is illustrated 
in Figure \ref{fig_fit}. 
After subtracting the components of the power law and iron template, we obtained the Mg {\sc ii} light curves by integrating the fluxes of the residual spectra within the range of 2775 – 2825\,\AA, where the Mg {\sc ii} features in the RMS spectra are most significant (see Figure \ref{fig_mean_rms}). The continuum light curves were generated by measuring the median fluxes in the range of 3000 – 3100\,\AA\ from the power-law components. The continuum light curves derived from the spectroscopic data generally agree well with the photometric light curves, indicating their reliability. Several obvious outliers in the light curves caused by bad weather conditions or inaccurate slit alignments were discarded.

\subsection{Intercalibration of Continuum Light Curves}
\label{sec:intercalib}

Due to the different apertures used in the spectroscopic and photometric
observations, the continuum light curves at 3000 \AA\ from the spectroscopic
data and the ZTF r-band continuum light curves must be intercalibrated. The
intercalibration was performed using the Bayesian package
{\textsc{PyCALI}}\footnote{\url{https://github.com/LiyrAstroph/PyCALI}}
\citep{li2024pycali}. This package assumes that the light curves can be modeled
by a damped random walk and determines the optimal multiplicative and additive
factors by exploring the posterior probability distribution using a diffusive
nested sampling algorithm (details can be found in \citealt{Li2014}). The
photometric and spectroscopic continuum light curves after intercalibration are
shown in Figure \ref{fig_lcs} and Appendix \ref{app}. The consistency between the photometric and spectroscopic
continuum light curves demonstrates the effectiveness of the intercalibration
process. The light curves after the intercalibration for all 18 targets are provided in Table \ref{tab_lc}. For subsequent analysis, the photometric and spectroscopic light curves were combined by averaging the observations taken during the same nights.

\subsection{Variability}
\label{sec_var}

The variability amplitudes and the corresponding uncertainties of the light curves can be quantified by 
\begin{equation}
   F_{\rm var} = \frac{\left(\sigma^2 - \Delta^2\right)^{1/2}}{\left\langle F \right\rangle},
\end{equation}
and
\begin{equation}
   \sigma_{\rm Fvar} = \frac{1}{\left(2N\right)^{1/2}F_{\rm var}} \frac{\sigma^2}{\left\langle F \right\rangle^2},
\end{equation}
where $\sigma$, $\Delta$, $\left\langle F \right\rangle$, and $N$ are the
standard deviation, mean uncertainty, mean flux, and the number of data points
of the light curve, respectively \citep[e.g.,][]{Rodriguez1997, Edelson2002}. 
The variability amplitudes and uncertainties of the continuum and Mg {\sc ii}
light curves are listed in Table \ref{lc_statistics}. The Mg {\sc ii} variabilities of the present sample are generally lower than $\sim$8\%.

\subsection{Time-lag Analysis}
\label{sec_lag}

We employed the interpolated cross-correlation function (ICCF; e.g.,
\citealt{Gaskell1987}) and the MICA algorithm \citep{Li2016} to determine the
time lags of the Mg {\sc ii} lines in the present sample. Details of both
methods can be found in the aforementioned references, and we provide a brief
introduction here.

\textit{ICCF:} ICCF is a commonly used method for determining time lags in RM
analysis. We adopted the centroid of the CCF above a threshold of 80\% of the
peak value as the time lag (referred to as the centroid time lag) and utilized
 the cross-correlation centroid distributions (CCCDs) through ``flux randomization/random subset sampling (FR/RSS)'' method
\citep{Peterson1998, Peterson2004} to estimate the uncertainties. 

\textit{MICA:} MICA is a Bayesian-based non-parametric method that infers the
one-dimensional transfer function from the light curves of the continuum and
emission lines. This approach models the transfer function as a combination of
one or more Gaussian or Gamma functions. To derive the posterior distributions
of the parameters, it employs the diffusive nested sampling technique. We
assumed that the transfer function for the Mg {\sc ii} lines could be
represented by a Gaussian function, and we used MICA to compute the
corresponding transfer function. The time lag of Mg {\sc ii} was then determined
from the centroids of these transfer functions. Ultimately, the median value of
the centroid distribution obtained through nested sampling serves as the
estimate of the time lag, with its uncertainty assessed using the 68.3\%
confidence interval.

It is evident that the time lags of the Mg {\sc ii} lines for some of our
targets were not well constrained due to their relatively low variability and
S/N ratios. We adopted a criterion of a maximum ICCF coefficient (\(r_{\rm
max}\)) of 0.5 to determine whether the lags could be reliably measured; below
this threshold, we consider the measurements of the Mg {\sc ii} line lag to be
unreliable. Eight of our targets meet the criterion of $r_{\rm max} > 0.5$,
and their light curves, along with time lag measurements from both ICCF and
MICA, are presented in Figure \ref{fig_lcs}. The light curves for the remaining ten
targets, for which we believe the lag measurements were unsuccessful, are shown
in Appendix \ref{app}. 

\begin{figure*}
\centering
\includegraphics[width=\columnwidth]{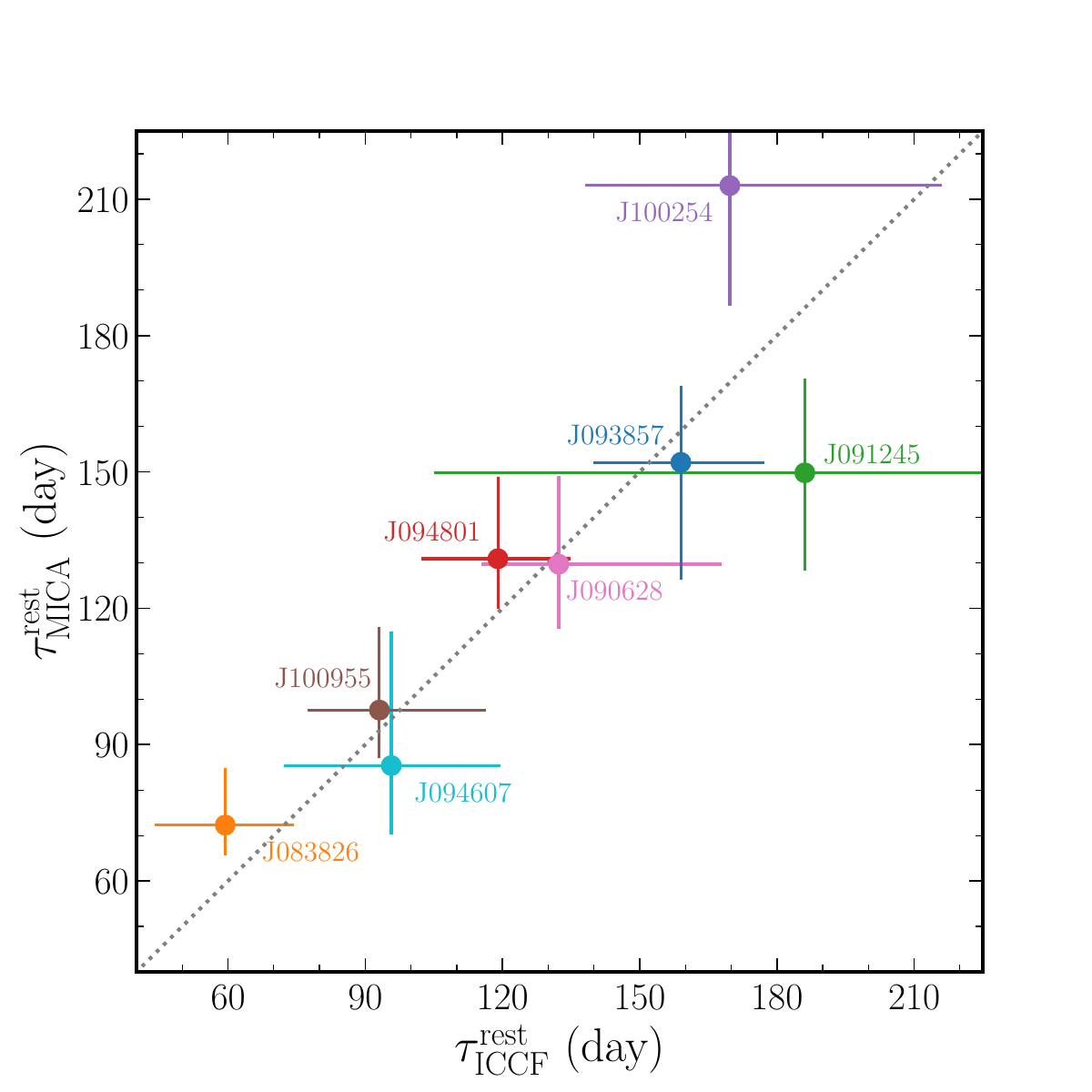}
\caption{Comparison between
$\tau_{\rm ICCF}^{\rm rest}$ and $\tau_{\rm MICA}^{\rm rest}$.}
\label{fig_mica_ccf}
\end{figure*}

The time lags measured by ICCF and MICA in the observed frame (denoted as $\tau_{\rm ICCF}^{\rm obs}$ and $\tau_{\rm MICA}^{\rm obs}$) and in the rest frame (denoted as $\tau_{\rm ICCF}^{\rm rest}$ and $\tau_{\rm MICA}^{\rm rest}$) for the eight successful objects are listed in Table \ref{measurement}. The comparison between $\tau_{\rm ICCF}^{\rm rest}$ and $\tau_{\rm MICA}^{\rm rest}$ is illustrated in Figure \ref{fig_mica_ccf}. Generally, $\tau_{\rm ICCF}^{\rm rest}$ and $\tau_{\rm MICA}^{\rm rest}$ are consistent with one another. The time lags of the Mg {\sc ii} emission for the current sample in the observed frames typically range from 100 to 300 days. Considering that MICA employs the damped random walk model to reconstruct the light curves, it is expected to provide more reliable lag measurements, especially in cases with seasonal gaps. Therefore, we adopted the time lag results from MICA for the subsequent black hole mass measurements and analyses.

\subsection{Line Width Measurements}
\label{sec_width}

To measure the masses of the SMBHs in these AGNs, it is essential to determine
the line widths of the Mg {\sc ii} lines. We measured the full width at half
maximum (FWHM) and the line dispersion ($\sigma_{\rm line}$; e.g.,
\citealt{Peterson2004}) from the line profiles in the mean spectra of the eight
successful objects. The contributions from the iron emission beneath the Mg {\sc
ii} lines were removed by the spectral fitting method outlined in Section
\ref{sec_lightcurve}. In this paper, we do not provide line width measurements from
their root-mean-square (rms) spectra, as the Mg {\sc ii} profiles in the rms
spectra are not well-defined due to low S/N ratios and variability
amplitudes.

The uncertainties of the line widths were estimated using a bootstrap method. We
first generated a subsample by randomly selecting $N$ spectra from the original
$N$ spectra (with replacement) and then created the mean spectra for line width
measurements based on this subsample. This procedure was repeated 500 times. The
uncertainties of the line widths were determined from the 16th and 84th
percentiles of the generated distributions. The instrumental broadening (FWHM
$\sim 1200\, {\rm km s^{-1}}$ for Lijiang and $\sim 1000\, {\rm km s^{-1}}$ for
CAHA), was then subtracted
from the measured widths of both FWHM and $\sigma_{\rm line}$. The instrumental broadening was estimated from the spectra of the comparison stars, by comparing the observed spectra of these stars with the high-resolution stellar spectra from the Indo-U.S. Library of Coud\'{e} Feed Stellar Spectra (\citealt{Valdes2004}, more details can be found in \citealt{Du2016VI}).

The Mg {\sc ii} emission line is actually a doublet (2796.35 \AA\ and 2803.53
\AA, see, e.g., \citealt{Morton1991}). Although we used two Gaussian functions
in the fitting process (see Section \ref{sec_lightcurve}), these Gaussians do not
precisely represent the doublet since we did not fix their wavelengths (or the
wavelength separation) to 2796.35 \AA\ and 2803.53 \AA\ (or $\sim 7.18$ \AA).
The widths and velocity shifts of the two Gaussians were treated as free parameters during
the fitting. 
Therefore, the line widths measured from the combination of the two Gaussians (here we practically measured the line widths from the residuals after subtracting the power law and Fe {\sc ii} components) may slightly deviate from the intrinsic velocity broadening of each component of the doublet. We estimated the influence of the line width measurements from the wavelength shift of the doublet by assuming that the flux ratio of the doublets is close to 1 \citep[e.g.,][]{Marziani2013,popovic2019} and that the intrinsic FWHM is 3000\,${\rm km\,s^{-1}}$ (which is similar to the typical value for our sample). This leads to the line widths measured from the combination of the doublets being larger than the intrinsic values by approximately 140\,${\rm km\,s^{-1}}$ for FWHM and approximately 60\,${\rm km\,s^{-1}}$ for $\sigma_{\rm line}$, respectively. Consequently, these values were included as additional uncertainties in the lower error bars.

\subsection{Measurement of SMBH Masses}
\label{sec_BHmass}

The SMBH masses can be determined using Equation (\ref{equation1}), given the Mg
{\sc ii} time lags and the measured line widths. Additionally, the virial factor
$f$ is required. The value of the virial factor for the H$\beta$ emission line
has been extensively investigated. It can be calibrated based on the
relationships between SMBH masses and the properties (e.g., stellar velocity
dispersion, luminosity, or stellar mass) of the bulges of host galaxies in
inactive galaxies \citep[e.g.,][]{Onken2004, Ho2014, Woo2015, yang2024}.
Alternatively, it can be determined through dynamical modeling of the BLR using
reverberation mapping (RM) observations \citep[e.g.,][]{pancoast2011,
pancoast2014, Li2018, villafana2023}, interferometric observations
\citep[e.g.,][]{GRAVITY2018, GRAVITY2020, GRAVITY2021}, or by joint analysis of
RM and interferometry \citep[e.g.,][]{Wang2020, Li2022, Li2025}. 

However, there is no well-established virial factor for the Mg {\sc ii} line.
Considering that the average virial factor corresponding to the FWHM of the
H$\beta$ line is close to 1 \citep[e.g.,][]{Onken2004, Ho2014, Woo2015,
yang2024}, and acknowledging that the average line widths of Mg {\sc ii} and
H$\beta$ are similar for the same object \citep[e.g.,][]{Wang2009, Ho2012, Bahk2019}, we
adopted $f = 1$ as the virial factor corresponding to the FWHM of the Mg {\sc
ii} line. It is
important to note that the virial factor $f$ may vary from one object to another
\citep[e.g.,][]{villafana2023}. A more accurate determination of the virial
factor $f$ and, consequently, a more precise measurement of $M_{\bullet}$ are
beyond the scope of this work. The uncertainties in the SMBH masses due to the
variability of the virial factor are not considered here. 
We acknowledge the significant uncertainty \citep[e.g.,][]{Grier2013, Woo2015, Fausnaugh2016} introduced by the virial factor $f$.
The measured SMBH masses range from $8.1 < \log~(M_\bullet/M_{\odot}) < 8.7$ and are listed in Table \ref{measurement}.

\section{Size-Luminosity Relation and Dependency on Accretion Rate}
\label{sec_results}

\subsection{$R_{\rm MgII}-L_{3000}$ Relation}
\label{sec_rl}

\begin{figure*}
\centering
\includegraphics[width=0.8\textwidth]{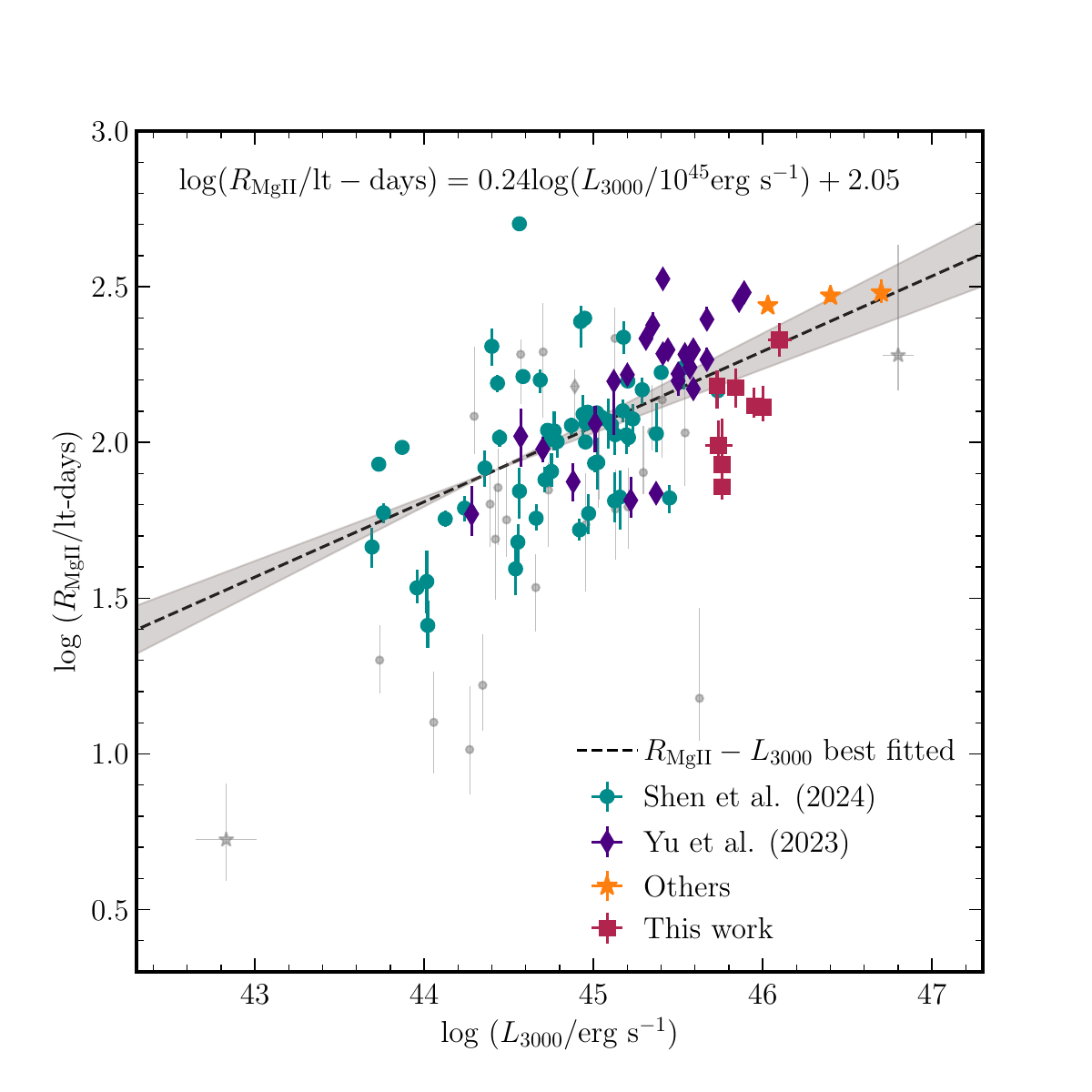}
\caption{$R_{\rm MgII}-L_{3000}$ relation. The red squares represent the measurements from this paper. The green circles and purple diamonds correspond to the SDSS sample from \cite{Shen2024} and the OzDES sample from \cite{Yu2023}, respectively. The orange stars indicate additional measurements collected from \cite{MOP2006}, \cite{Lira2018}, \cite{Zajacek2020}, \cite{Zajacek2021}, and \cite{Prince2022}. Colors of objects rejected for their large measurement errors are replaced with gray (see Section \ref{sec_rl}). The black dashed line represents the best-fitting $R_{\rm MgII}-L_{3000}$ relation, while the gray shaded region indicates the 1$\sigma$ confidence band derived from the LINMIX algorithm.}
\label{fig_RL}
\end{figure*}

We present the $R_{\rm MgII}-L_{3000}$ relation for our sample in conjunction
with previous works in Figure \ref{fig_RL}. For the SDSS-RM project, we did not adopt the entire Mg {\sc ii} sample from \cite{Shen2024} due to their relatively larger measurement errors in time lags. Instead, we selected objects based on the same criteria used in the present study. Specifically, we only included objects from \cite{Shen2024} whose peak correlation coefficients from the ICCF are greater than 0.5 ($r_{\rm max} > 0.5$). Even after applying this criterion, the error bars for the time lags of some remaining objects are still significantly larger than those of the SEAMBH sample in this paper. 
We further excluded objects with lag uncertainties greater than 1.5 times the mean lag uncertainties of the SEAMBH sample in logarithmic space (relative errors). We also verified that employing a slightly softer criterion does not alter the conclusions of subsequent analysis. More details can be found in Appendix \ref{appB}.

It is evident that our results
significantly augment the sample at high luminosities. To parameterize the
$R_{\rm MgII}-L_{3000}$ relation, we adopt the following formalism:
\begin{equation}
    \log\left(\frac{R_{\rm MgII}}{\text{lt-days}}\right) = 
        \alpha \log\left(\frac{L_{3000}}{10^{45}~{\rm erg~s^{-1}}}\right) + \beta,
\end{equation}
where $\alpha$ and $\beta$ represent the slope and intercept of the
relation. We perform the linear regression using the
LINMIX\footnote{\url{https://github.com/jmeyers314/linmix}} algorithm
\citep{Kelly2007}, which employs a hierarchical Bayesian approach to linear
regression while accounting for measurement errors in both the \(X\) and \(Y\)
axes. An intrinsic random scatter, \(\sigma_{\rm int}\), is also included in the model. Combining all available samples from the literature with our own, the
best-fitting parameters are \(\alpha = 0.24 \pm 0.03\), \(\beta = 2.05 \pm
0.02\), and \(\sigma_{\rm int} = 0.04 \pm 0.01\). 

The total scatter is defined by 
\begin{equation}
   \sigma_{R_{\rm MgII}-L_{3000}} = \sqrt{\frac{1}{N_{\rm s}}\sum^{N_{\rm s}}_{i=1} 
             \left( \log~R_{\rm MgII}^i - \log~R_{R_{\rm MgII}-L_{3000}}^i \right)^2},
\end{equation}
where $N_{\rm s}$ is the sample size, $i$ is the index of the object, and
$R_{R_{\rm MgII}-L_{3000}}$ is the predicted radius from the best-fitting
relation. Given the best parameters of the linear regression, the total 
scatter of the current sample is $\sigma_{R_{\rm MgII}-L_{3000}} = 0.20$.

The slope of the ${R_{\rm MgII}-L_{3000}}$ relation reported by the OzDES-RM project \citep{Yu2023}, which includes the "gold sample" from \cite{Homayouni2020}, is $0.39 \pm 0.08$. In contrast, the slope reported by the SDSS-RM project is $0.31 \pm 0.06$ \citep{Shen2024}. After incorporating our samples, the slope decreases slightly but remains consistent with the values reported by \cite{Yu2023} and \cite{Shen2024} within the uncertainties. Furthermore, the intercept of the ${R_{\rm MgII}-L_{3000}}$ relation reported here is also consistent with the values found in these references.

\subsection{Dependency on Accretion Rate}
\label{sec_accretion_rate}

For AGNs exhibiting higher accretion rates, the observed H$\beta$ time lags are
systematically shorter than those predicted by the standard $R_{\rm
H\beta}$--$L_{5100}$ relation \citep[e.g.,][]{Du2015, Du2016, Du2018}. While
still preliminary, analogous behavior has been reported for Mg\,\textsc{ii}
emission lines in the context of the $R_{\rm MgII}$--$L_{3000}$ relation by
\cite{MA2020} and \cite{Yu2021}. The targets investigated in this study are
candidate super-Eddington accretors, thereby constituting an ideal sample for
probing this phenomenon.

\begin{figure*}
\centering
\includegraphics[width=\textwidth]{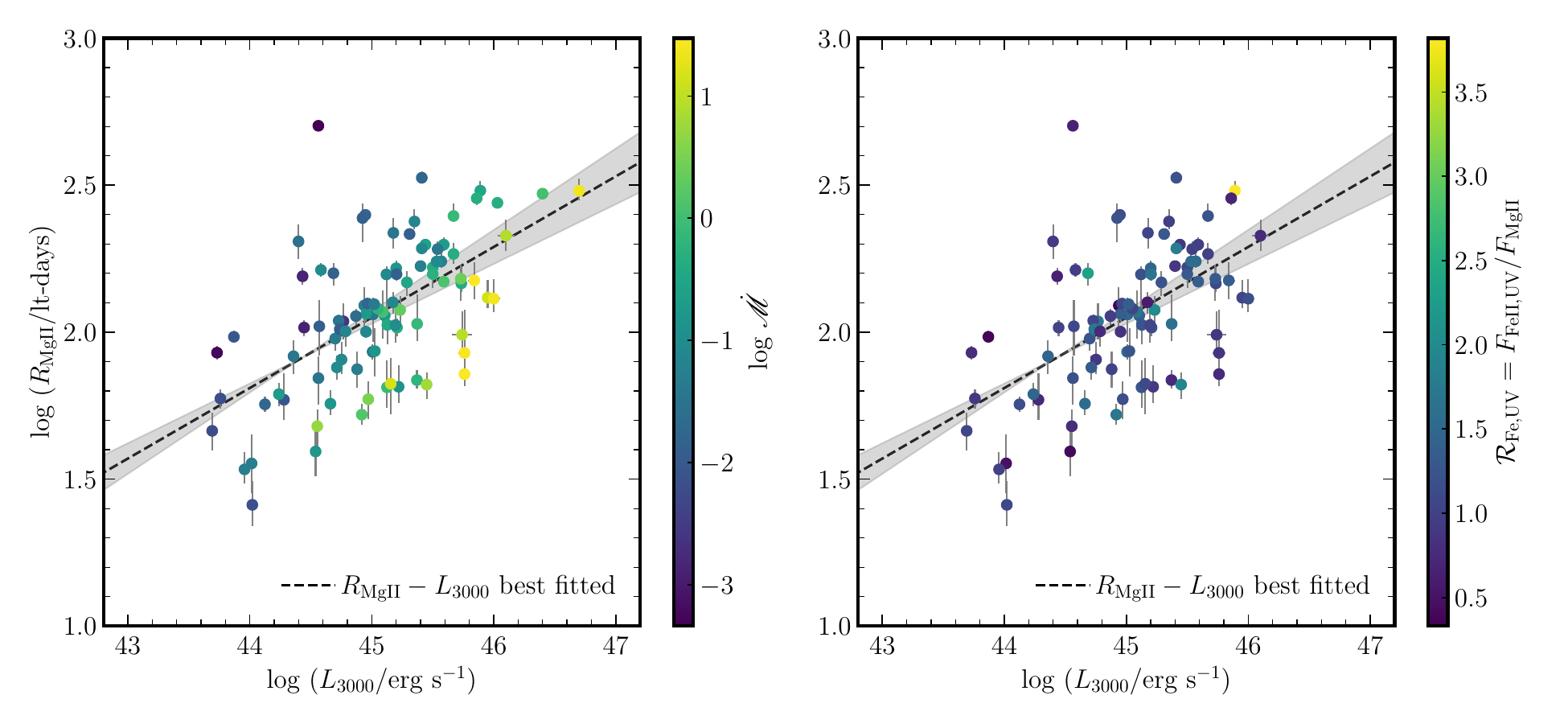}
\caption{$R_{\rm MgII}-L_{3000}$ relation color-coded by both the dimensionless accretion rate $\dot{\mathscr{M}}$ (left panel) and the UV iron strength $\mathcal{R}_{\rm Fe,UV}$ (right panel). The dashed line and shaded region have the same meaning as in Figure~\ref{fig_RL}, representing the best-fit relation and its $1\sigma$ uncertainty, respectively. Three targets not from the SDSS and OzDES samples are absent for the $\mathcal{R}_{\rm Fe}^{\rm UV}$ measurement, as we do not obtain their corresponding spectra for a unified analysis, consistent with the presentation in subsequent figures.}
\label{fig_RL_colorcode}
\end{figure*}

As shown in Figure~\ref{fig_RL}, our targets systematically lie below the
best-fit $R_{\rm MgII}$--$L_{3000}$ relation compared to other AGNs. To
further illustrate this trend, Figure~\ref{fig_RL_colorcode} displays a
color-coded distribution of sources in the $R_{\rm MgII}$--$L_{3000}$ plane,
clearly indicating that AGNs with higher accretion rates deviate significantly
below the canonical best-fit relation. To quantify this offset, we define the deviation
parameter:
\begin{equation}
    \Delta R_{\rm MgII} = \log R_{\rm MgII} - \log R_{R_{\rm MgII}-L_{3000}},
\end{equation}
where $R_{R_{\rm MgII}-L_{3000}}$ is the expected radius derived from the
best-fit $R_{\rm MgII}$--$L_{3000}$ relation. The errors of $\Delta R_{\rm MgII}$ are merely from errors of time delays and any other errors are not considered.  Figure~\ref{diff_Mdot_Rfe}
presents the correlation between $\Delta R_{\rm MgII}$ and the dimensionless
accretion rate $\dot{\mathscr{M}}$, revealing a clear anti-correlation: higher
accretion rates correspond to more negative $\Delta R_{\rm MgII}$ values. The Pearson correlation coefficient ($\rho$) for the correlation is -0.51, with the corresponding two-sided $p$-value for a null hypothesis test of $2.8\times10^{-7}$.
The Mg {\sc ii} time lags in SEAMBHs can be shortened by nearly an order of 
magnitude compared to those in AGNs with the lowest accretion rates. 
This behavior closely mirrors the shortened H$\beta$ lags reported in \cite{Du2015,
Du2016, Du2018}. Since the accretion rate depends on the lag measurement, larger values of $\dot{\mathscr{M}}$ may arise from shorter measured lags, thus leading to spurious relevance. To address this, we conduct a partial correlation analysis with respect to the lag, which yields a more robust anti-correlation. The results of the partial correlation analysis are presented in Appendix \ref{appC}.

\subsection{Correlation with UV Iron Strength}
\label{sec:Rfe}
The relative strength of optical iron emission, quantified as $\mathcal{R}_{\rm
Fe}^{\rm opt} = F_{\rm FeII}^{\rm opt}/F_{\rm H\beta}$, has been established as
a reliable indicator of accretion rate in AGNs \citep[e.g.,][]{boroson1992,
Boroson2002, marziani2001, shen2014}. Building upon this, \cite{DW2019}
developed an improved $R_{\rm H\beta}$--$L_{5100}$ relation by incorporating
$\mathcal{R}_{\rm Fe}^{\rm opt}$, which significantly reduces the scatter in
time lag estimates. 

 To investigate whether a similar correlation exists for the UV iron emission, we define the UV iron strength parameter: 
\begin{equation} 
    \mathcal{R}_{\rm Fe}^{\rm UV} = \frac{F_{\rm FeII}^{\rm UV}}{F_{\rm MgII}}, 
\end{equation} where
 $F_{\rm MgII}$ is the Mg\,\textsc{ii}
line flux, obtained by integrating the residual spectra from spectral fitting within 2700-2900\,\AA\ in the rest frame.
$F_{\rm FeII}^{\rm UV}$ represents the integrated iron flux between
2250--2650\,\AA\ in the rest frame. For objects lacking spectral
coverage in this wavelength range, we employed extrapolation using our iron
template during the spectral fitting process.

For the measurement of $\mathcal{R}_{\rm Fe}^{\rm UV}$, we performed spectral fitting following the procedure described in Section~\ref{sec_lightcurve}, but extended the fitting window to cover 2260--3400\,\AA. This broader range helps mitigate degeneracy between the UV iron emission and the power-law continuum, leading to a more robust determination of the Fe\,\textsc{ii} strength. For the Mg\,\textsc{ii} light curves and lag measurements, however, we found that expanding the window has no significant effect. As demonstrated in Appendix \ref{app_window}, the narrower window produces light curves with marginally lower scatter while yielding Mg\,\textsc{ii} time lags fully consistent with those from the broader window. We therefore retained the narrower window for light-curve construction, but used the extended window exclusively for measuring $\mathcal{R}_{\rm Fe}^{\rm UV}$.

The final $\mathcal{R}_{\rm Fe}^{\rm UV}$ 
values represent the mean measurements across all available epochs, with 
uncertainties calculated as:
\begin{equation} 
    \sigma_{\mathcal{R}_{\rm Fe}^{\rm UV}} =
        \frac{S_{\mathcal{R}_{\rm Fe}^{\rm UV}}}{\sqrt{N}}, 
\end{equation}
where $S_{\mathcal{R}_{\rm Fe}^{\rm UV}}$ is the standard deviation and $N$ is
the number of epochs. Both the dimensionless accretion rates
$\dot{\mathscr{M}}$ and UV iron strengths $\mathcal{R}_{\rm Fe}^{\rm UV}$ for
our sample are presented in Table~\ref{measurement}.

\begin{figure*}
\centering
\includegraphics[width=\columnwidth]{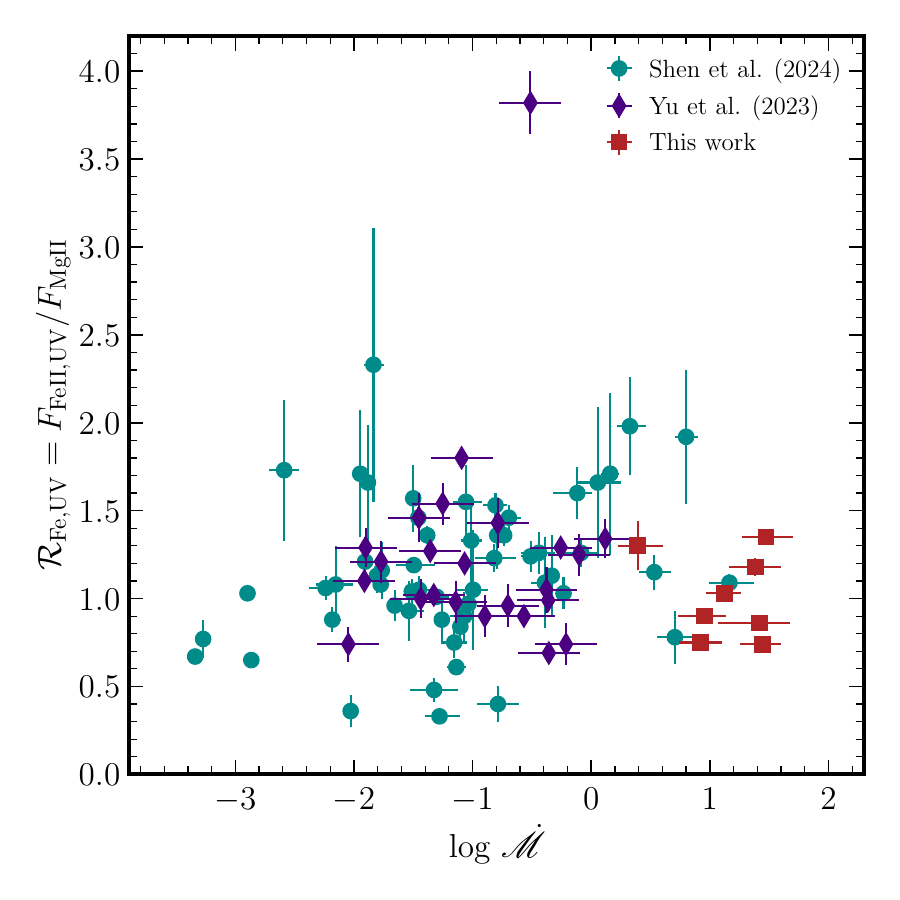}
\caption{Correlation of the UV iron strength $\mathcal{R}_{\rm Fe,UV}$ with the dimensionless accretion rate $\dot{\mathscr{M}}$.}
\label{fig_Mdot_Rfe}
\end{figure*}

Given that different studies have adopted distinct iron templates and spectral fitting windows to measure $\mathcal{R}_{\rm Fe}^{\rm UV}$ (e.g., \citealt{Yu2023}; \citealt{Shen2024}), a direct combination of their values could introduce systematic biases. To ensure consistency, we collected the mean spectra from the RM campaigns presented in \cite{Yu2023} and \cite{Shen2024}, and re-measured their $\mathcal{R}_{\rm Fe}^{\rm UV}$ values using the same iron template and fitting window as applied to our own sample. Similarly, we present the $R_{\rm MgII}-L_{3000}$ relation color-coded by
$\mathcal{R}_{\rm Fe}^{\rm UV}$, along with the correlation between $\Delta
R_{\rm MgII}$ and $\mathcal{R}_{\rm Fe}^{\rm UV}$ in Figures
\ref{fig_RL_colorcode} and \ref{diff_Mdot_Rfe}, respectively. Additionally, we present the correlation between $\mathcal{R}_{\rm Fe}^{\rm UV}$ and $\dot{\mathscr{M}}$ in Figure \ref{fig_Mdot_Rfe}. The Pearson correlation coefficient for $\Delta R_{\rm MgII}$--$\mathcal{R}_{\rm Fe}^{\rm UV}$ correlation is 0.06, with the corresponding two-sided $p$-value for a null hypothesis test of 0.59. The
anticorrelation between $\Delta R_{\rm MgII}$ and $\mathcal{R}_{\rm Fe}^{\rm
UV}$ is not as significant as the anticorrelation between $\Delta R_{\rm MgII}$
and $\dot{\mathscr{M}}$, which is consistent with the findings of
\cite{Yu2023}. A potential explanation is that the correlation between the accretion rate (or Eddington ratio) and $\mathcal{R}_{\rm Fe}^{\rm UV}$ is relatively weaker than the correlation between the accretion rate (or Eddington ratio) and $\mathcal{R}_{\rm Fe}^{\rm opt}$ \citep[e.g.,][]{MA2020,Khadka2022,  pan2025}. This is further supported by the substantial scatter in the correlation between $\mathcal{R}_{\rm Fe}^{\rm opt}$ and $\mathcal{R}_{\rm Fe}^{\rm UV}$, as well as their equivalent widths \citep[e.g.,][]{Kovacevic2015, pan2025}. Additionally, the current Mg\,\textsc{ii} RM sample remains relatively small and likely incomplete; expanding this sample in future campaigns will be essential to clarify these trends. A larger sample with better quality may enable a more thorough investigation of the $\Delta R_{\rm MgII}$--$\mathcal{R}_{\rm Fe}^{\rm UV}$
correlation.
\section{Discussions}
\label{sec_dis}

\subsection{Mg {\sc ii} Lag vs. H$\beta$ Lag}
\label{sec_MgHbeta}

\begin{figure*}
\centering
\includegraphics[width=\columnwidth]{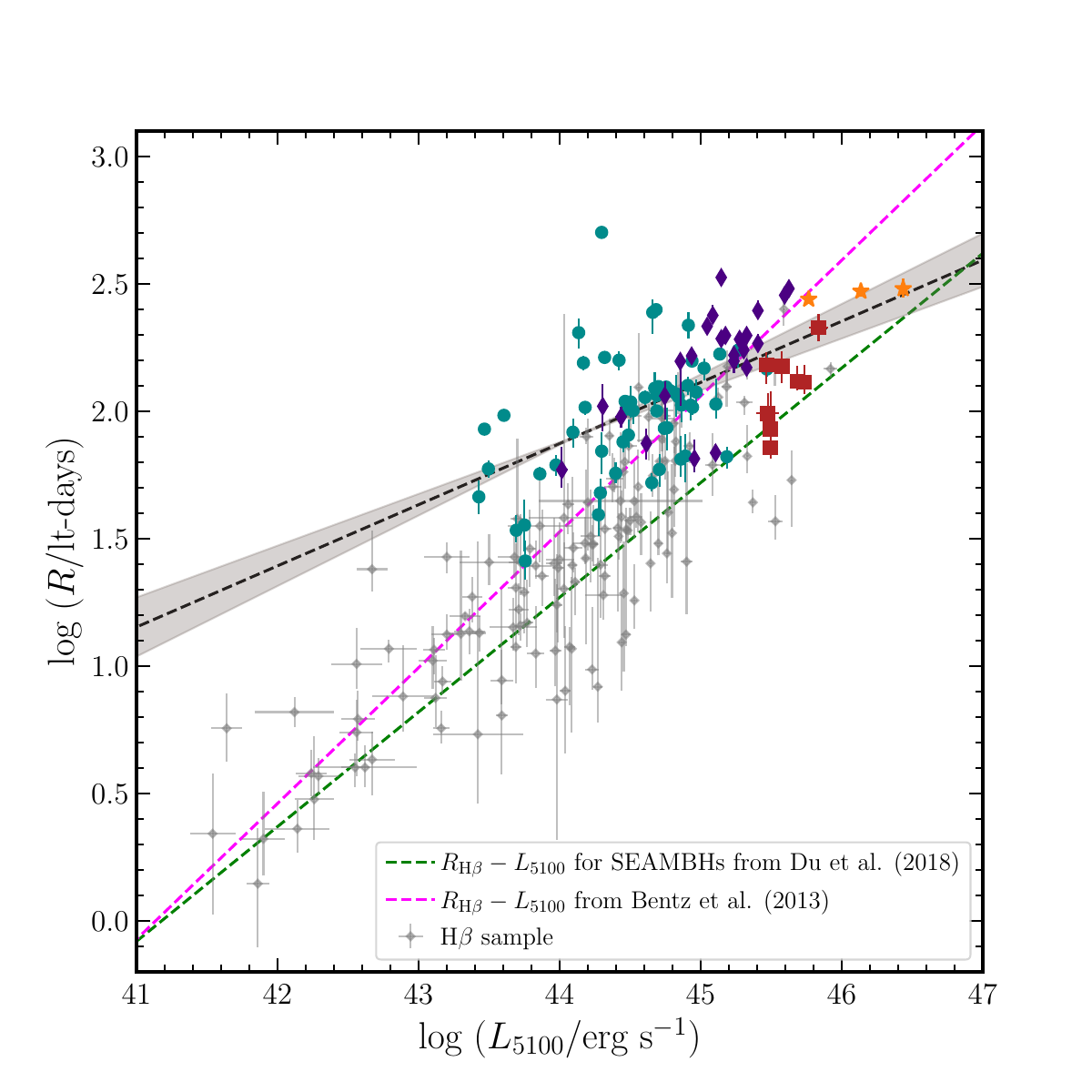}
\caption{Comparison between Mg {\sc ii} and H$\beta$ time lags. Symbols follow the same color-coding as in Figure~\ref{fig_RL}, with gray diamonds representing H$\beta$ lags. The black dashed line and gray shaded region show the best-fit $R_{\rm MgII}-L_{5100}$ relation and its uncertainties in Figure~\ref{fig_RL}. The magenta dashed line shows the $R_{\rm H\beta}-L_{5100}$ relation from \cite{Bentz2013}, while the green dashed line shows the $R_{\rm H\beta}-L_{5100}$ relation for SEAMBHs from \cite{Du2018}.}
\label{fig_RL_hb}
\end{figure*}

Given the lower ionization potential of Mg$^+$ (7.6 eV), it is expected that the time lags of Mg {\sc ii} lines should be longer than those of H$\beta$ lines in AGNs with the same luminosities. Here, we present a comparison between the Mg {\sc ii} and H$\beta$ lags in Figure~\ref{fig_RL_hb}. The $L_{3000}$ values for the Mg {\sc ii} sample were converted to $L_{5100}$ using the flux ratio at these two wavelengths from the quasar mean spectrum provided in \cite{Richards2006}. 

For the H$\beta$ lags, we utilized the compilation from \cite{DW2019} (the collection was updated by \citealt{chen2023} based on more recent observations), which includes samples from \cite{Peterson1998}, \cite{Kaspi2000}, \cite{Du2014}, \cite{Bentz2009}, and others, as well as data from some Palomar-Green quasars presented in \cite{Hu2021} and low-redshift galaxies from \cite{U2022}. Additionally, we augmented this collection with more recent results from \cite{Woo2024}\footnote{Seven objects with unreliable measurements are excluded; see further details in \cite{Woo2024}.} and \cite{Hu2025}.
The $R_{\rm H\beta}-L_{5100}$ relation from \cite{Bentz2013} and the relation for SEAMBHs from \cite{Du2018} are also shown in Figure~\ref{fig_RL_hb}. It is evident that the Mg {\sc ii} lags are indeed longer than the H$\beta$ lags. Assuming that the Mg {\sc ii} and H$\beta$ emitters at the same luminosities have the same SMBH masses, the line widths of Mg {\sc ii} should be narrower than those of H$\beta$, which is consistent with observational studies regarding line widths \citep[e.g.,][]{Hu2008, Wang2009, Shen2012, Marziani2013, Mejia-Restrepo2016}.

Furthermore, it is noteworthy that, although the luminosity coverage of the present Mg {\sc ii} sample is still limited, the slope of the $R_{\rm MgII}-L_{5100}$ relation is shallower than that of the $R_{\rm H\beta}-L_{5100}$ relation. This trend may be associated with the fact that the Mg {\sc ii} line widths are more consistent with those of H$\beta$ when the line widths are narrower \citep[e.g.,][]{Wang2009, Marziani2013}. Larger samples of high-quality Mg {\sc ii} measurements across a wide luminosity range and more detailed BLR dynamical modeling are needed for further investigation of this point in the future.

\subsection{Physical Mechanism for Shortened Lags}
\label{sec_phy}

\begin{figure*}
\centering
\includegraphics[width=\textwidth]{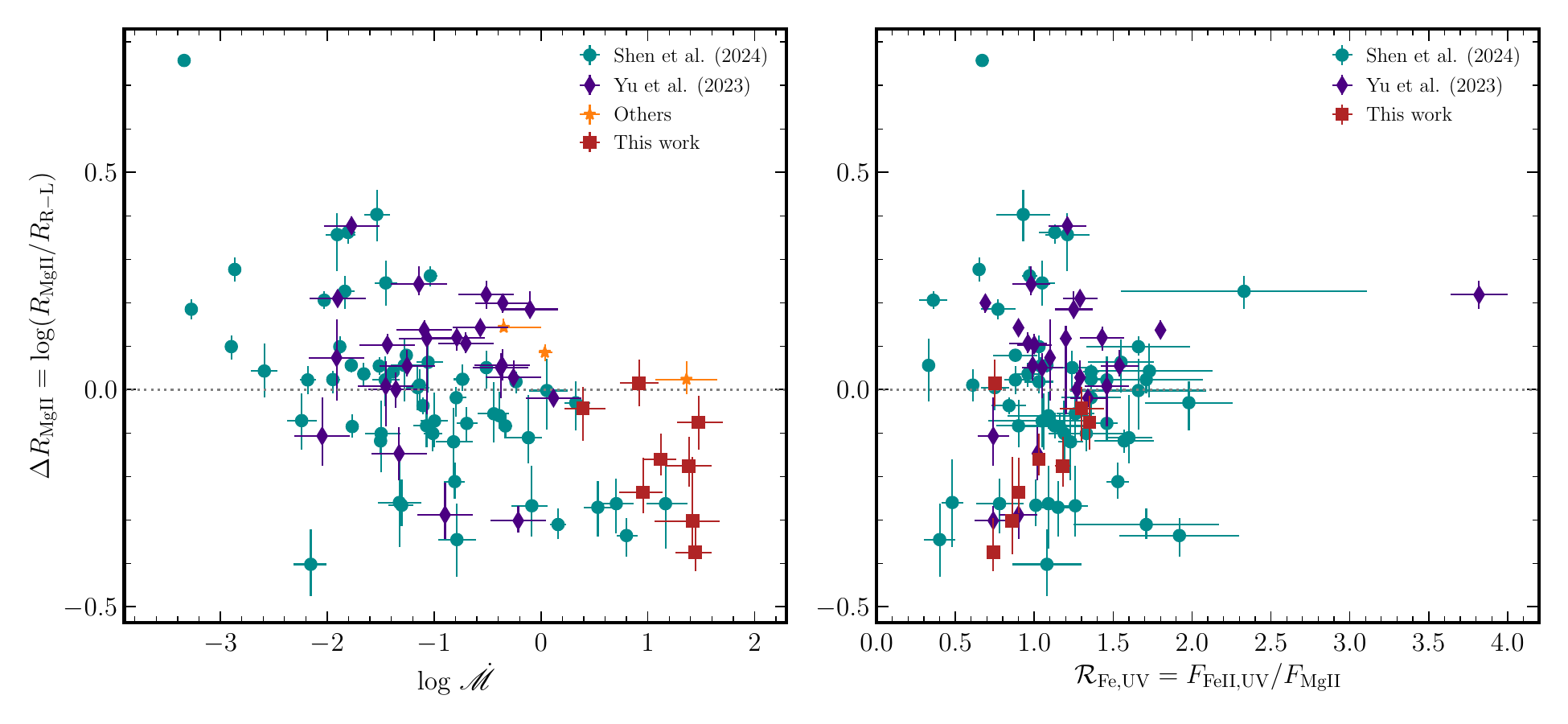}
\caption{Correlations of $\Delta R_{\rm MgII}$ with the dimensionless accretion rate $\dot{\mathscr{M}}$ (left panel) and the UV iron strength $\mathcal{R}_{\rm Fe,UV}$ (right panel). $\Delta R_{\rm MgII}$ represents the residual Mg {\sc ii} lags relative to the $R_{\rm MgII}-L_{3000}$ relation. The black dotted line indicates the locus where the observed Mg {\sc ii} lags exactly match the predictions from the $R_{\rm MgII}-L_{3000}$ relation.}
\label{diff_Mdot_Rfe}
\end{figure*}

It has been demonstrated that AGNs with high accretion rates exhibit shorter Mg {\sc ii} time lags compared to AGNs with normal accretion rates (see Figure \ref{diff_Mdot_Rfe}). One possible explanation for this phenomenon is the self-shadowing effects of slim accretion disks in SEAMBHs \citep{Wang2014}. The inner region of the accretion disk is puffed up, thus the ionizing radiation emitted from the innermost region is shielded and cannot reach the gas in the BLRs. This results in a reduced radius of the BLRs that respond to the varying continuum. Furthermore, due to the lower ionization potential of Mg$^+$, the radius of the accretion disk emitting photons that ionize Mg {\sc ii} is larger than that for H$\beta$. Consequently, the degree of self-shadowing for Mg {\sc ii} lines may be slightly weaker than for H$\beta$ lines. However, as shown in Figure \ref{diff_Mdot_Rfe}, the shortening of Mg {\sc ii} time lags remains significant. More detailed calculations comparing the self-shadowing effects in Mg {\sc ii} and H$\beta$, as well as comparisons with observational data, are needed in future studies.

\section{Summary}
\label{sec_summary}

In this 15th paper of the SEAMBH series, we present the results of a 7-year reverberation mapping (RM) campaign that monitored the Mg {\sc ii} emission lines in eighteen high-luminosity SEAMBHs, conducted from 2017 to 2024 using the Lijiang 2.4 m and CAHA 2.2 m telescopes. The key findings of this study can be summarized as follows:

\begin{enumerate}
    \item We successfully measured the Mg {\sc ii} time lags for eight of the eighteen objects. By incorporating the measurements of their line widths, we were able to determine the masses of their central SMBHs. The Mg {\sc ii} time lags typically range from 100 to 300 days in the observed frames, while the SMBH masses span a range of $8.1 < \log(M_\bullet/M_\sun) < 8.7$.
    
    \item By combining our objects with previous Mg {\sc ii} RM samples, we updated the $R_{\rm MgII}-L_{3000}$ relation. The slope and intercept of the $R_{\rm MgII}-L_{3000}$ relation obtained here are $\alpha = 0.24 \pm 0.03$ and $\beta = 2.05 \pm 0.02$, respectively, which are consistent with those reported in the existing literature.
    
    \item We found that SEAMBHs exhibit shortened Mg {\sc ii} time lags compared to AGNs with normal accretion rates. This phenomenon closely mirrors the shortened H$\beta$ lags reported in SEAMBHs. Furthermore, we examined the relationship between the deviation in the $R_{\rm MgII}-L_{3000}$ relation and the accretion rate, as well as the relative strength of UV iron emissions. The deviation shows a clear correlation with the accretion rate but no significant correlation with the strength of UV iron emissions. 
\end{enumerate}

In the future, larger samples of high-quality Mg {\sc ii} RM data will not only facilitate a more precise refinement of the $R_{\rm MgII}-L_{3000}$ relation but also deepen our understanding of the accretion physics in SMBHs. This will ultimately allow us to establish more reliable SMBH mass indicators for AGNs at high redshifts and enhance our comprehension of the formation and evolution of SMBHs in the early universe.

\section*{acknowledgements}

We thank Zhefu Yu for kindly sharing the mean spectra of OzDES RM objects and the anonymous referee for useful comments. We acknowledge the support of the staff at the Lijiang 2.4 m telescope. Funding for the telescope has been provided by the Chinese Academy of Sciences (CAS) and the People's Government of Yunnan Province. This work is also based on observations collected at the Centro Astron\'omico Hispano en Andaluc\'ia (CAHA) at Calar Alto, which is operated jointly by the Andalusian Universities and the Instituto de Astrof\'isica de Andaluc\'ia (CSIC). This research is supported by the National Key R\&D Program of China (2021YFA1600404 and 2023YFA1607903), the National Science Foundation of China through grants NSFC-12521005, -12333003, -12122305, -12022301, -11991050, and -11833008, as well as the National Science and Technology Major Project (2024ZD0300303). Additionally, it is funded by the China Manned Space Project (CMS-CSST-2025-A07 and CMS-CSST-2021-A06). L.C.H. was supported by the National Science Foundation of China (12233001) and the China Manned Space Program (CMS-CSST-2025-A09).

This work used archival data obtained with the Samuel Oschin Telescope 48-inch
and the 60-inch Telescope at the Palomar Observatory as part of the Zwicky
Transient Facility project. ZTF is supported by the National Science Foundation
under Grants No.AST-1440341 and AST-2034437 and a collaboration including
current partners Caltech, IPAC, the Weizmann Institute for Science, the Oskar
Klein Center at Stockholm University, the University of Maryland, Deutsches
Elektronen-Synchrotron and Humboldt University, the TANGO Consortium of Taiwan,
the University of Wisconsin at Milwaukee, Trinity College Dublin, Lawrence
Livermore National Laboratories, IN2P3, University of Warwick, Ruhr University
Bochum, Northwestern University and former partners the University of
Washington, Los Alamos National Laboratories, and Lawrence Berkeley National
Laboratories. Operations are conducted by COO, IPAC, and UW.

\software{{\textsc{linmix}}(\citealt{Kelly2007}),
{\textsc{PyCALI}} (\citealt{Li2014,li2024pycali}),
{\textsc{MICA}} (\citealt{Li2016,li2024mica}),
{\textsc{DASpec}} (\citealt{Du2024})
}

\appendix

\section{Photometric light curves of Comparison stars}\label{appD}
For the comparison stars, their photometric light curves are presented in Figure \ref{fig_comparison_lc}. The magnitudes are shown after subtracting their mean values (differential magnitudes). During the monitoring period, the variations of comparison stars were consistently low.

\section{The choice of iron template}\label{app_template}
 
\begin{figure*}
\centering
\includegraphics[width=\columnwidth]{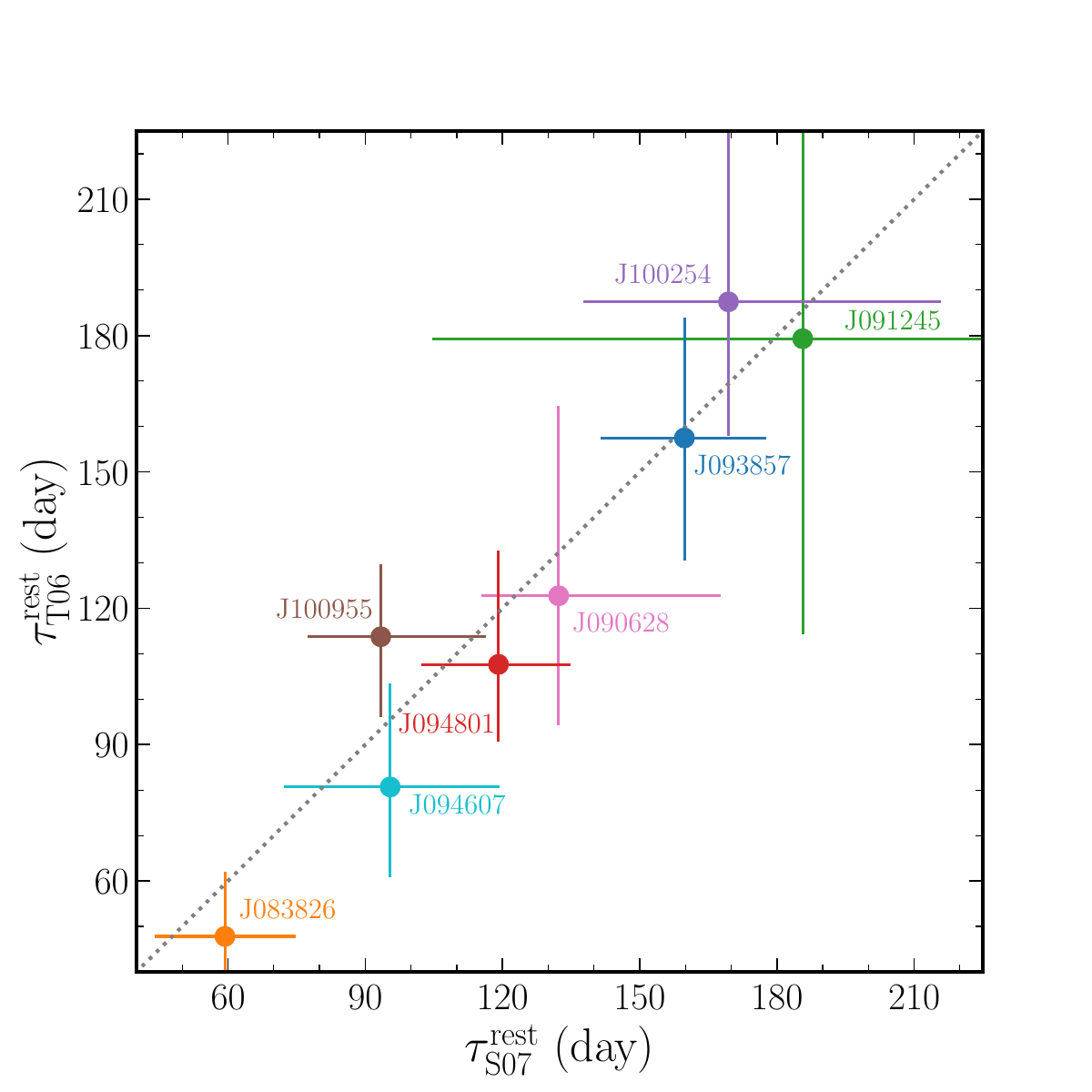}
\caption{Comparison between the lag in the rest frame derived from the light curve by employing \cite{Salviander2007} 
($\tau_{\rm S07}^{\rm rest}$) and \cite{Tsuzuki2006} ($\tau_{\rm T06}^{\rm rest}$) iron templates , via ICCF method.}
\label{fig:lag_Template}
\end{figure*}

Discrepancies in Mg\,\textsc{ii} and Fe\,\textsc{ii} flux measurements arising from different iron templates have been reported in the literature \citep[e.g.,][]{Woo2018, Shin2019}. In our analysis, the iron template used in the main text is the composite template compiled by \cite{Shen2019}, which incorporates the \cite{Salviander2007} template across the Mg\,\textsc{ii} wavelength region (see Section \ref{sec_lightcurve}). To examine whether our light curves and lag measurements are sensitive to the choice of iron template, we performed a test by replacing the \cite{Salviander2007} segment with the \cite{Tsuzuki2006} template. The latter is based on synthetic spectra from photoionization calculations and has been recommended as an improved template \citep[e.g.,][]{Wang2009}. We re-derived the continuum and Mg\,\textsc{ii} light curves and remeasured the time lags using the ICCF method (Section~\ref{sec_lag}) for the eight objects with previously reliable lag measurements. The resulting light curves are shown in Figure~\ref{fig:compare_Template}, and a comparison of the Mg\,\textsc{ii} lags obtained with the \cite{Salviander2007} and \cite{Tsuzuki2006} templates is presented in Figure~\ref{fig:lag_Template}. We find that the time lags derived from the two templates are consistent within the uncertainties, in agreement with a similar test presented by \cite{Yu2021}.

\begin{figure*}
\centering
\includegraphics[width=\columnwidth]{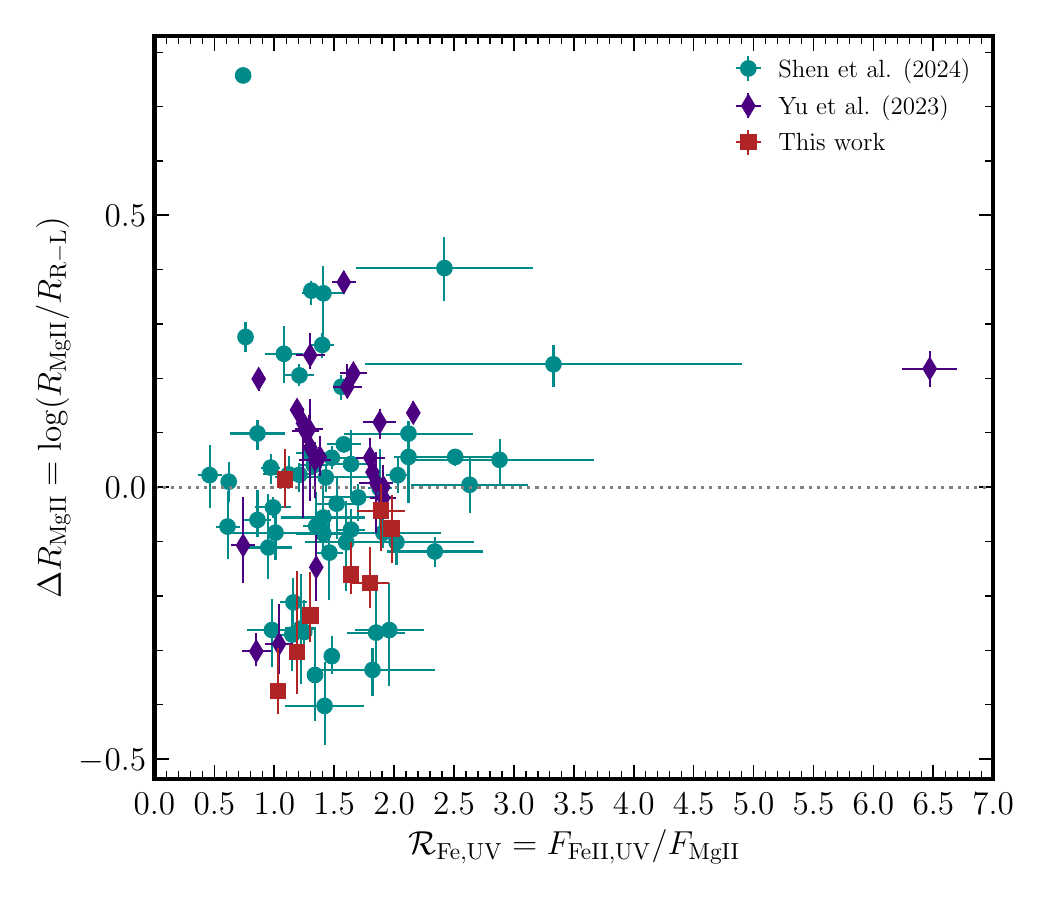}
\caption{Correlation of $\Delta R_{\rm MgII}$ with the UV iron strength $\mathcal{R}_{\rm Fe,UV}$ by employing iron template from \cite{Tsuzuki2006}. The figure is plotted in the same manner as Figure \ref{diff_Mdot_Rfe}.}
\label{fig_diff_Rfe_t06}
\end{figure*}

Regarding $\mathcal{R}_{\rm Fe}^{\rm UV}$, we also tested the results using template from \cite{Tsuzuki2006}. In this case, the relation between $\mathcal{R}_{\rm Fe}^{\rm UV}$ and the deviation parameter $\Delta R_{\rm MgII}$ is shown in Figure \ref{fig_diff_Rfe_t06}. The correlation remains statistically ambiguous.

\section{Light Curves of Objects with Unsuccessful Lag Measurements}\label{app}

We present the light curves of ten targets for which reliable lag measurements could not be obtained in Figure \ref{fig_con_lc}. During our time-series analysis, the peak values of the cross-correlation functions ($r_{\rm max}$) for these light curves did not exceed 0.5.

\section{Refine Samples Based on Lag Measurement Errors}\label{appB}
The error of the time lag in logarithmic space is defined as
\begin{equation}
    \sigma_{\rm log \tau} = \frac{1}{\ln 10} \cdot \frac{\sigma_{\tau}}{\tau},
\end{equation}

where $\sigma_{\tau}$ is the measurement error of the time lag in linear space. We calculate the mean error of lag in logarithmic space for SEAMBH sample in this paper as $\sigma_{\rm critical}=0.068$. For all available samples from the literature (where the sample from \citealt{Shen2024} is limited to objects with $r_{\rm max} > 0.5$), objects with errors in logarithmic space larger than 1.5$\sigma_{\rm critical}$ are discarded. We have verified that the results of the present paper do not change significantly if we instead discard objects  whose errors exceed 2$\sigma_{\rm critical}$.

\section{Partial Correlation Analysis for Dependence on Accretion Rate}\label{appC}

An artificial anti-correlation between $\Delta R_{\rm MgII}$ and $\dot{\mathscr{M}}$ may arise because $\dot{\mathscr{M}}$ is anti-correlated with the time lag, while $\Delta R_{\rm MgII}$ is positively correlated with the time lag. To investigate whether the observed anti-correlation between $\Delta R_{\rm MgII}$ and $\dot{\mathscr{M}}$ is due to this effect, we conducted a partial correlation analysis. The results of the partial correlation analysis are shown in Table \ref{tab_partial}. The correlation coefficient between $\Delta R_{\rm MgII}$ and $\dot{\mathscr{M}}$ is -0.51, while the partial correlation coefficient with respect to the time lag $\tau_{\rm MgII}$ is $-0.85$, indicating a more significant anti-correlation.

Furthermore, $\dot{\mathscr{M}}$ is also positively correlated with luminosity. We conducted another partial correlation analysis between $\Delta R_{\rm MgII}$ and $\dot{\mathscr{M}}$ while controlling for $L_{3000}$. The anti-correlation remains substantial, as the partial correlation coefficient is $-0.70$.

\section{The fitting window for Mg {\sc ii} light curve measurement}\label{app_window}

\begin{figure*}
\centering
\includegraphics[width=\columnwidth]{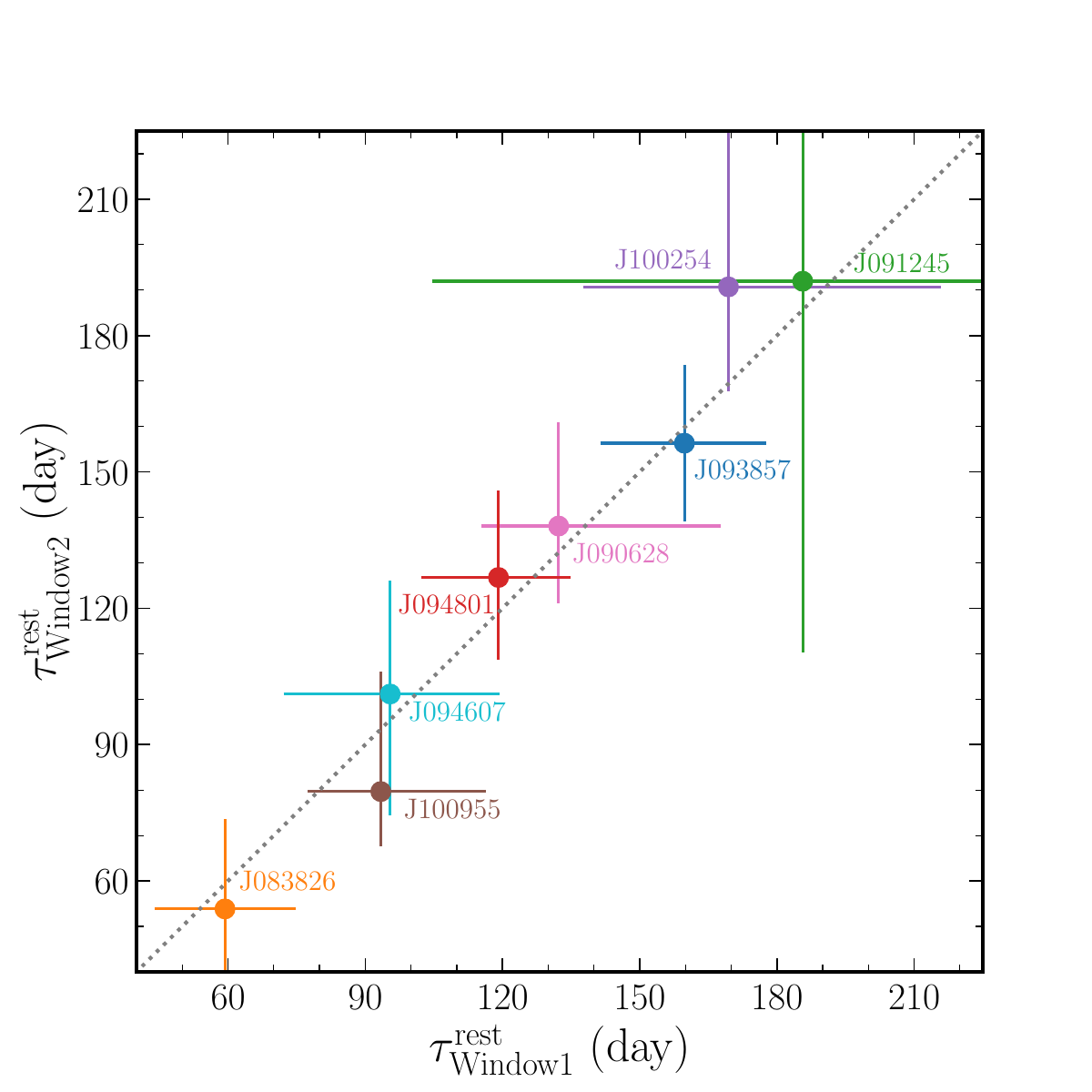}
\caption{Comparison between the lag in the rest frame derived from the light curve using fitting range Window 1 
($\tau_{\rm Window1}^{\rm rest}$) and Window 2 ($\tau_{\rm Window2}^{\rm rest}$), via ICCF method.}
\label{fig:lag_range}
\end{figure*}

In measuring $\mathcal{R}_{\rm Fe}^{\rm UV}$, we employed a broader fitting window of 2260--3400\,\AA. For the Mg\,\textsc{ii} flux measurements used in constructing light curves, however, we adopted a narrower range of 2260--3050\,\AA. To assess the sensitivity of our results to this choice, we compared the Mg\,\textsc{ii} light curves and time lags derived using the two windows: the narrower Window 1 (2260--3050\,\AA) and the broader Window 2 (2260--3400\,\AA). As shown in Figures~\ref{fig:compare_range} and~\ref{fig:lag_range} for the eight targets with reliable lag measurements, the resulting light curves and ICCF-derived time lags are generally consistent within uncertainties. We note that the narrower window yields several light curves with marginally reduced scatter (see, e.g., J093857). Therefore, we adopted the results from the narrower fitting window as our final measurements in the main text.


% \newpage

\begin{figure*}
\centering
\includegraphics[width=\textwidth]{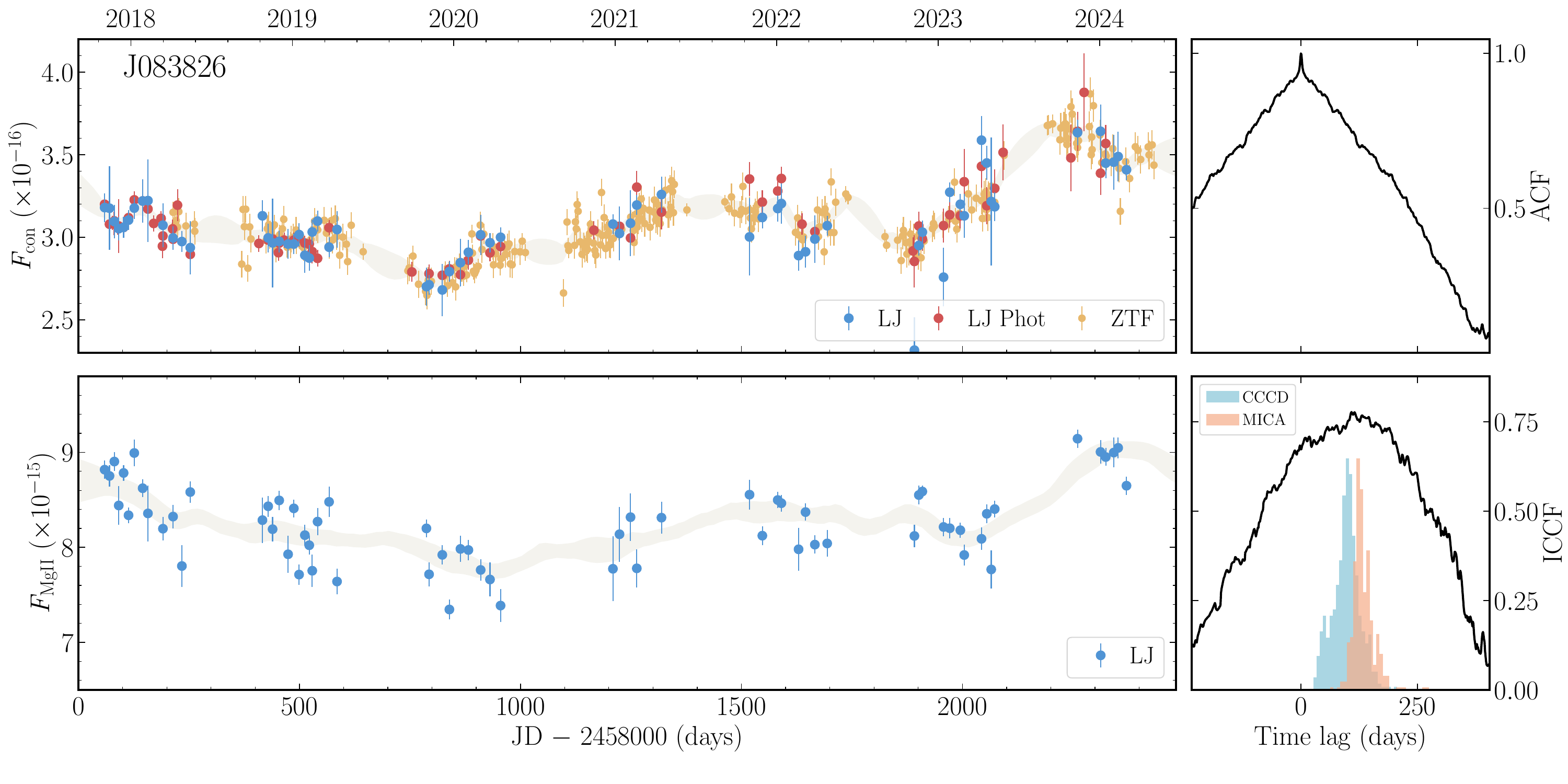}
\caption{Light curves and time-lag measurements. The left panels show the 3000 \AA\ continuum light curves after intercalibration alongside the Mg {\sc ii} light curves. ``LJ/CAHA'', ``LJ/CAHA Phot'', and ``ZTF'' represent the Lijiang/CAHA spectroscopic data, Lijiang/CAHA photometric data, and ZTF photometric data points, respectively. The continuum light curves are in units of $10^{-16} \, {\rm erg} \, {\rm s}^{-1} \, {\rm cm}^{-2} \, \text{\AA}^{-1}$, while the Mg {\sc ii} light curves are in units of $10^{-15} \, {\rm erg} \, {\rm s}^{-1} \, {\rm cm}^{-2}$. The gray shaded regions represent the reconstructions from the MICA model, incorporating additional systematic errors that were automatically included during the reconstruction process.
The upper-right panel displays the auto-correlation function (ACF) of the continuum light curve. The lower-right panel shows the ICCF (black line), CCCD (blue histogram), and time-lag distribution obtained from MICA (orange histogram) in the observed frame. The complete figure set (8 images) is available in the online article.}
\label{fig_lcs}
\end{figure*}

\begin{figure*}
\centering
\includegraphics[width=\textwidth]{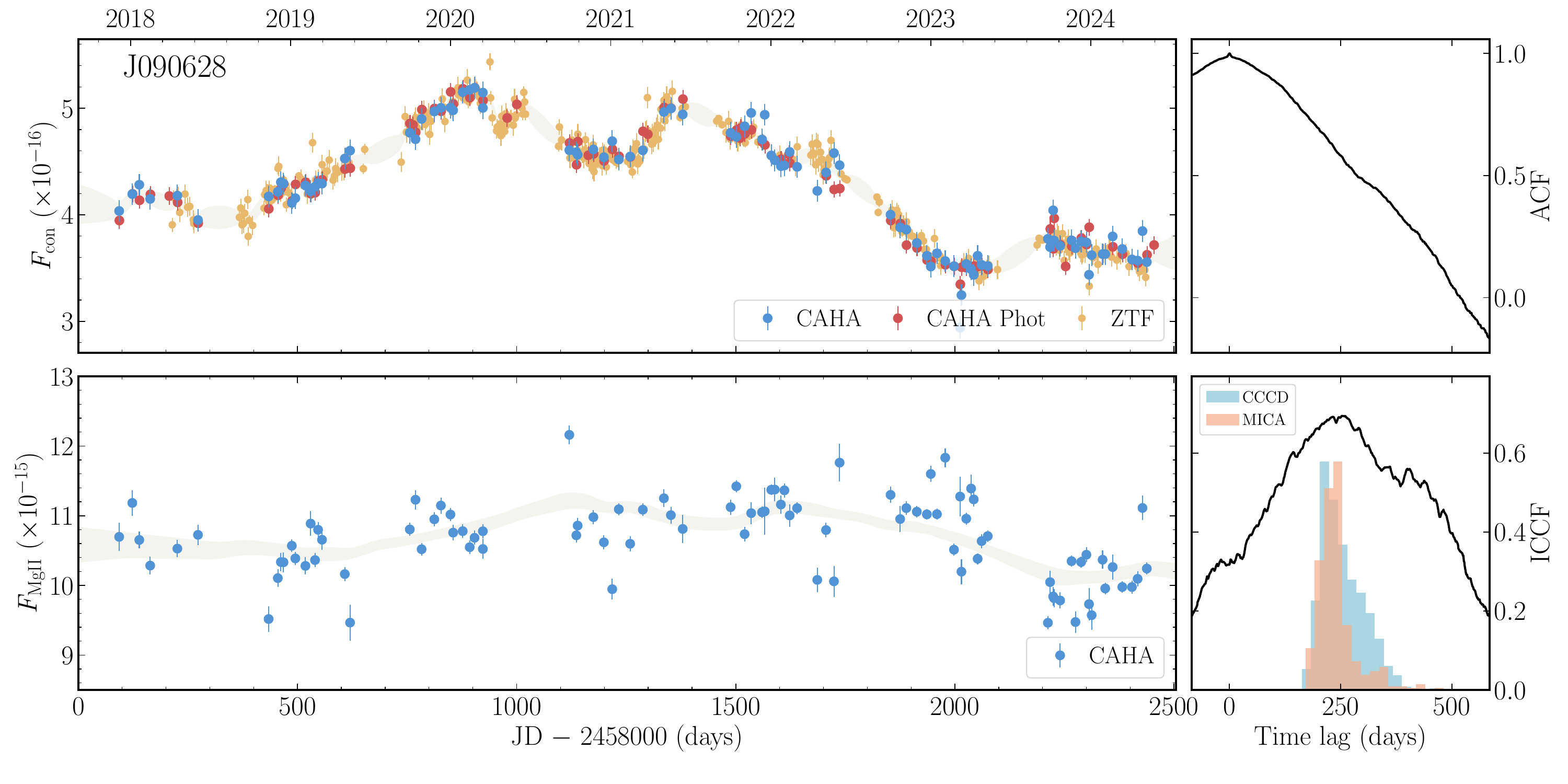}
\addtocounter{figure}{-1}
\caption{(Continued.) }
\end{figure*}
\begin{figure*}
\centering
\includegraphics[width=\textwidth]{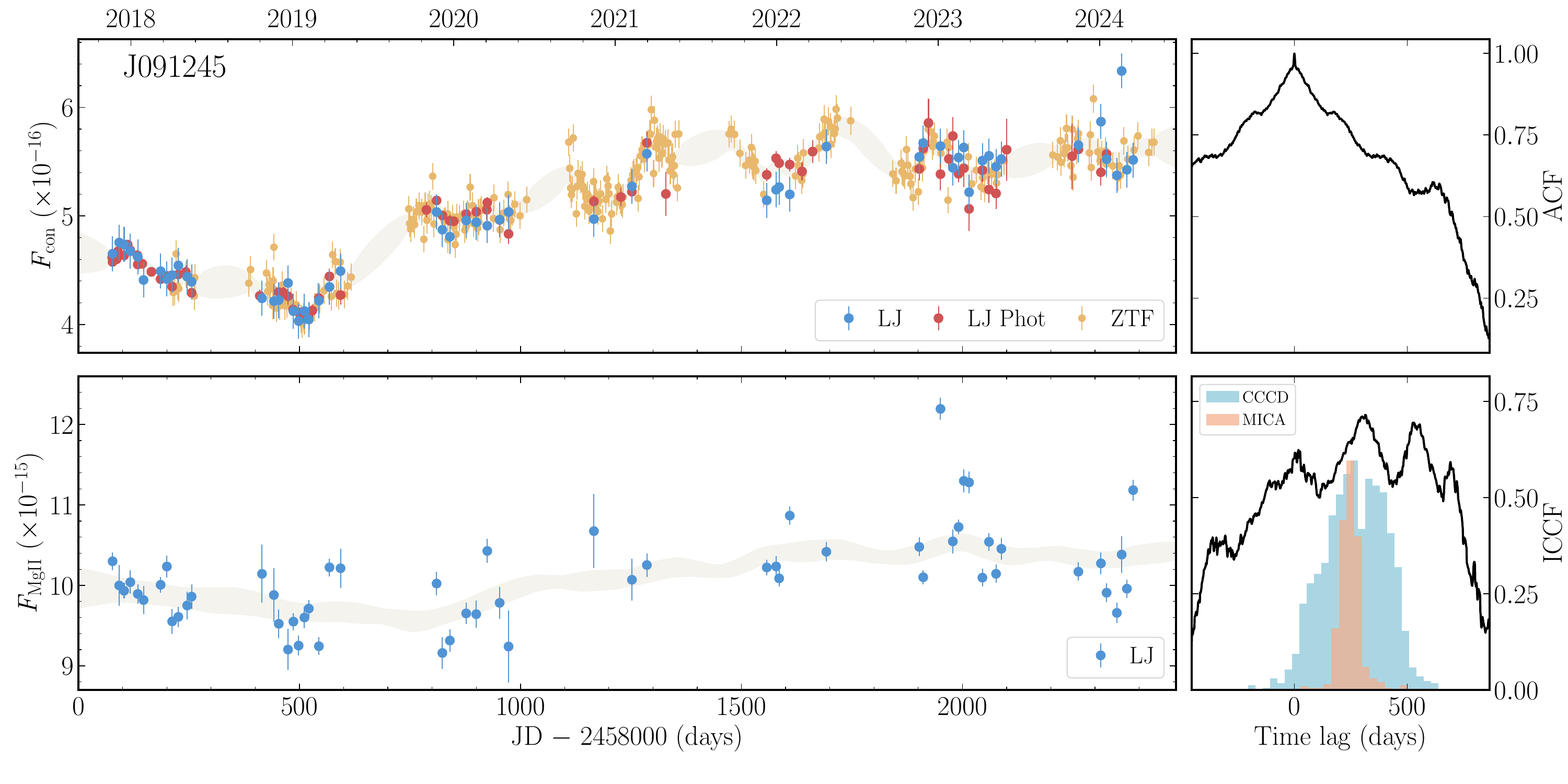}
\addtocounter{figure}{-1}
\caption{(Continued.) }
\end{figure*}
\begin{figure*}
\centering
\includegraphics[width=\textwidth]{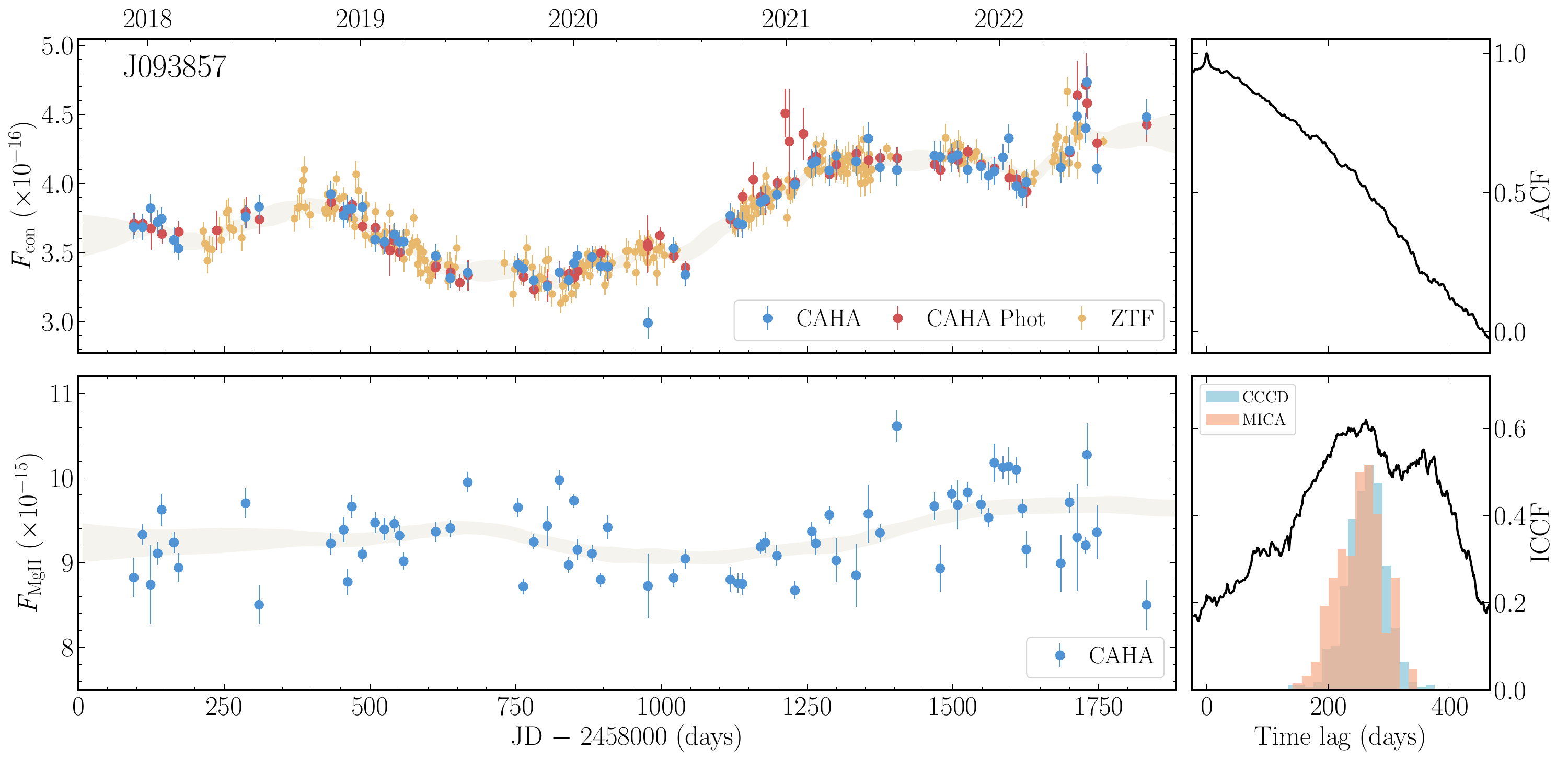}
\addtocounter{figure}{-1}
\caption{(Continued.) }
\end{figure*}
\begin{figure*}
\centering
\includegraphics[width=\textwidth]{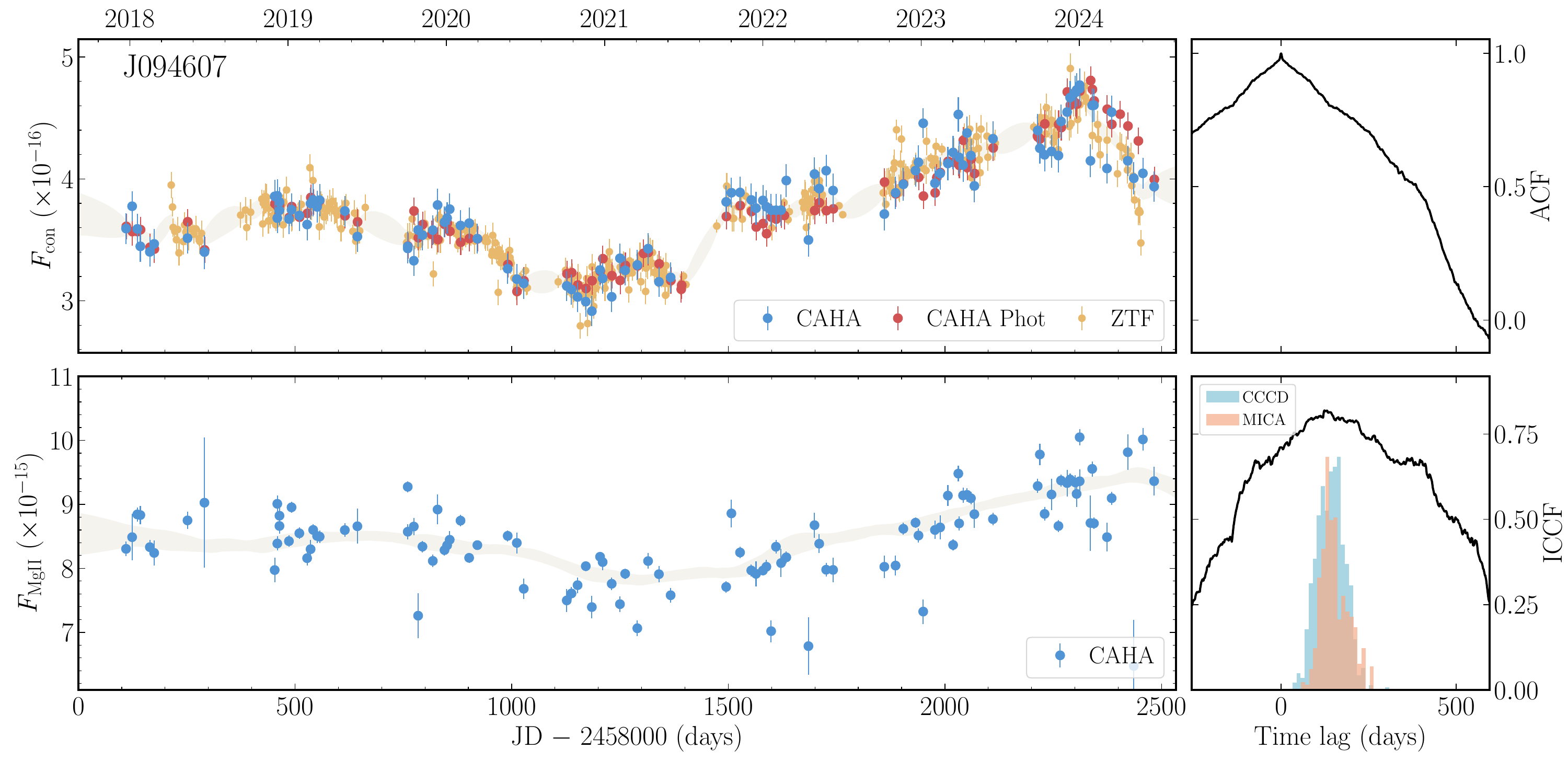}
\addtocounter{figure}{-1}
\caption{(Continued.) }
\end{figure*}
\begin{figure*}
\centering
\includegraphics[width=\textwidth]{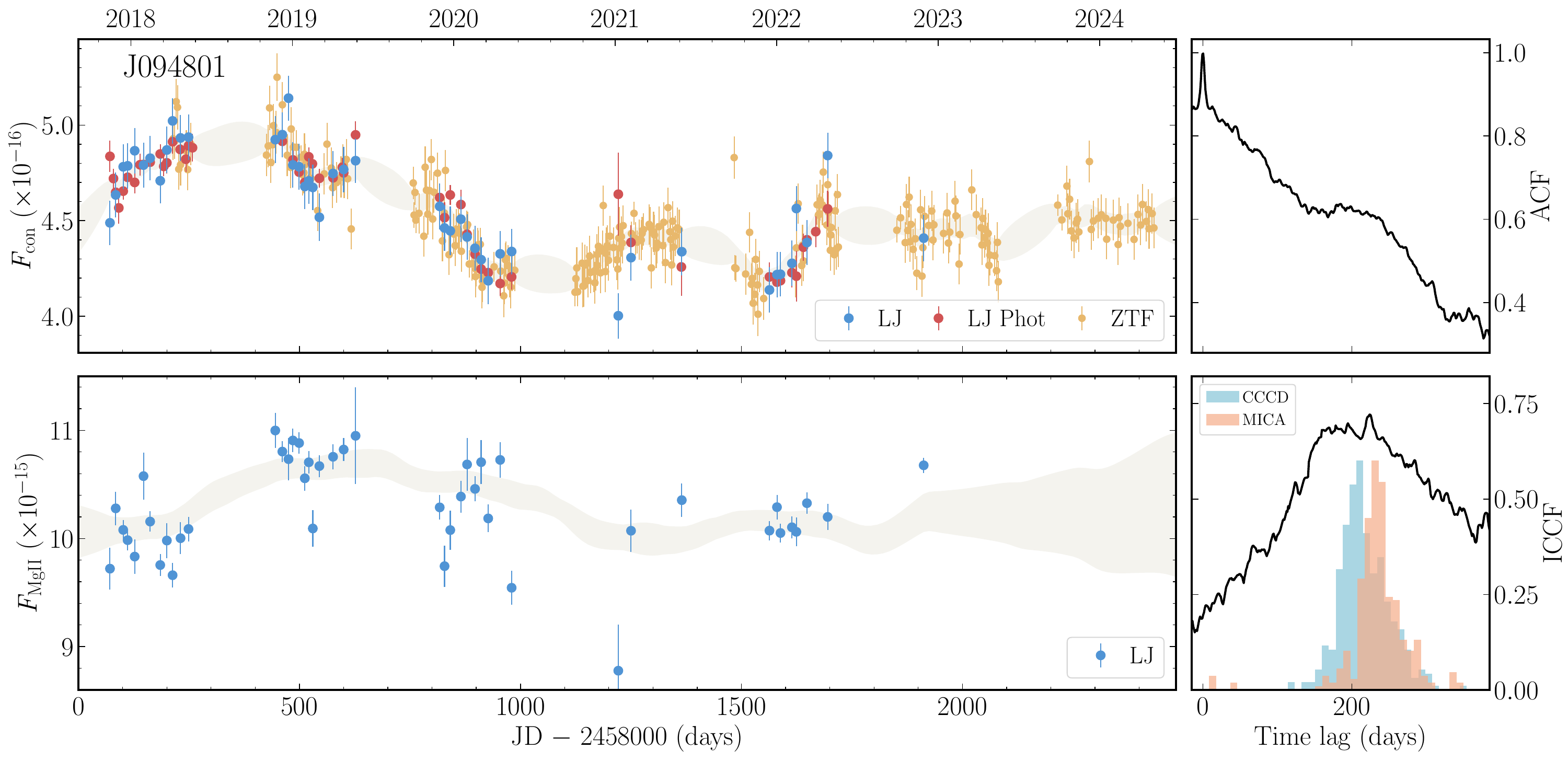}
\addtocounter{figure}{-1}
\caption{(Continued.) }
\end{figure*}
\begin{figure*}
\centering
\includegraphics[width=\textwidth]{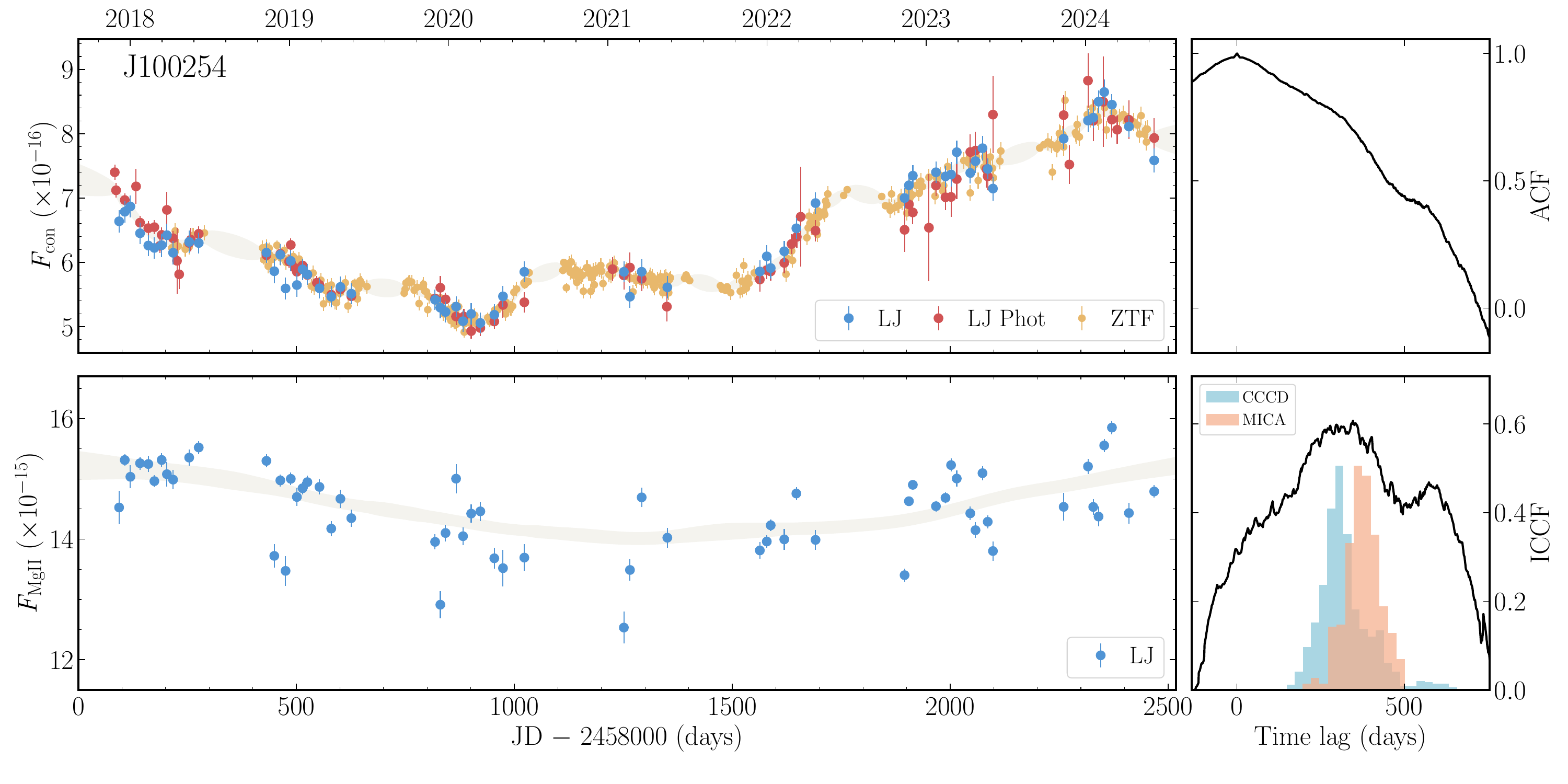}
\addtocounter{figure}{-1}
\caption{(Continued.) }
\end{figure*}
\begin{figure*}
\centering
\includegraphics[width=\textwidth]{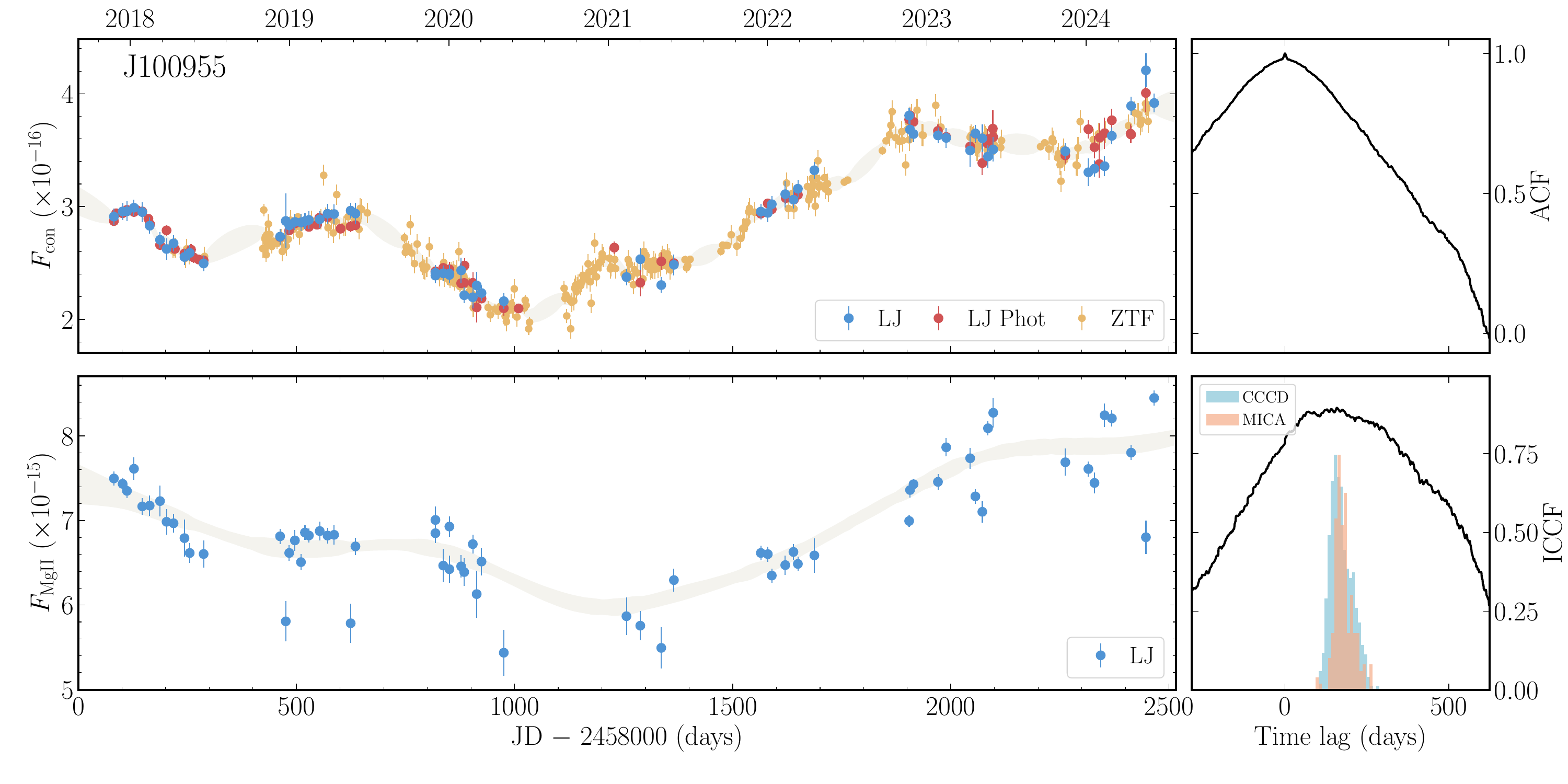}
\addtocounter{figure}{-1}
\caption{(Continued.) }
\end{figure*}

% \newpage

\newpage

\begin{deluxetable*}{cccccccccc}[t]%\label{tab_obs}
   \tabletypesize{\footnotesize}
   \tablecaption{Objects and Observations \label{Object_Observation}}
   \tablecolumns{8}
   \tablewidth{0.95\textwidth}
   \tablehead{
   \colhead{Name} & \colhead{Redshift}  & \colhead{Observatory} & \colhead{Filter} & 
   \colhead{Monitoring Period} & \colhead{$N_{\rm spec}$} & \colhead{$T_{\rm median}$} & \colhead{Duration} &\multicolumn{2}{c}{Comparison Star}\\ 
   \cline{9-10}
   \colhead{}   & & &  & \colhead{(yyyy/mm-yyyy/mm)} & \colhead{} & \colhead{(day)} & \colhead{(day)}& \colhead{$\alpha_{2000}$}&\colhead{$\delta_{2000}$}\\
   \colhead{(1)}   & \colhead{(2)} & \colhead{(3)} & \colhead{(4)}  & \colhead{(5)} & \colhead{(6)} & \colhead{(7)} & \colhead{(8) }& \colhead{(9) }& \colhead{(10)}
   }
   \startdata
       SDSS J082338.49+592648.4 &  0.76 & Lijiang & SDSS $\rm r^{\prime}$ & 2017/10 -- 2021/12 & 42  & 17 & 1508 &08 23 41.75&+59 27 03.60\\
       SDSS J083525.39+513829.7 &  0.82 & Lijiang & SDSS $\rm r^{\prime}$ & 2017/10 -- 2024/04 & 67  & 21 & 2361&08 35 38.72&+51 35 34.42\\
       SDSS J083826.25+253010.8 &  0.82 & Lijiang & SDSS $\rm r^{\prime}$ & 2017/11 -- 2024/03 & 65  & 18 & 2312&08 38 19.94&+25 31 48.93\\
       SDSS J084141.66+405237.6 &  0.77 & CAHA    & V                  & 2018/01 -- 2022/09 & 56  & 20 & 1749&08 41 38.37&+40 53 28.51\\
       SDSS J084808.36+083208.1 &  0.71 & Lijiang & SDSS $\rm r^{\prime}$ & 2017/11 -- 2021/12 & 36  & 21 & 1494&08 48 16.06&+08 33 03.16\\
       SDSS J085557.11+561534.7 &  0.72 & Lijiang & SDSS $\rm r^{\prime}$ & 2017/11 -- 2022/12 & 62  & 14 & 1853&08 56 15.66&+56 15 26.17\\
       SDSS J090628.04+434225.0 &  0.89 & CAHA    & V                  & 2017/12 -- 2024/05 & 95  & 15 & 2345&09 06 29.88& +43 43 05.13\\
       SDSS J091245.42+213139.1 &  0.74 & Lijiang & SDSS $\rm r^{\prime}$ & 2017/11 -- 2024/04 & 57  & 18 & 2309&09 12 55.78&+21 32 44.73\\
       SDSS J092835.01+623633.1 &  0.75 & Lijiang & SDSS $\rm r^{\prime}$ & 2017/11 -- 2024/04 & 73  & 17 & 2346&09 27 49.04&+62 38 08.61\\
       SDSS J093857.97+614019.6 &  0.74 & CAHA    & V                  & 2017/12 -- 2022/09 & 71  & 17 & 1737&09 39 15.42& +61 40 11.37\\
       SDSS J094009.76+455250.8 &  0.83 & CAHA    & V                  & 2017/12 -- 2024/06 & 103 & 15 & 2373&09 40 05.51&+45 55 26.10\\
       SDSS J094607.30+495412.8 &  0.76 & CAHA    & V                  & 2017/12 -- 2024/06 & 107 & 14 & 2373&09 46 17.48 &+49 51 41.58\\
       SDSS J094801.95+032118.9 &  0.83 & Lijiang & SDSS $\rm r^{\prime}$ & 2017/11 -- 2022/11 & 45  & 17 & 1841&09 48 01.61& +03 22 51.07\\
       SDSS J100254.51+324039.0 &  0.83 & Lijiang & SDSS $\rm r^{\prime}$ & 2017/12 -- 2024/06 & 63  & 18 & 2375&10 02 50.99&+32 37 59.37\\
       SDSS J100955.45+302734.3 &  0.82 & Lijiang & SDSS $\rm r^{\prime}$ & 2017/11 -- 2024/06 & 64  & 18 & 2385&10 09 49.87&+30 27 55.30\\
       SDSS J101227.21+345515.4 &  0.77 & Lijiang & SDSS $\rm r^{\prime}$ & 2017/12 -- 2022/11 & 42  & 18 & 1820&10 12 46.38& +34 54 05.28\\
       SDSS J101622.60+470643.3 &  0.82 & CAHA    & V                  & 2017/12 -- 2024/07 & 98  & 19 & 2408&10 16 38.36&+47 07 56.24\\
       SDSS J101730.21+474000.0 &  0.79 & Lijiang & SDSS $\rm r^{\prime}$ & 2017/11 -- 2021/12 & 39  & 14 & 1467&10 17 16.76& +47 41 48.55\\ 
   \enddata
   \tablecomments{Columns (1)–(3) provide the name, redshift, and observatory for each object. Column (4) presents the filter used for the photometric observations. Column (5) lists the start and end dates of observations at Lijiang and CAHA. Columns (6)–(8) indicate the number of spectroscopic epochs, the median sampling cadence, and the total duration of the entire spectroscopic monitoring period. Finally, Columns (9) and (10) present the coordinates of the comparison star used in the spectroscopic observations.}
   \end{deluxetable*}

\begin{deluxetable*}{rlcrllcrlcrll}
\tablecolumns{13}
\setlength{\tabcolsep}{5pt}
\tablewidth{0pc}
\tablecaption{Light Curves \label{tab_lc}}
\tabletypesize{\footnotesize}
\tablehead{
      \multicolumn{6}{c}{J082338}        &
      \colhead{}                         &
      \multicolumn{6}{c}{J083525}        \\ \cline{1-6}\cline{8-13}
      \multicolumn{2}{c}{Photometry}     &
      \colhead{}                         &
      \multicolumn{3}{c}{Spectra}        &
      \colhead{}                         &
      \multicolumn{2}{c}{Photometry}     &
      \colhead{}                         &
      \multicolumn{3}{c}{Spectra}        \\ \cline{1-2}\cline{4-6}\cline{8-9}\cline{11-13}
      \colhead{JD}                       &
      \colhead{$F_{\rm phot}$}                      &
      \colhead{}                         &
      \colhead{JD}                       &
      \colhead{$F_{3000}$}               &
      \colhead{$F_{\rm MgII}$}         &
      \colhead{}                         &
      \colhead{JD}                       &
      \colhead{$F_{\rm phot}$}                      &
      \colhead{}                         &
      \colhead{JD}                       &
      \colhead{$F_{3000}$}               &
      \colhead{$F_{\rm MgII}$}
         }
\startdata
53.343 & $2.216\pm 0.039$  & &   53.353 & $2.176\pm 0.072$ & $4.388\pm 0.216$ & &   55.316 & $4.934\pm 0.036$  & &   55.343 & $ 4.937\pm 0.113$ & $ 13.642\pm 0.099$ \\
54.349 & $2.214\pm 0.038$  & &   54.323 & $2.272\pm 0.073$ & $4.534\pm 0.058$ & &   76.391 & $4.900\pm 0.032$  & &   67.316 & $ 4.875\pm 0.095$ & $ 13.204\pm 0.142$ \\
66.272 & $2.280\pm 0.058$  & &   66.291 & $2.325\pm 0.073$ & $4.292\pm 0.200$ & &   79.235 & $4.861\pm 0.034$  & &   79.262 & $ 4.806\pm 0.114$ & $ 12.952\pm 0.117$ \\
76.336 & $2.263\pm 0.032$  & &   78.381 & $2.296\pm 0.073$ & $4.419\pm 0.082$ & &   84.291 & $4.878\pm 0.032$  & &   88.331 & $ 5.029\pm 0.128$ & $ 13.820\pm 0.122$ \\
78.360 & $2.324\pm 0.027$  & &   87.299 & $2.305\pm 0.073$ & $4.121\pm 0.173$ & &   99.284 & $4.905\pm 0.031$  & &   99.311 & $ 4.844\pm 0.112$ & $ 13.162\pm 0.090$ 
\enddata
\tablecomments{Julian date (JD) is from 2,458,000. $F_{\rm phot}$, $F_{3000}$, and $F_{\rm Mg II}$ are the fluxes of photometric continuum, spectroscopic continuum, and Mg {\sc ii} line, respectively. The continuum fluxes and uncertainties are the values after the {\textsc{PyCALI}} intercalibration. The continuum and Mg {\sc ii} fluxes are in units of $10^{-16} {\rm erg}~{\rm s}^{-1} {\rm cm}^{-2} \text{\AA}^{-1}$ and $10^{-15} \, {\rm erg} \, {\rm s}^{-1} \, {\rm cm}^{-2}$, respectively.}
\end{deluxetable*}

\begin{deluxetable*}{ccccccccc}
\tabletypesize{\footnotesize}
\tablecaption{Light-curve Statistics \label{lc_statistics}}
\tablecolumns{6}
\tablewidth{0.95\textwidth}
\tablehead{
\colhead{Object} &\multicolumn{2}{c}{3000 {\AA}} &  &\multicolumn{2}{c}{\ion{Mg}{2}} \\ \cline{2-3}\cline{5-6}
\colhead{}   & Flux & $F_{\rm var}$ & & Flux & $F_{\rm var}$\\
\colhead{}   & ($10^{-16} {\rm erg}~{\rm s}^{-1} {\rm cm}^{-2} \text{\AA}^{-1}$) &(\%)  & & ($10^{-15} {\rm erg}~{\rm s}^{-1} {\rm cm}^{-2}$) & (\%) 
}
\startdata
    J082338  & $2.402\pm{0.111}$ &$4.6\pm{0.6}$ & & $4.533\pm{0.226}$&$4.4\pm{0.6}$\\
    J083525  & $4.876\pm{0.231}$ &$4.6\pm{0.4}$ & & $13.288\pm{0.530}$&$3.8\pm{0.4}$\\
    J083826  & $3.075\pm{0.231}$ &$7.4\pm{0.5}$ & & $8.256\pm{0.408}$&$4.7\pm{0.5}$\\
    J084141 & $3.960\pm{0.180}$ &$4.6\pm{0.3}$ & & $13.441\pm{0.441}$&$3.0\pm{0.3}$\\
    J084808  & $4.720\pm{0.179}$ &$3.8\pm{0.8}$ & & $8.317\pm{0.491}$&$5.6\pm{0.8}$\\
    J085557 & $6.993\pm{0.481}$ &$6.9\pm{0.5}$ & & $16.526\pm{0.917}$&$5.4\pm{0.5}$\\
    J090628  & $4.226\pm{0.540}$ &$12.9\pm{0.4}$ & & $10.666\pm{0.558}$&$5.1\pm{0.4}$\\
    J091245  & $4.976\pm{0.557}$ &$11.2\pm{0.6}$ & & $10.084\pm{0.568}$&$5.4\pm{0.6}$\\
    J092835  & $3.823\pm{0.281}$ &$7.1\pm{0.7}$ & & $7.473\pm{0.597}$&$7.6\pm{0.7}$\\
    J093857 & $3.833\pm{0.359}$ &$9.4\pm{0.5}$ & & $9.335\pm{0.446}$&$4.3\pm{0.5}$ \\
    J094009  & $3.724\pm{0.216}$ &$5.8\pm{0.3}$ & & $11.028\pm{0.472}$&$4.1\pm{0.3}$\\
    J094607  & $3.820\pm{0.437}$ &$11.5\pm{0.6}$ & & $8.474\pm{0.667}$&$7.6\pm{0.6}$\\
    J094801  & $4.597\pm{0.272}$ &$5.8\pm{0.5}$ & & $10.297\pm{0.441}$&$4.0\pm{0.5}$\\
    J100254  & $6.475\pm{0.989}$ &$15.4\pm{0.4}$ & & $14.513\pm{0.669}$&$4.5\pm{0.4}$\\
    J100955 & $2.967\pm{0.499}$ &$16.9\pm{0.9}$ & & $6.912\pm{0.665}$&$9.5\pm{0.9}$\\
    J101227 & $4.109\pm{0.258}$ &$6.3\pm{0.5}$ & & $7.013\pm{0.283}$&$3.4\pm{0.5}$\\
    J101622  & $4.347\pm{0.253}$ &$5.9\pm{0.4}$ & & $9.220\pm{0.498}$&$5.1\pm{0.4}$\\
    J101730  & $3.166\pm{0.054}$ &$1.4\pm{0.6}$ & & $7.625\pm{0.362}$&$4.4\pm{0.6}$\\    
\enddata
\tablecomments{The flux and its uncertainty listed represent the mean and standard deviation measured from the spectroscopic continuum and Mg {\sc ii} light curves. $F_{\rm var}$ denotes the fractional variability amplitude and its uncertainty, as defined in Section \ref{sec_var}.}
\end{deluxetable*}
\begin{deluxetable*}{cccccccccccc}\label{measurement}
\tabletypesize{\footnotesize}
\tablewidth{\textwidth}
\tablecaption{Time Lags}
\tablehead{
\colhead{Name}& $r_{\rm max}$ & $\tau_{\rm ICCF}^{\rm obs}$&$\tau_{\rm MICA}^{\rm obs}$ &$\tau_{\rm ICCF}^{\rm rest}$&$\tau_{\rm MICA}^{\rm rest}$ &FWHM&$\sigma_{\rm line}$& log($L_{3000}$/erg s$^{-1}$) &log($M_{\bullet} /M_{\sun}$)&$\mathcal{R}_{\rm Fe,UV}$&log$\dot{\mathscr{M}}$\\
\colhead{}          & & (day) &(day)& (day) &(day) & (km s$^{-1}$) & (km s$^{-1}$) & &\\
\colhead{(1)}     &   (2)     &(3) & (4) &(5) & (6) & (7)& (8) & (9)& (10)& (11)& (12)
}
\startdata
    J083826  & 0.79 & $108_{-28}^{+28}$ & $132_{-12}^{+23}$& $59_{-16}^{+15}$ & $72_{-7}^{+13}$  &  $3037_{-223}^{+172}$ & $3183_{-134}^{+144}$ & $45.76\pm{0.03}$ &$8.11_{-0.08}^{+0.09}$& $0.74\pm{0.02}$ & $1.444_{-0.186}^{+0.157}$\\
    J090628 & 0.69 & $250_{-32}^{+67}$ & $244_{-27}^{+37}$& $132_{-17}^{+36}$ & $130_{-14}^{+20}$ &  $2887_{-291}^{+255}$ & $2707_{-156}^{+155}$ & $46.00\pm{0.06}$&$8.32_{-0.10}^{+0.10}$& $1.18\pm{0.05}$  &$1.384_{-0.221}^{+0.213}$\\
    J091245  & 0.72 & $323_{-141}^{+141}$ & $260_{-37}^{+36}$& $186_{-81}^{+81}$ & $150_{-22}^{+21}$ &  $2222_{-229}^{+180}$ & $2108_{-145}^{+156}$ & $45.84\pm{0.05}$ &$8.16_{-0.11}^{+0.09}$&$1.35\pm{0.04}$ & $1.474_{-0.200}^{+0.231}$\\
    J093857  & 0.62 & $278_{-32}^{+31}$ & $264_{-45}^{+29}$& $159_{-19}^{+18}$ & $152_{-26}^{+17}$ &  $3737_{-310}^{+276}$ & $2772_{-211}^{+242}$ & $45.73\pm{0.04}$&$8.62_{-0.10}^{+0.08}$&$1.30\pm{0.14} $ &$0.392_{-0.171}^{+0.215}$\\
    J094607  & 0.82 & $168_{-41}^{+42}$ & $150_{-27}^{+52}$& $96_{-24}^{+24}$ & $85_{-15}^{+29}$ &  $2834_{-308}^{+273}$ & $2963_{-163}^{+129}$ & $45.76\pm{0.05}$&$8.13_{-0.12}^{+0.17}$& $0.86\pm{0.03}$ &$1.419_{-0.352}^{+0.255}$\\
    J094801  & 0.75 & $218_{-31}^{+29}$ & $240_{-20}^{+33}$& $119_{-17}^{+16}$ & $131_{-11}^{+18}$ &  $3200_{-221}^{+171}$ & $2670_{-173}^{+156}$ & $45.95\pm{0.03}$ &$8.42_{-0.07}^{+0.08}$& $1.03\pm{0.03}$ &$1.122_{-0.158}^{+0.144}$\\
    J100254  & 0.61 & $310_{-58}^{+85}$ & $389_{-48}^{+50}$& $170_{-32}^{+46}$ & $213_{-26}^{+27}$ &  $3210_{-222}^{+171}$ & $2809_{-133}^{+141}$ & $46.10\pm{0.07}$&$8.63_{-0.08}^{+0.07}$& $0.75\pm{0.02}$  &$0.919_{-0.178}^{+0.185}$\\
    J100955  & 0.90 & $170_{-29}^{+42}$ & $178_{-19}^{+33}$& $93_{-16}^{+23}$ & $98_{-11}^{+18}$ &  $3404_{-222}^{+172}$ & $3108_{-150}^{+141}$ & $45.74\pm{0.08}$&$8.34_{-0.07}^{+0.09}$& $0.90\pm{0.03}$&$0.955_{-0.221}^{+0.181}$\\   
\enddata
\tablecomments{Column (1) presents the names of the targets. Column (2) provides the peak values ($r_{\rm max}$) of the cross-correlation functions. Columns (3)–(6) list the time lags and uncertainties measured from the ICCF and MICA methods in both the observed and rest frames. Columns (7) and (8) indicate the FWHM and $\sigma_{\rm line}$ of the Mg {\sc ii} line, respectively. Column (9) presents the monochromatic continuum luminosities at 3000 \AA\ along with their uncertainties. Column (10) gives the black hole masses and their uncertainties, noting that the uncertainty in the $f$ factor is not included. Finally, Column (11) and (12) present the ultraviolet iron ratio $\mathcal{R}_{\rm Fe,UV}$ and the dimensionless accretion rate $\dot{\mathscr{M}}$ measured, respectively.}
\end{deluxetable*}

\begin{figure*}[t]
\centering
\includegraphics[width=\textwidth]{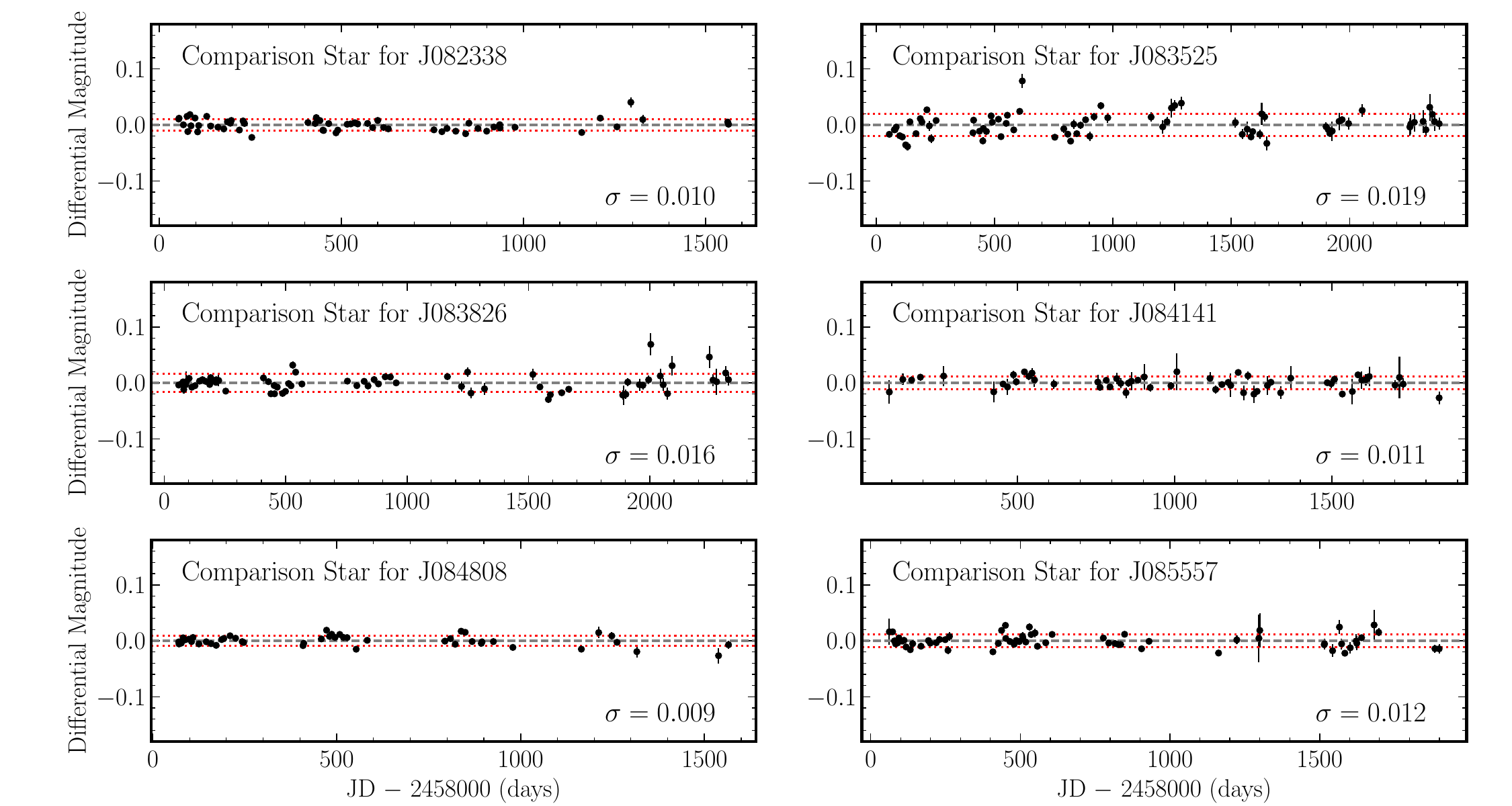}
\caption{Photometric light curves of comparison stars in the slits. The standard deviations are represented by red dotted lines and explicitly written as $\sigma$ in each panel. }
\label{fig_comparison_lc}
\end{figure*}

\begin{figure*}
\centering
\includegraphics[width=\textwidth]{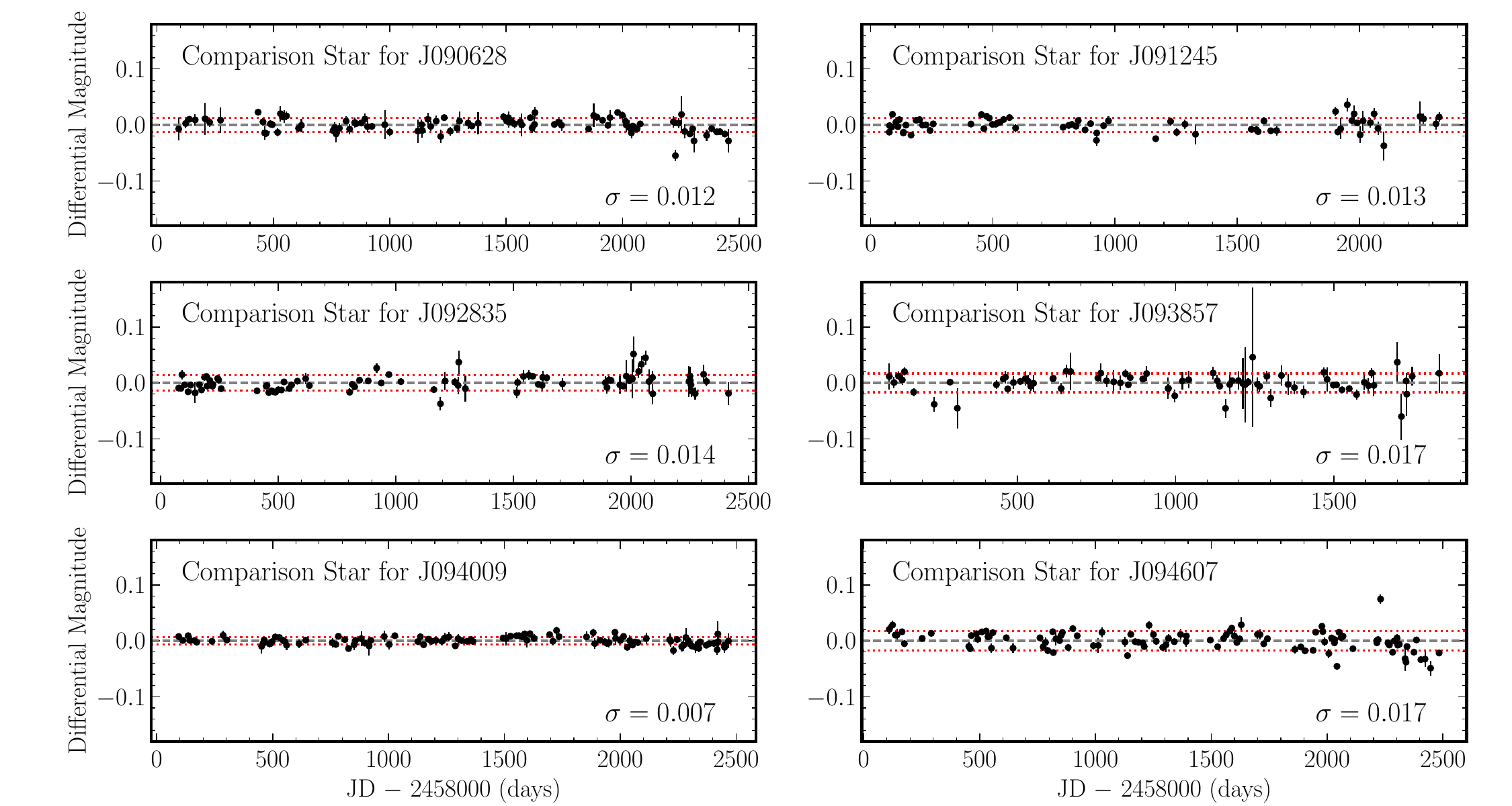}
\addtocounter{figure}{-1}
\caption{(Continued.) }
\end{figure*}

\begin{figure*}
\centering
\includegraphics[width=\textwidth]{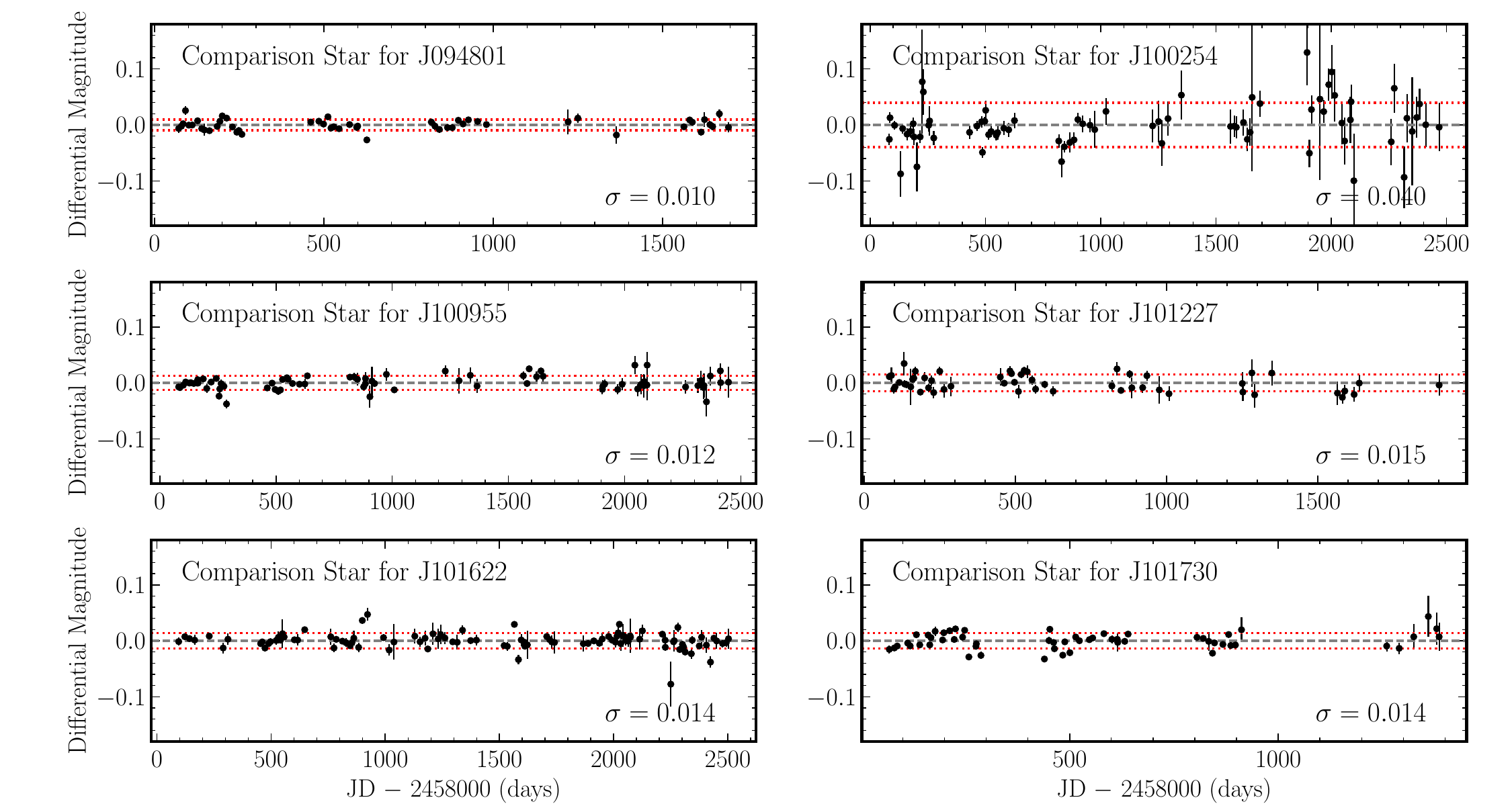}
\addtocounter{figure}{-1}
\caption{(Continued.) }
\end{figure*}

%%%%%%
\begin{figure*}
\centering
\includegraphics[width=\textwidth]{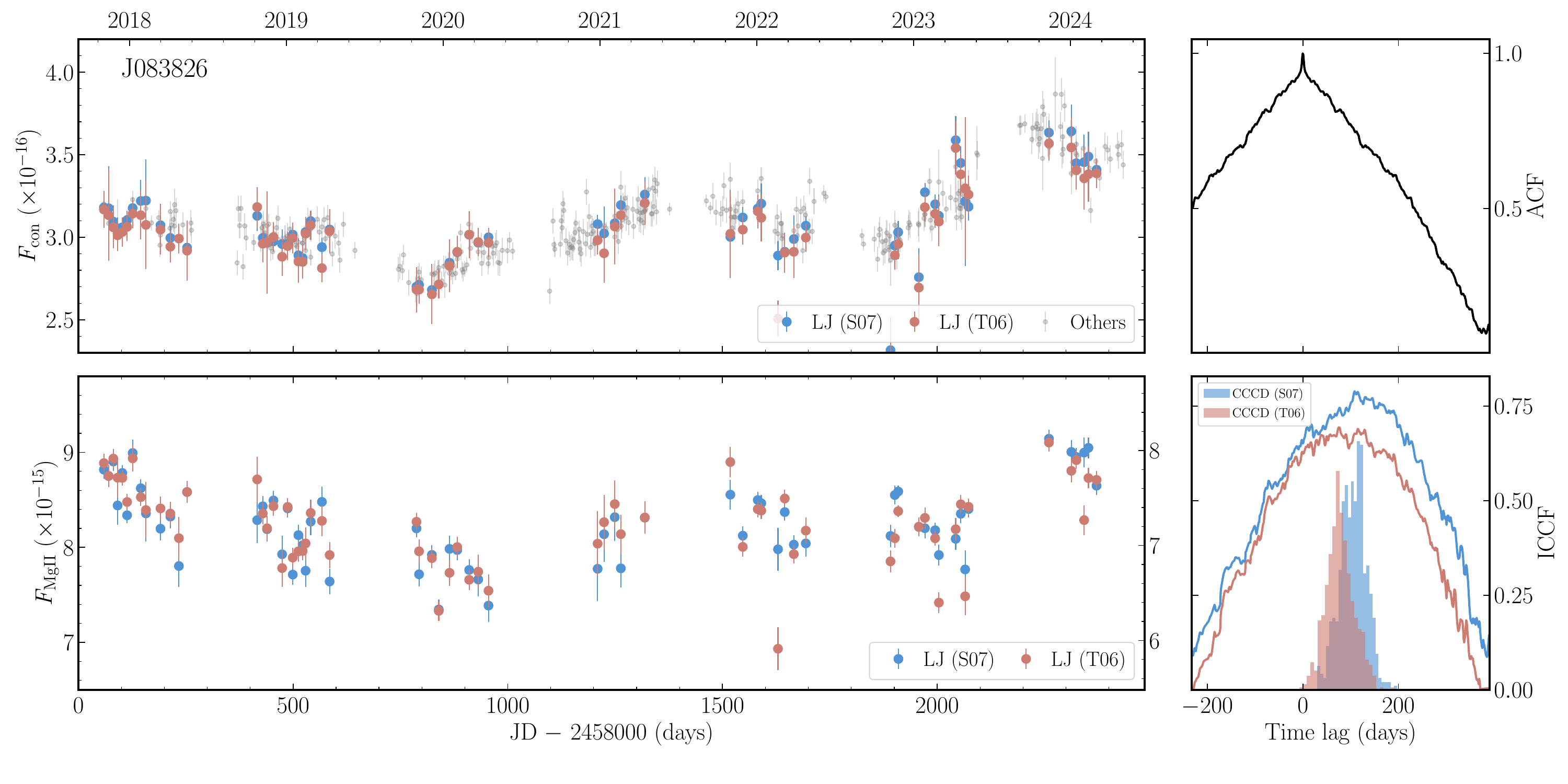}
\caption{Comparison between light curves and time-lag measurements for using two iron templates: \cite{Salviander2007} and \cite{Tsuzuki2006}. Meanings of panels are the same as Figure \ref{fig_lcs}. ``LJ/CAHA (S07/T06)'' represents the Lijiang/CAHA spectroscopic data measured using the iron template of \cite{Salviander2007} or \cite{Tsuzuki2006}. ``Others'' stands for the photometric data of ZTF or Lijiang/CAHA. Data points of the Mg\,{\sc ii} light curve corresponding to \cite{Salviander2007} and \cite{Tsuzuki2006} templates should be referenced against the tick labels on the left and right, respectively. The complete figure set (8 images) is available in the online article.}
\label{fig:compare_Template}
\end{figure*}

\begin{figure*}
\centering
\includegraphics[width=\textwidth]{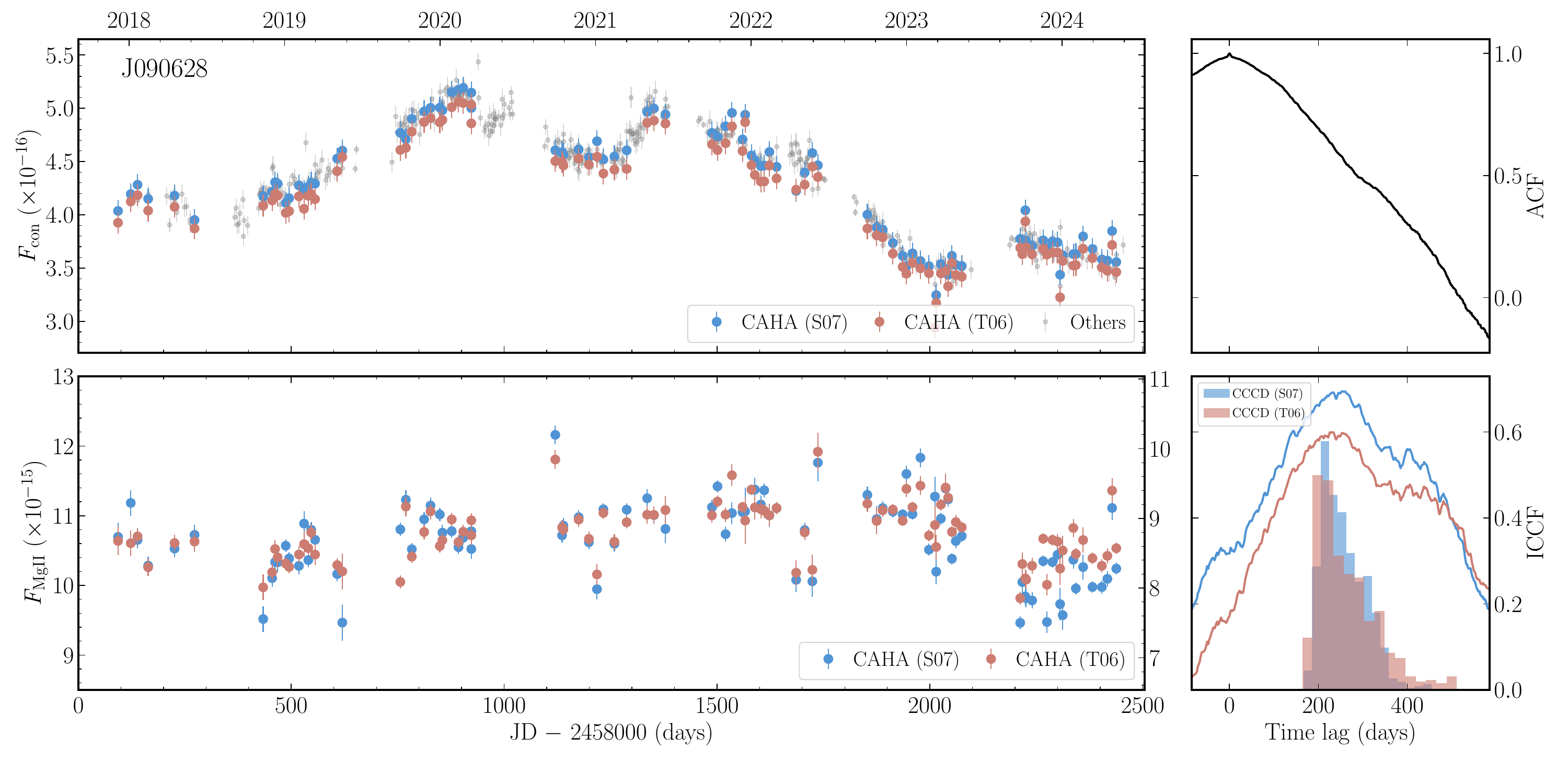}
\addtocounter{figure}{-1}
\caption{(Continued.) }
\end{figure*}
\begin{figure*}
\centering
\includegraphics[width=\textwidth]{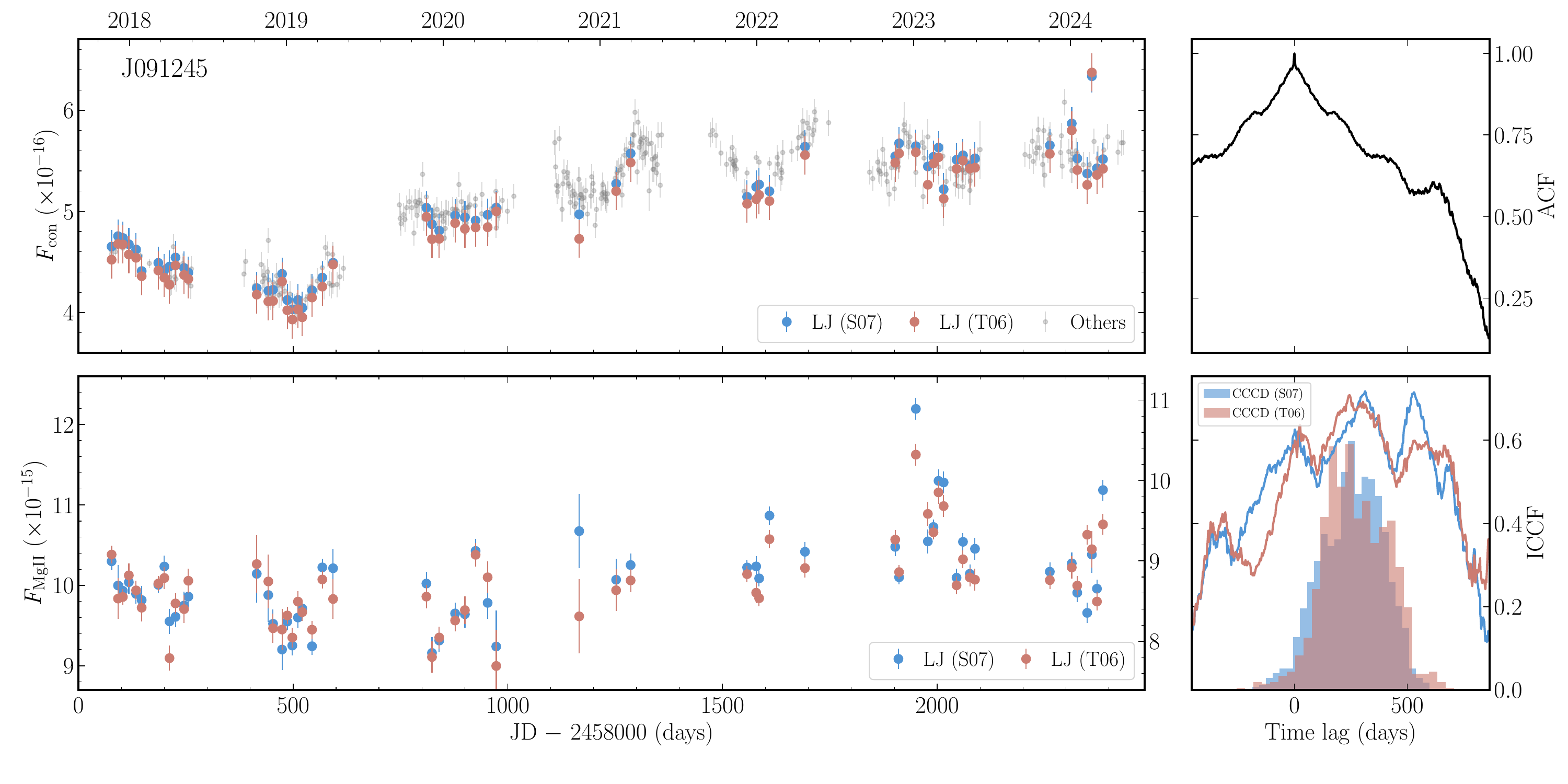}
\addtocounter{figure}{-1}
\caption{(Continued.) }
\end{figure*}
\begin{figure*}
\centering
\includegraphics[width=\textwidth]{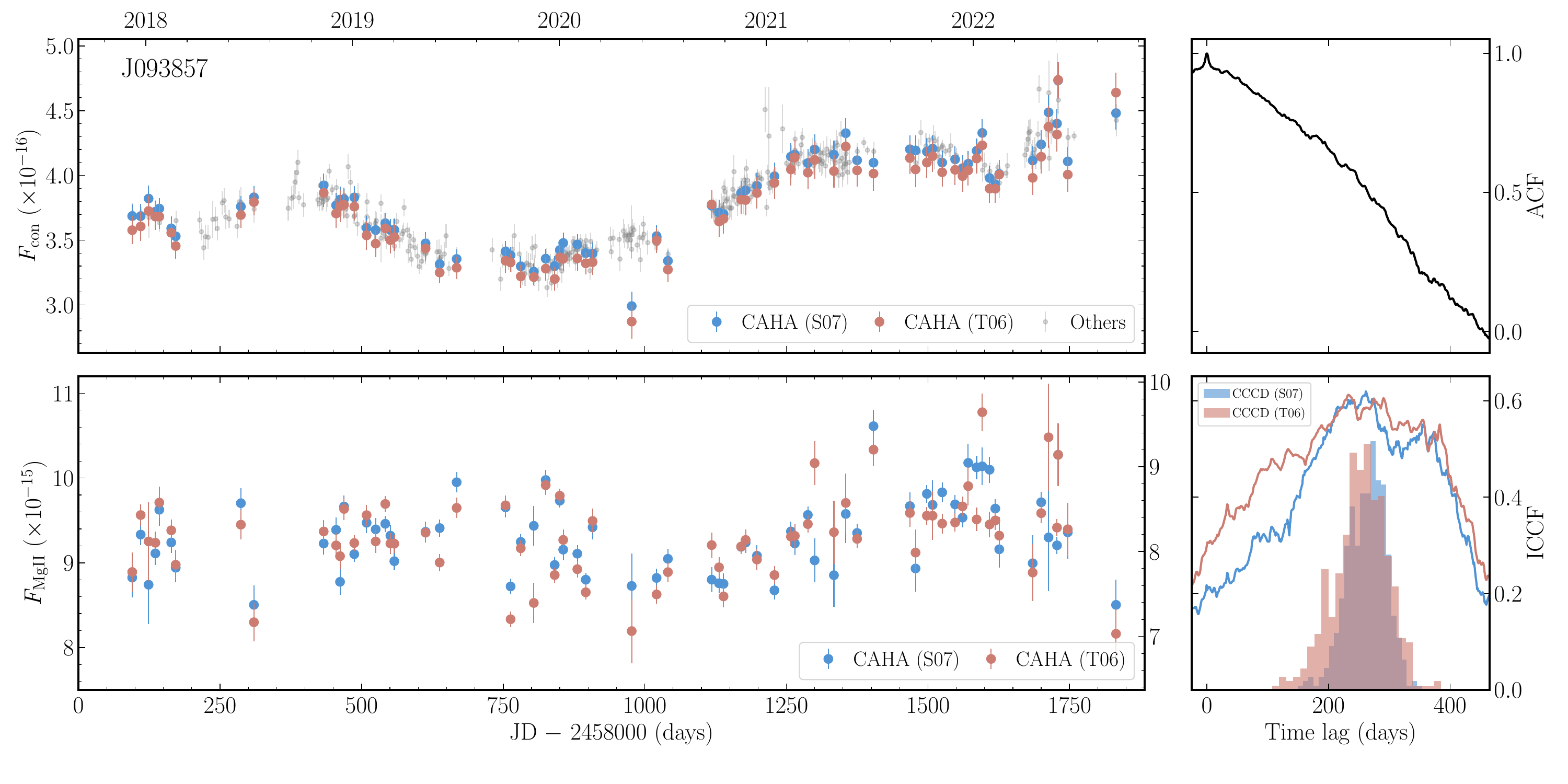}
\addtocounter{figure}{-1}
\caption{(Continued.) }
\end{figure*}
\begin{figure*}
\centering
\includegraphics[width=\textwidth]{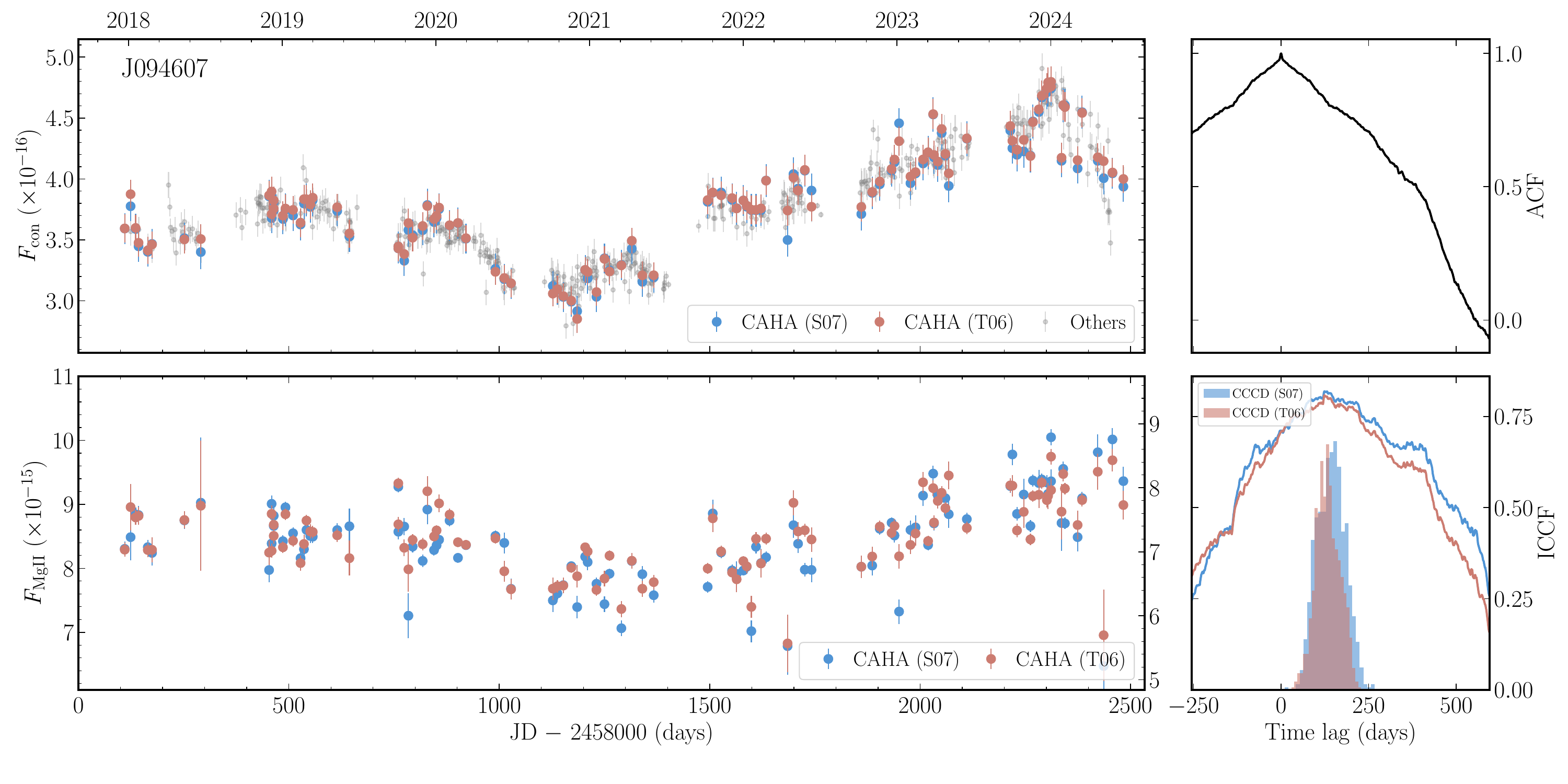}
\addtocounter{figure}{-1}
\caption{(Continued.) }
\end{figure*}
\begin{figure*}
\centering
\includegraphics[width=\textwidth]{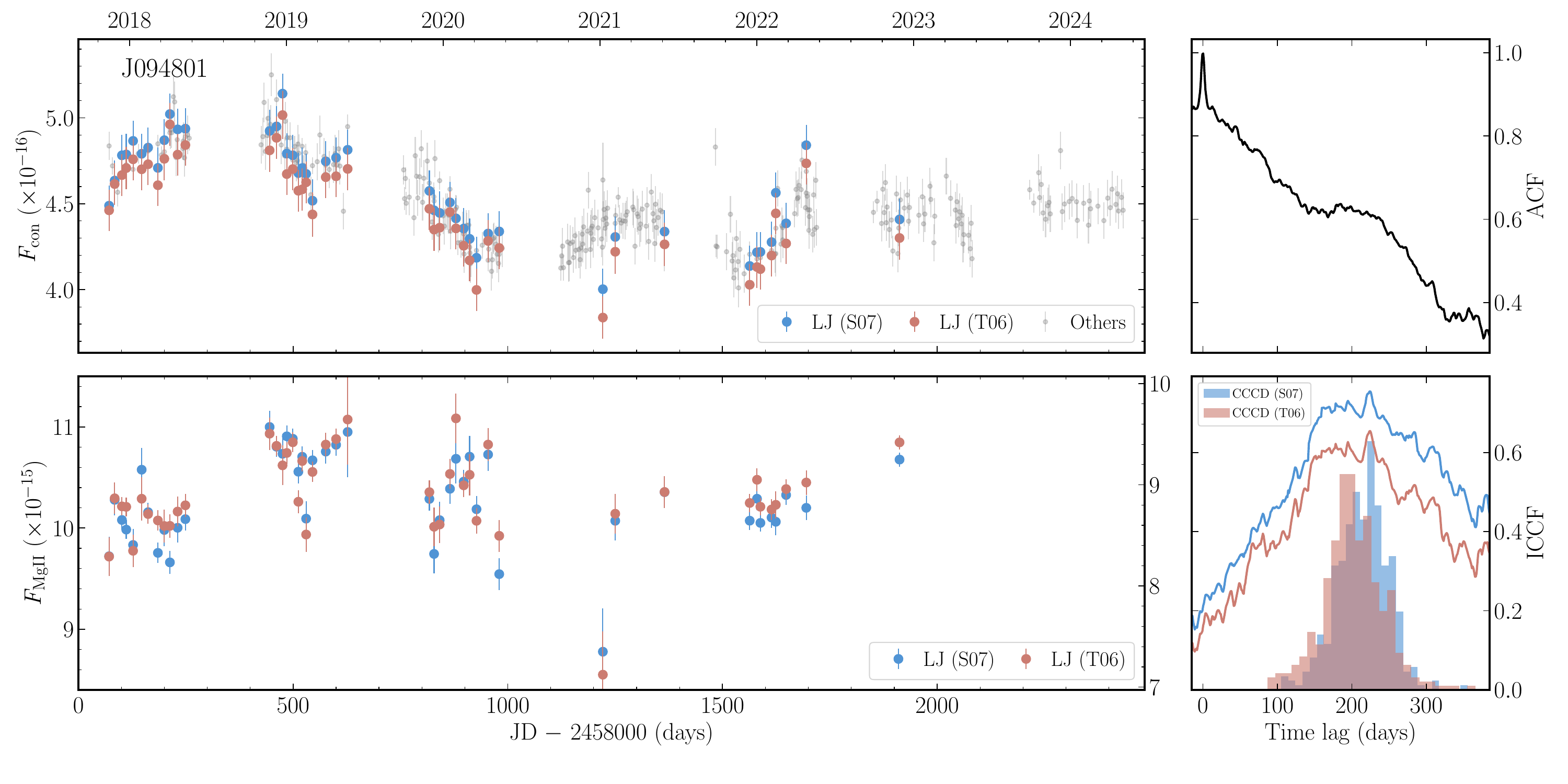}
\addtocounter{figure}{-1}
\caption{(Continued.) }
\end{figure*}
\begin{figure*}
\centering
\includegraphics[width=\textwidth]{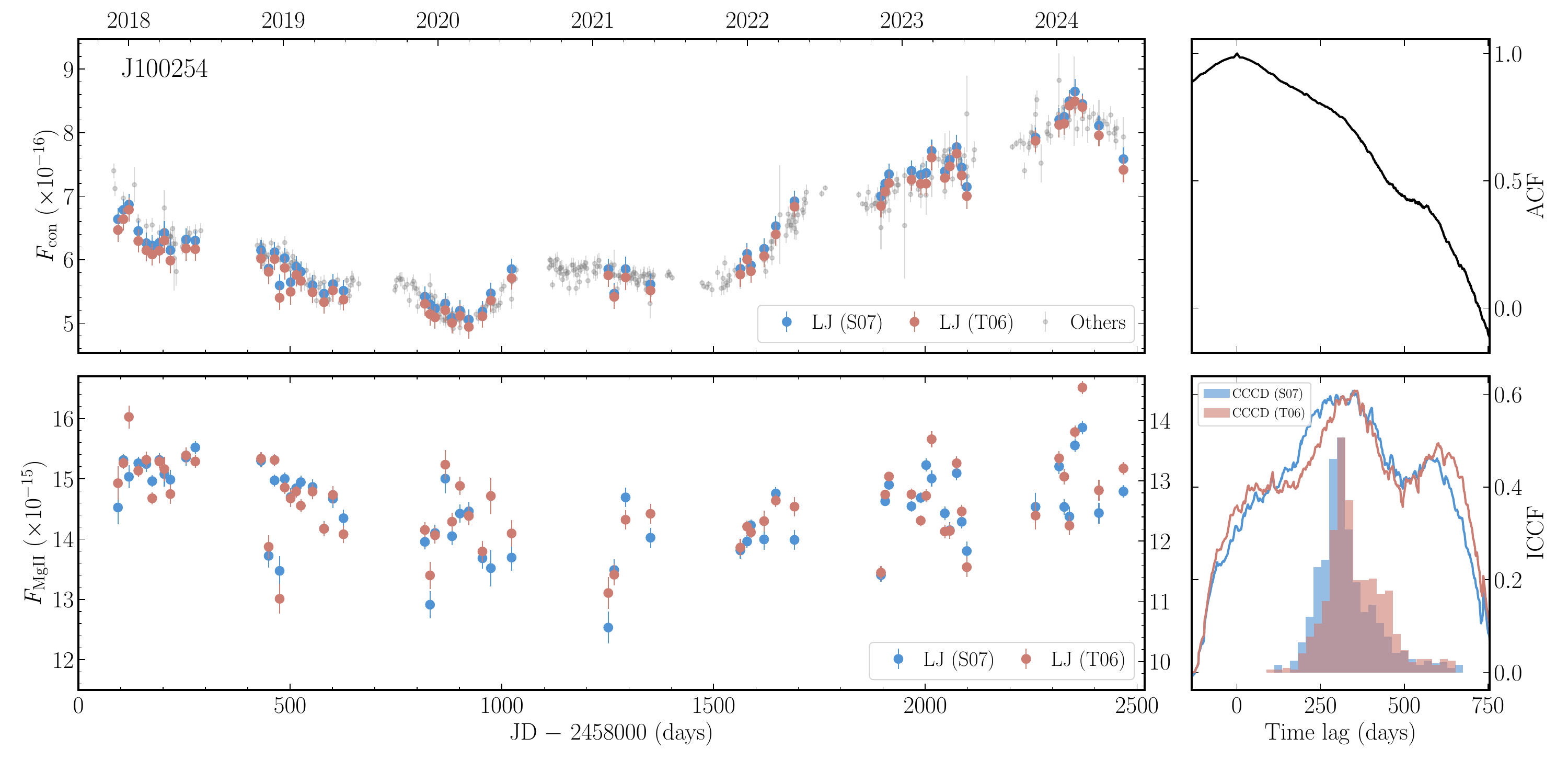}
\addtocounter{figure}{-1}
\caption{(Continued.) }
\end{figure*}
\begin{figure*}
\centering
\includegraphics[width=\textwidth]{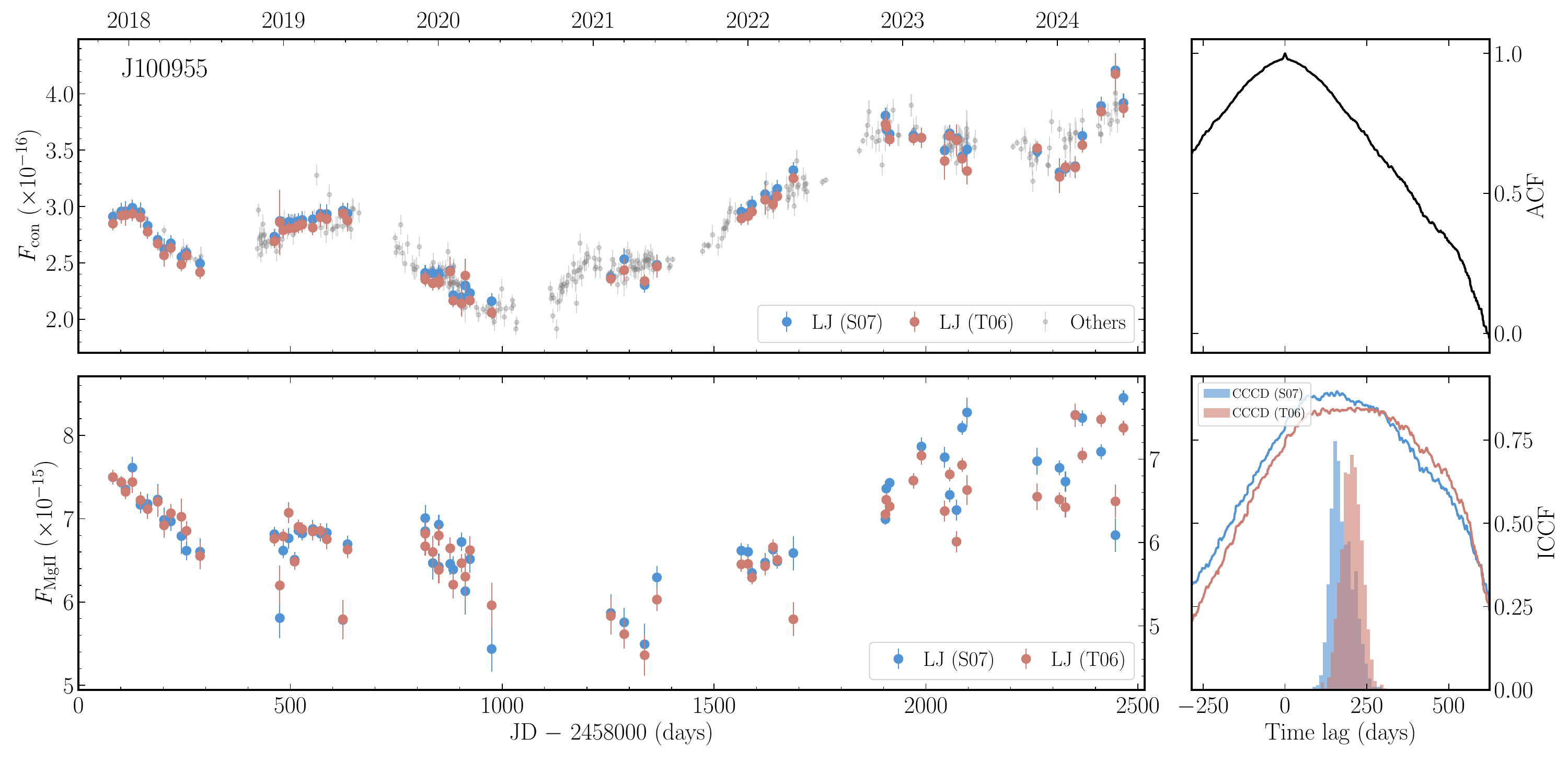}
\addtocounter{figure}{-1}
\caption{(Continued.) }
\end{figure*}

%%%%%%%%%
\begin{figure*}[t]
\centering
\includegraphics[width=\textwidth]{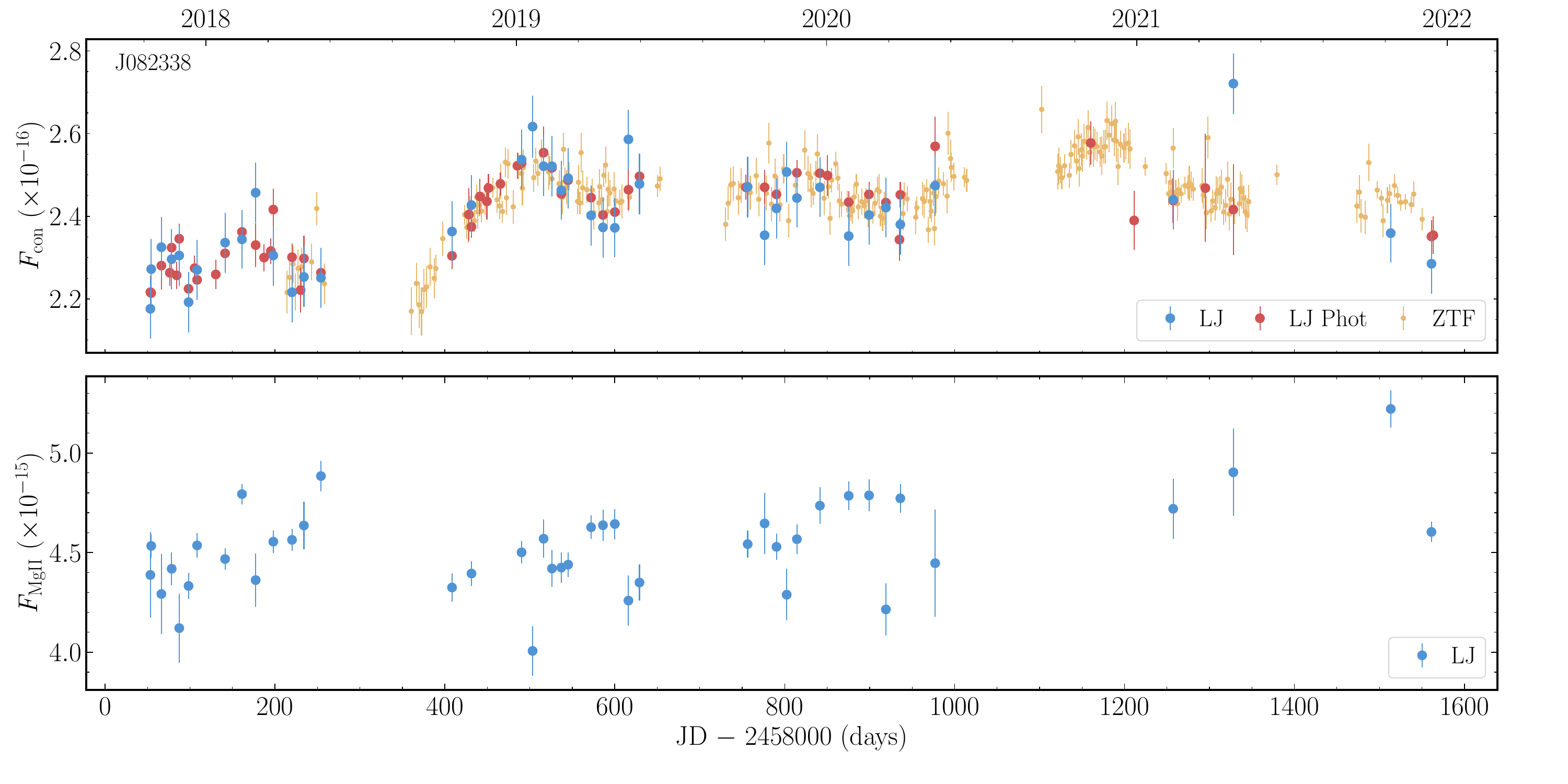}
\caption{Light curves of targets without successful lag measurements, plotted in the same manner as in Figure \ref{fig_lcs}. The complete figure set (10 images) is available in the online article.}
\label{fig_con_lc}
\end{figure*}

\begin{figure*}
\centering
\includegraphics[width=\textwidth]{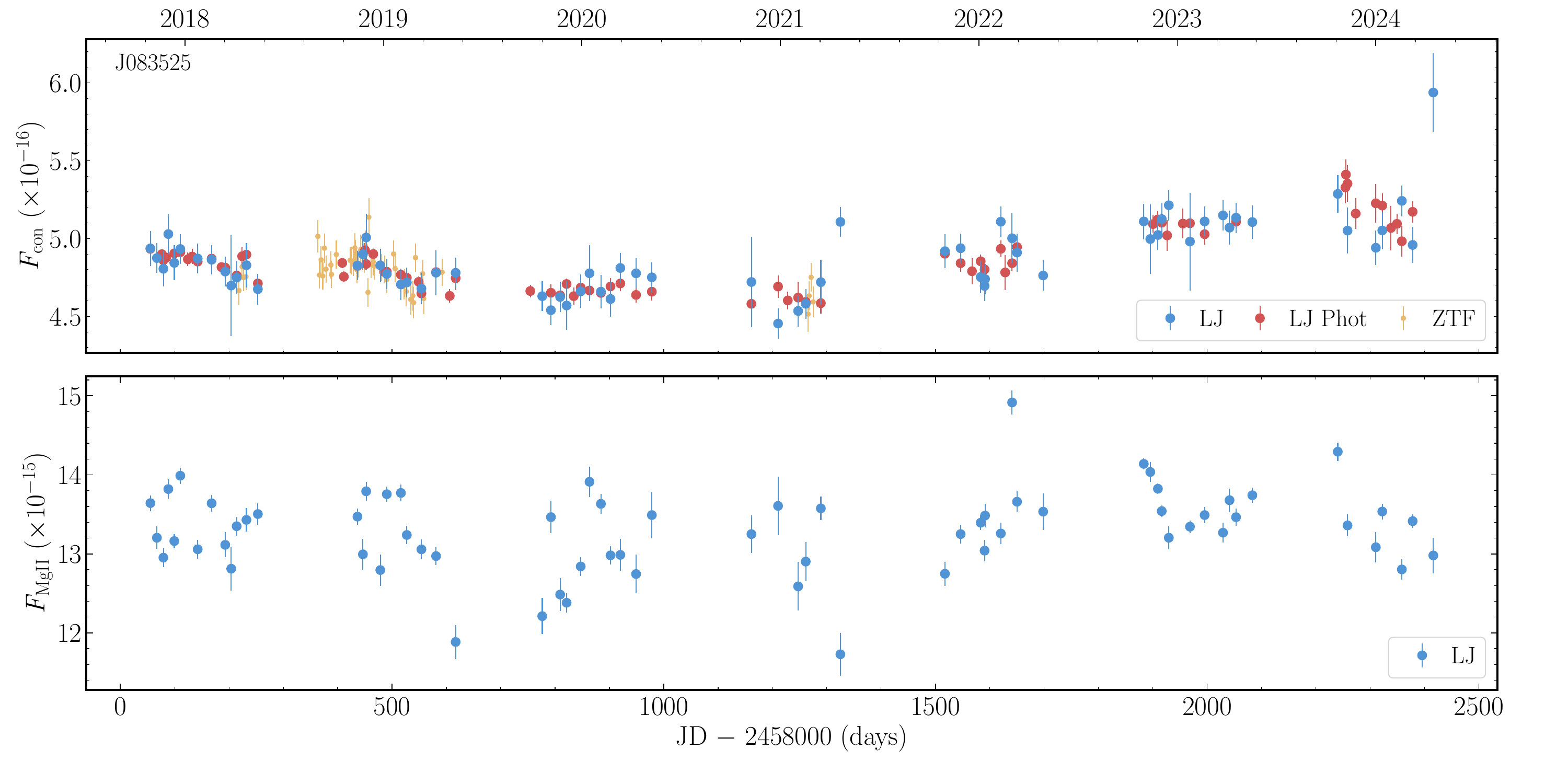}
\addtocounter{figure}{-1}
\caption{(Continued.) }
\end{figure*}

\begin{figure*}
\centering
\includegraphics[width=\textwidth]{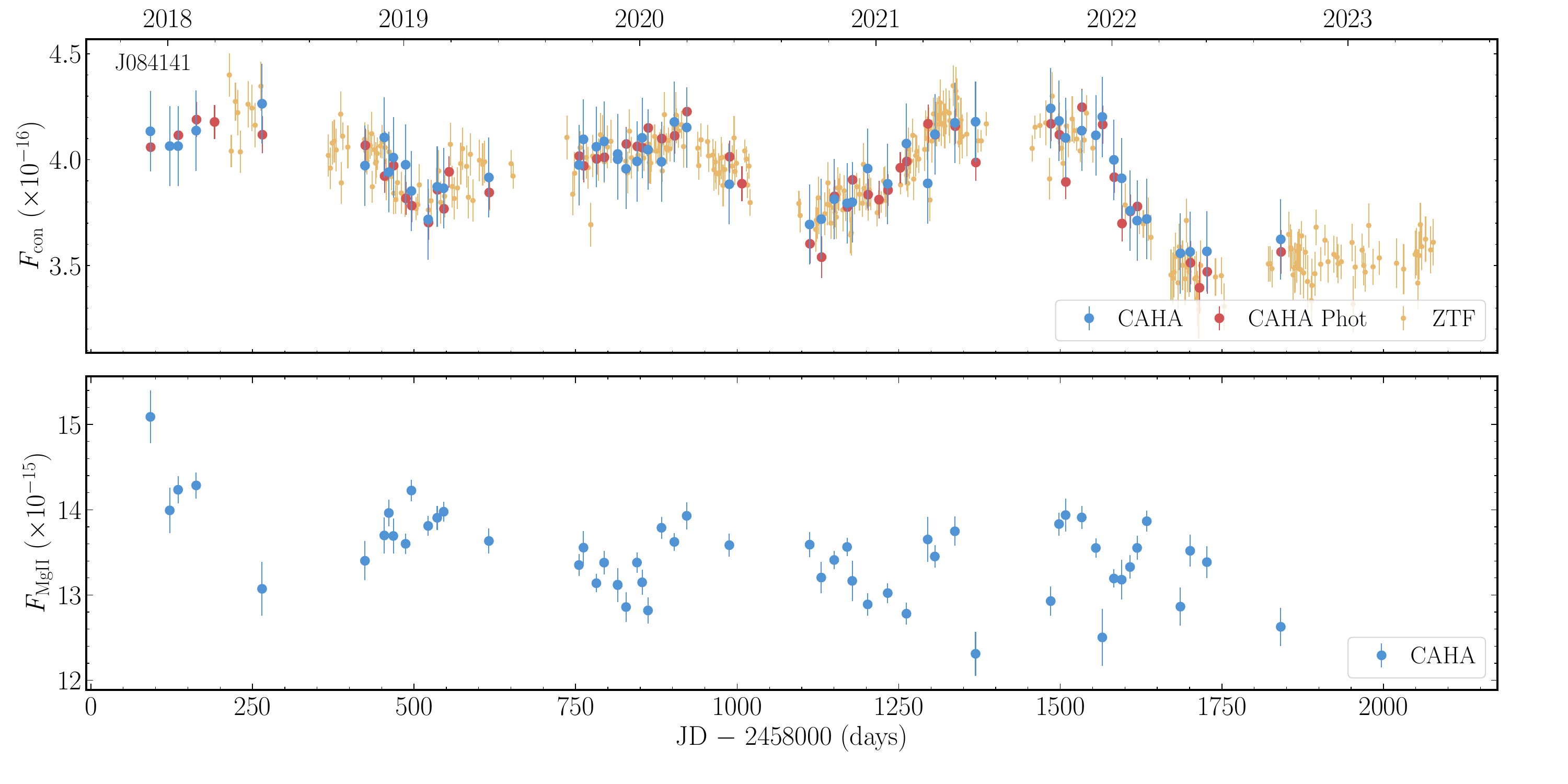}
\addtocounter{figure}{-1}
\caption{(Continued.) }
\end{figure*}

\begin{figure*}
\centering
\includegraphics[width=\textwidth]{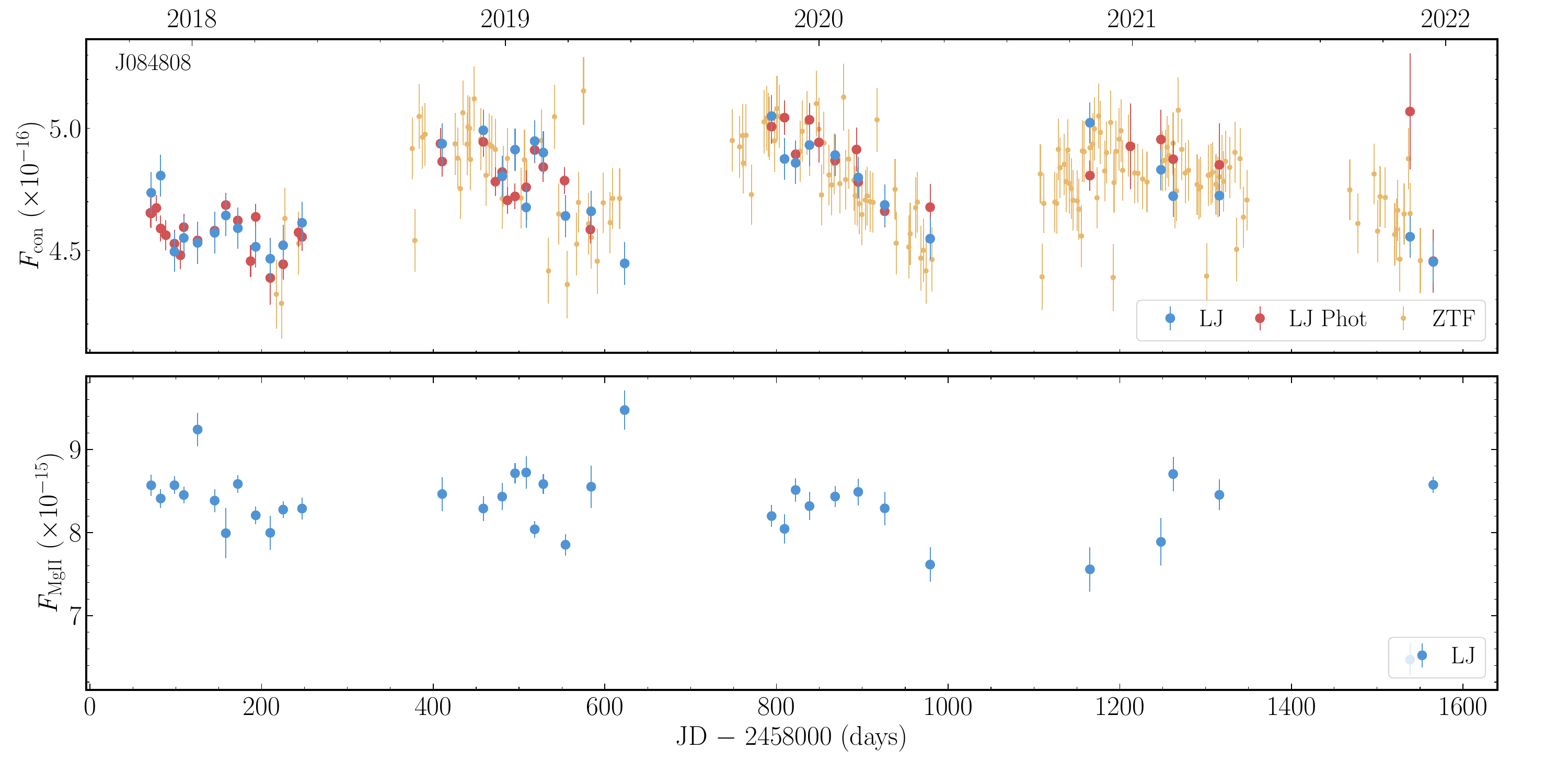}
\addtocounter{figure}{-1}
\caption{(Continued.) }
\end{figure*}

\begin{figure*}
\centering
\includegraphics[width=\textwidth]{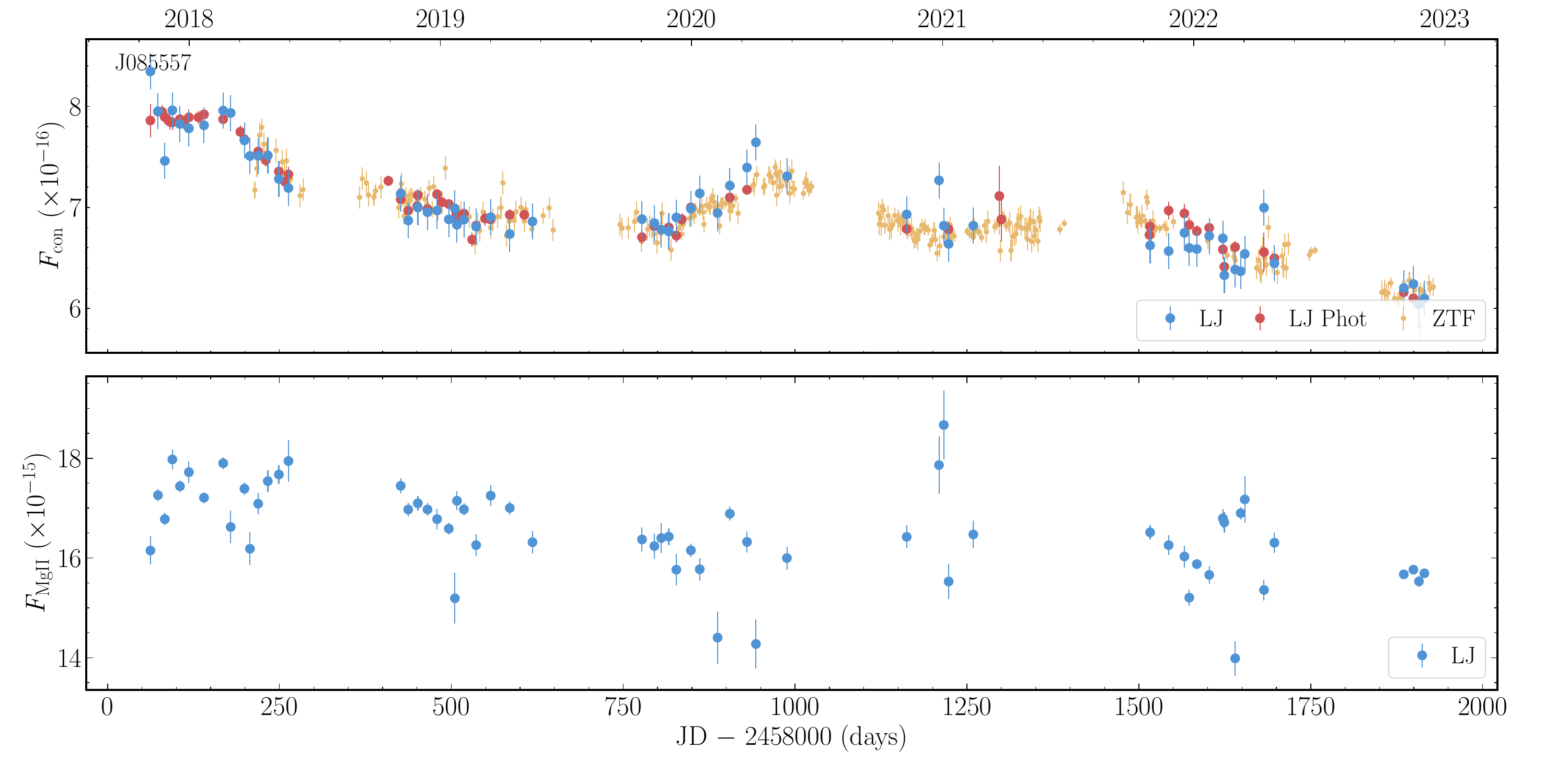}
\addtocounter{figure}{-1}
\caption{(Continued.) }
\end{figure*}

\begin{figure*}
\centering
\includegraphics[width=\textwidth]{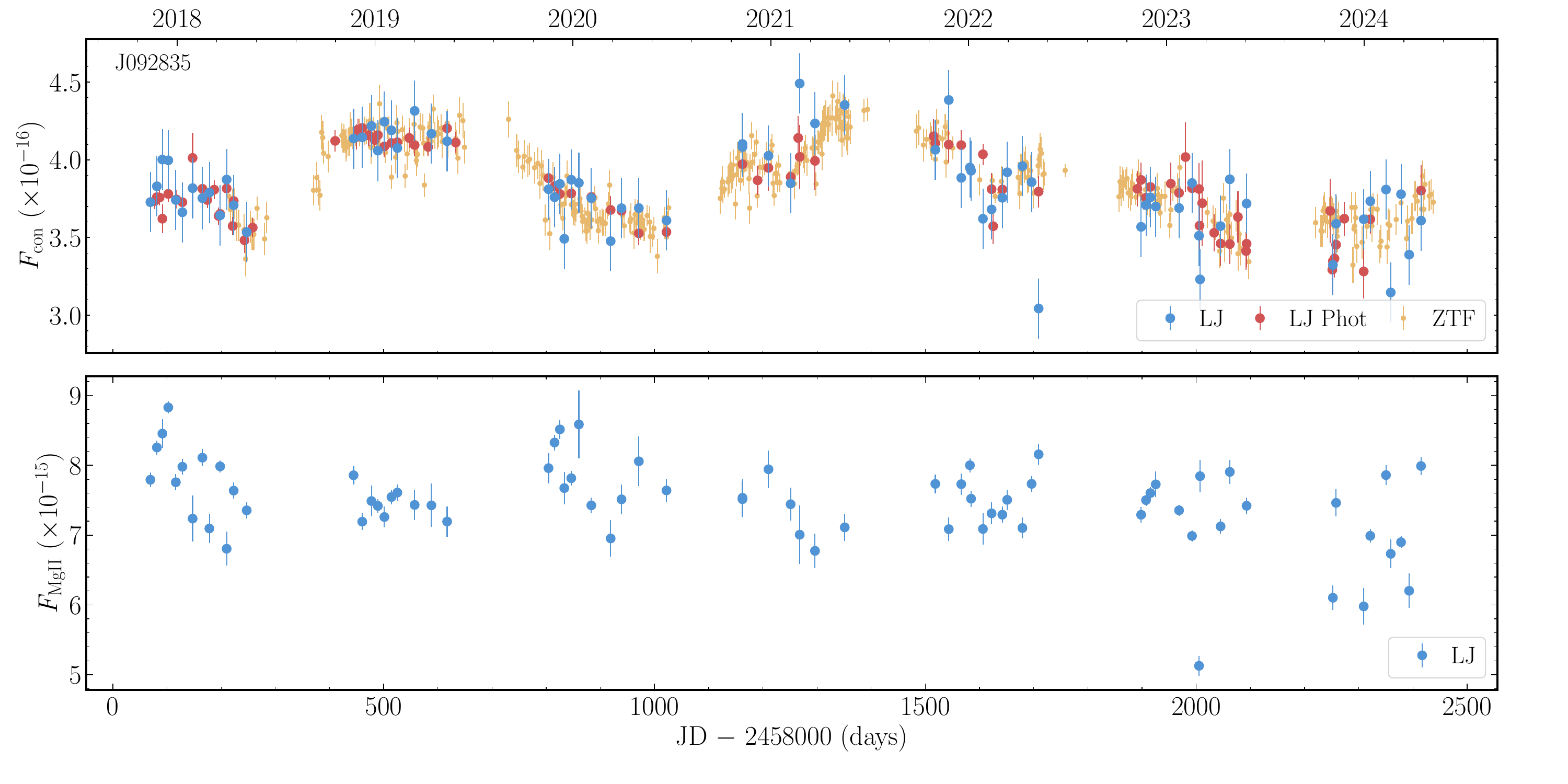}
\addtocounter{figure}{-1}
\caption{(Continued.) }
\end{figure*}

\begin{figure*}
\centering
\includegraphics[width=\textwidth]{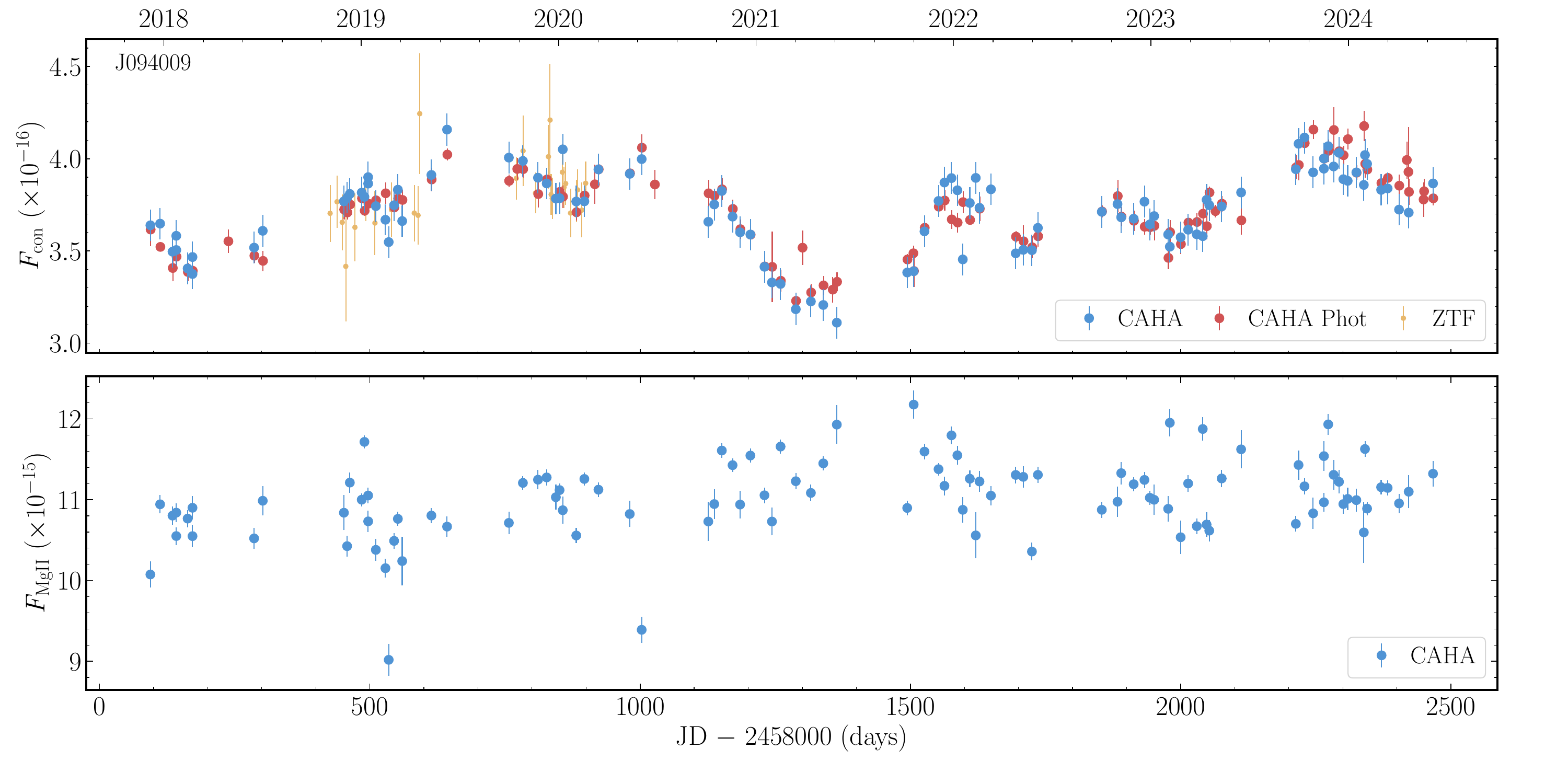}
\addtocounter{figure}{-1}
\caption{(Continued.) }
\end{figure*}

\begin{figure*}
\centering
\includegraphics[width=\textwidth]{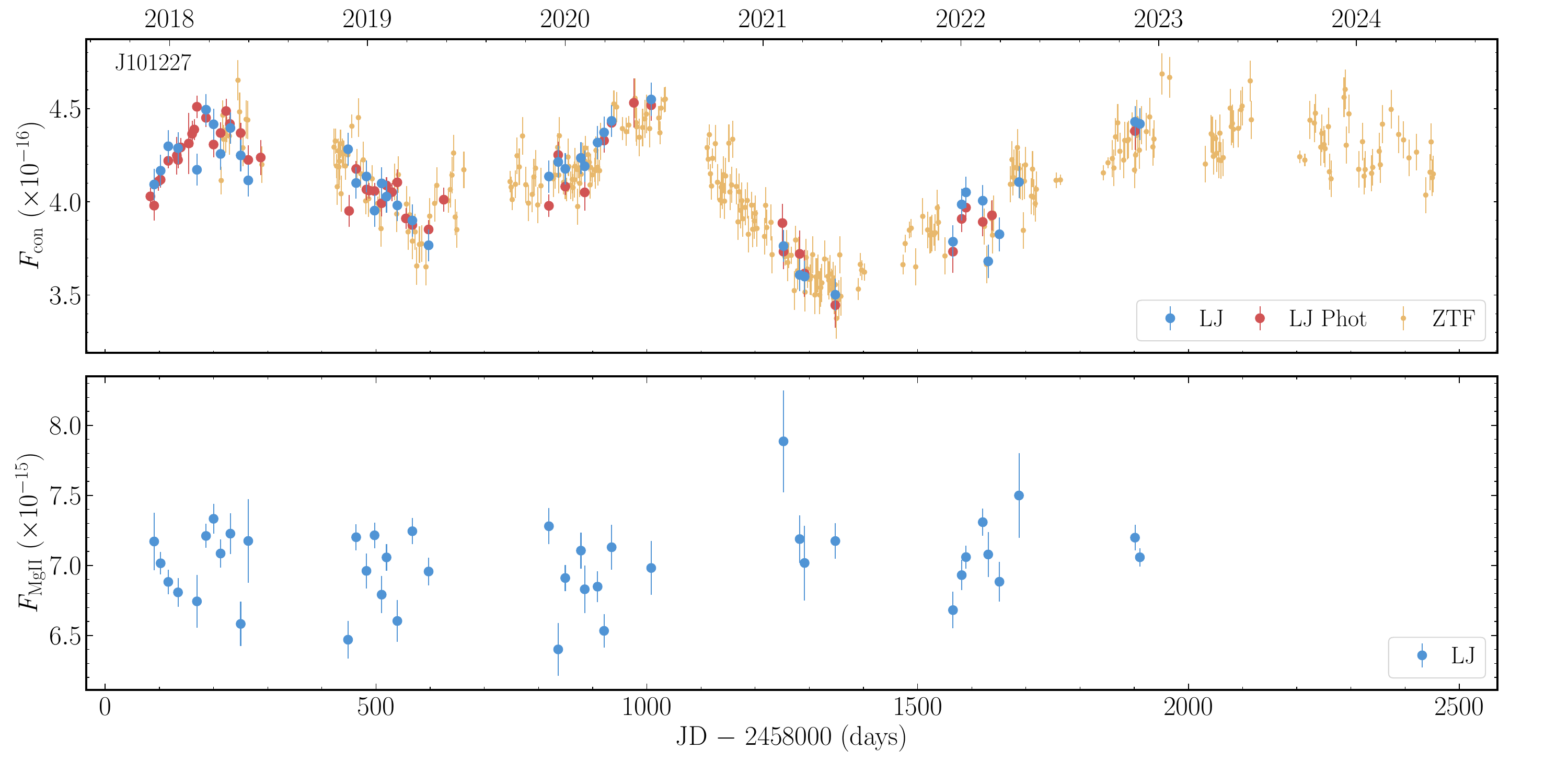}
\addtocounter{figure}{-1}
\caption{(Continued.) }
\end{figure*}

\begin{figure*}
\centering
\includegraphics[width=\textwidth]{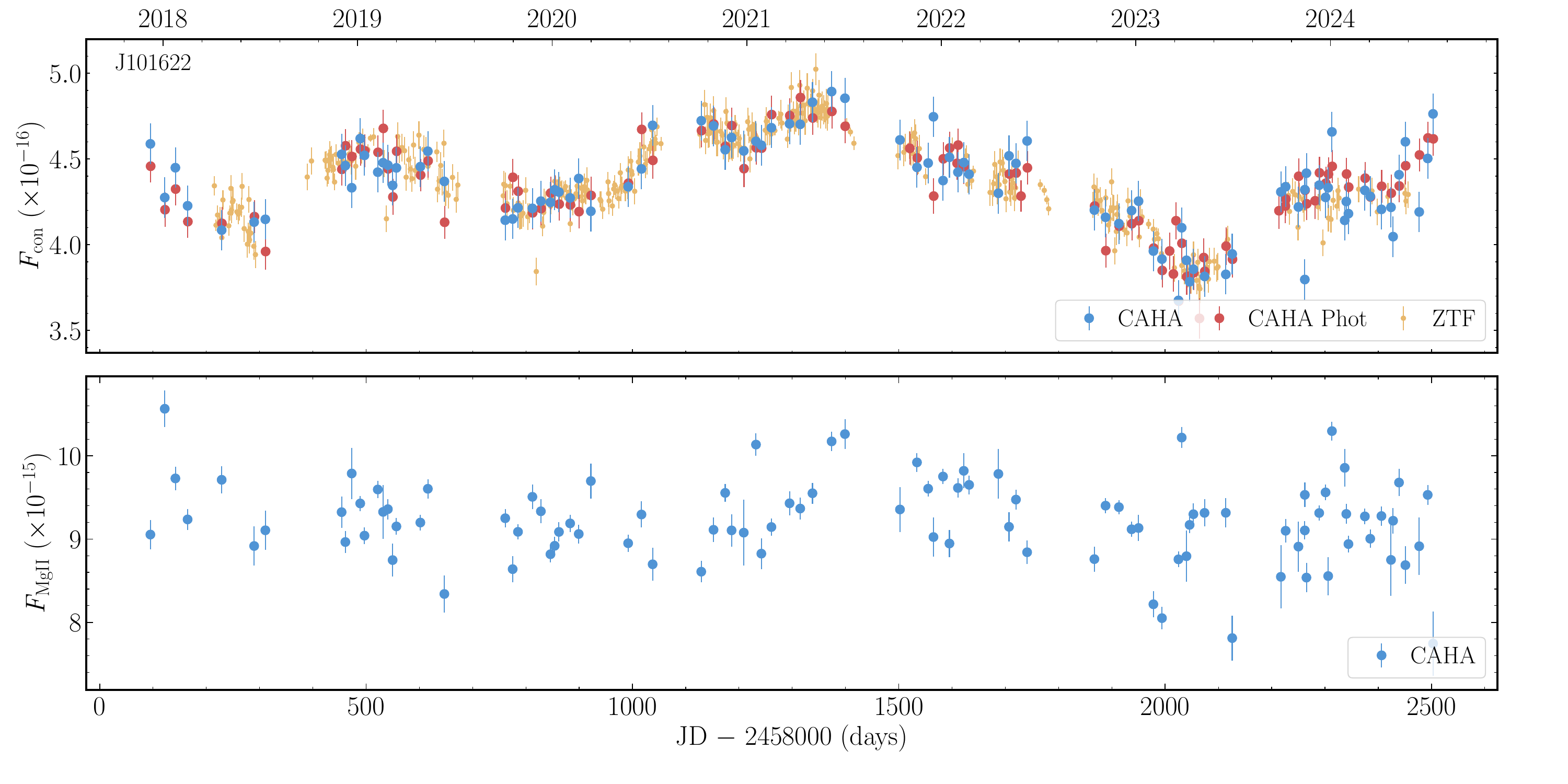}
\addtocounter{figure}{-1}
\caption{(Continued.) }
\end{figure*}

\begin{figure*}
\centering
\includegraphics[width=\textwidth]{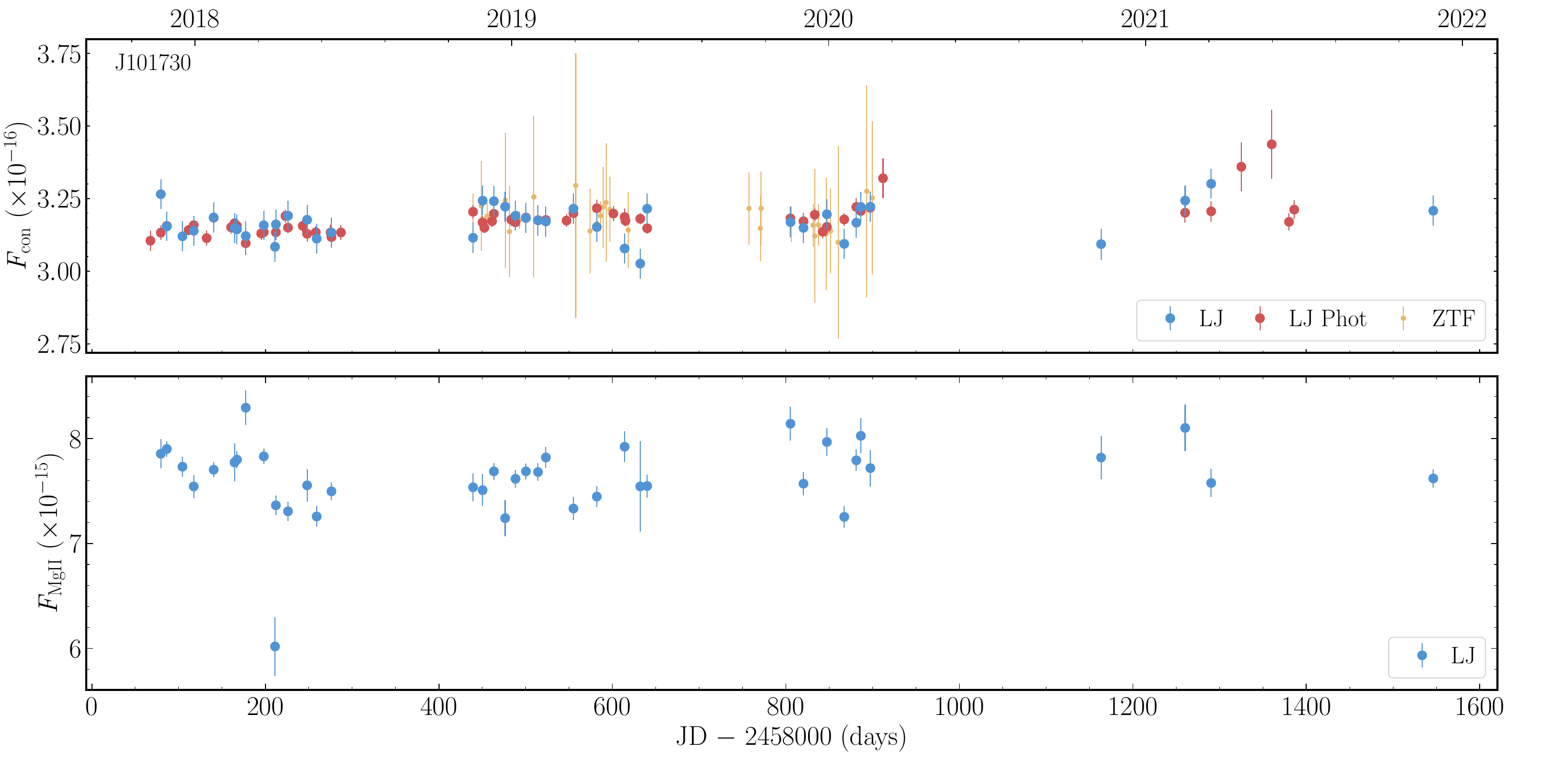}
\addtocounter{figure}{-1}
\caption{(Continued.) }
\end{figure*}

%%%%%%
\begin{figure*}
\centering
\includegraphics[width=\textwidth]{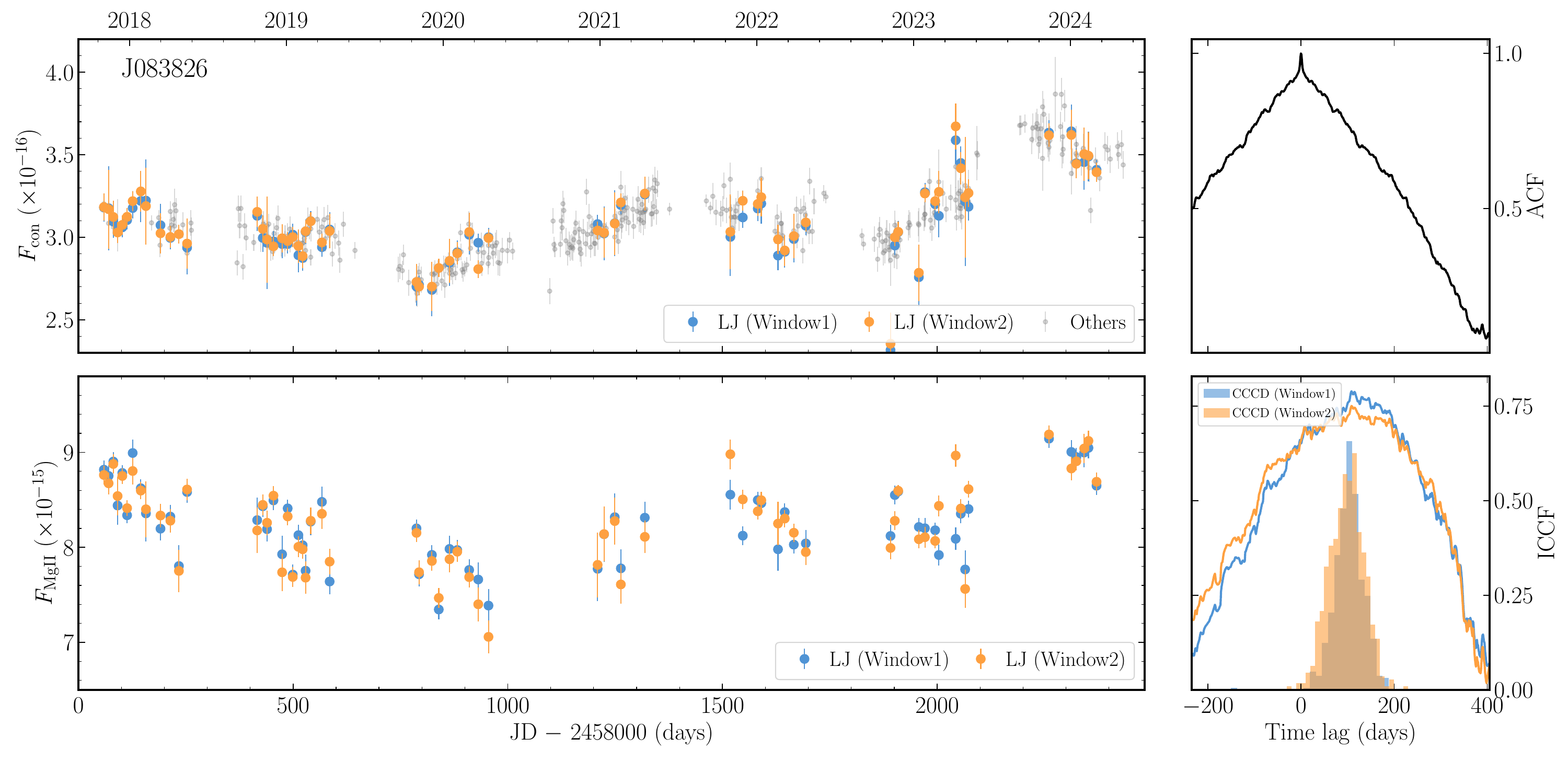}
\caption{Comparison between light curves and time-lag measurements for two fitting windows: Window 1 (2260-3050\,\text{\AA}) and Window 2 (2260-3400\,\text{\AA}). Meanings of panels are the same as Figure \ref{fig_lcs}. ``LJ/CAHA (Window1/2)'' represents the Lijiang/CAHA spectroscopic data using the corresponding fitting window. ``Others'' stands for the photometric data of ZTF or Lijiang/CAHA. The complete figure set (8 images) is available in the online article.}
\label{fig:compare_range}
\end{figure*}

\begin{figure*}
\centering
\includegraphics[width=\textwidth]{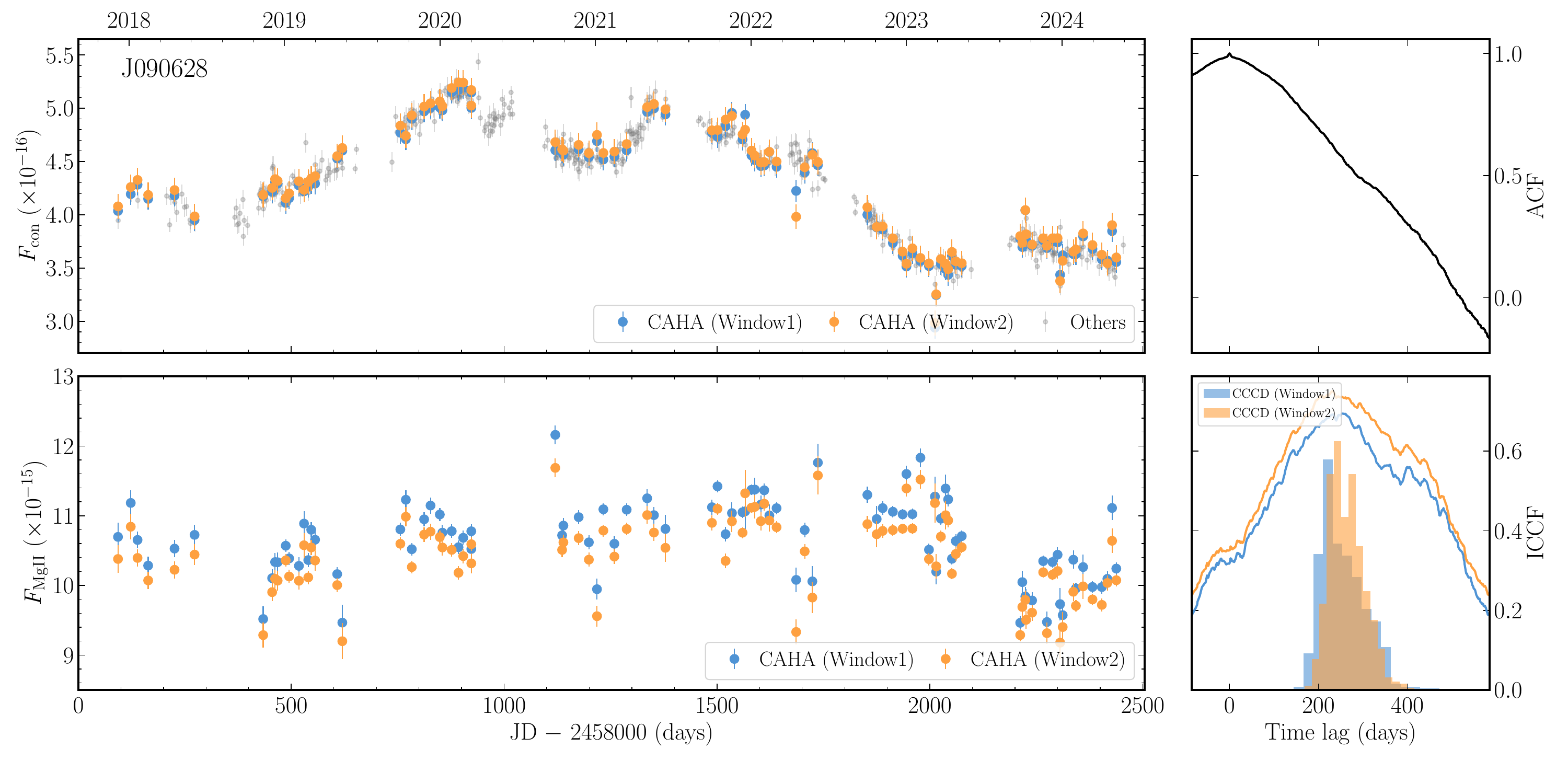}
\addtocounter{figure}{-1}
\caption{(Continued.) }
\end{figure*}
\begin{figure*}
\centering
\includegraphics[width=\textwidth]{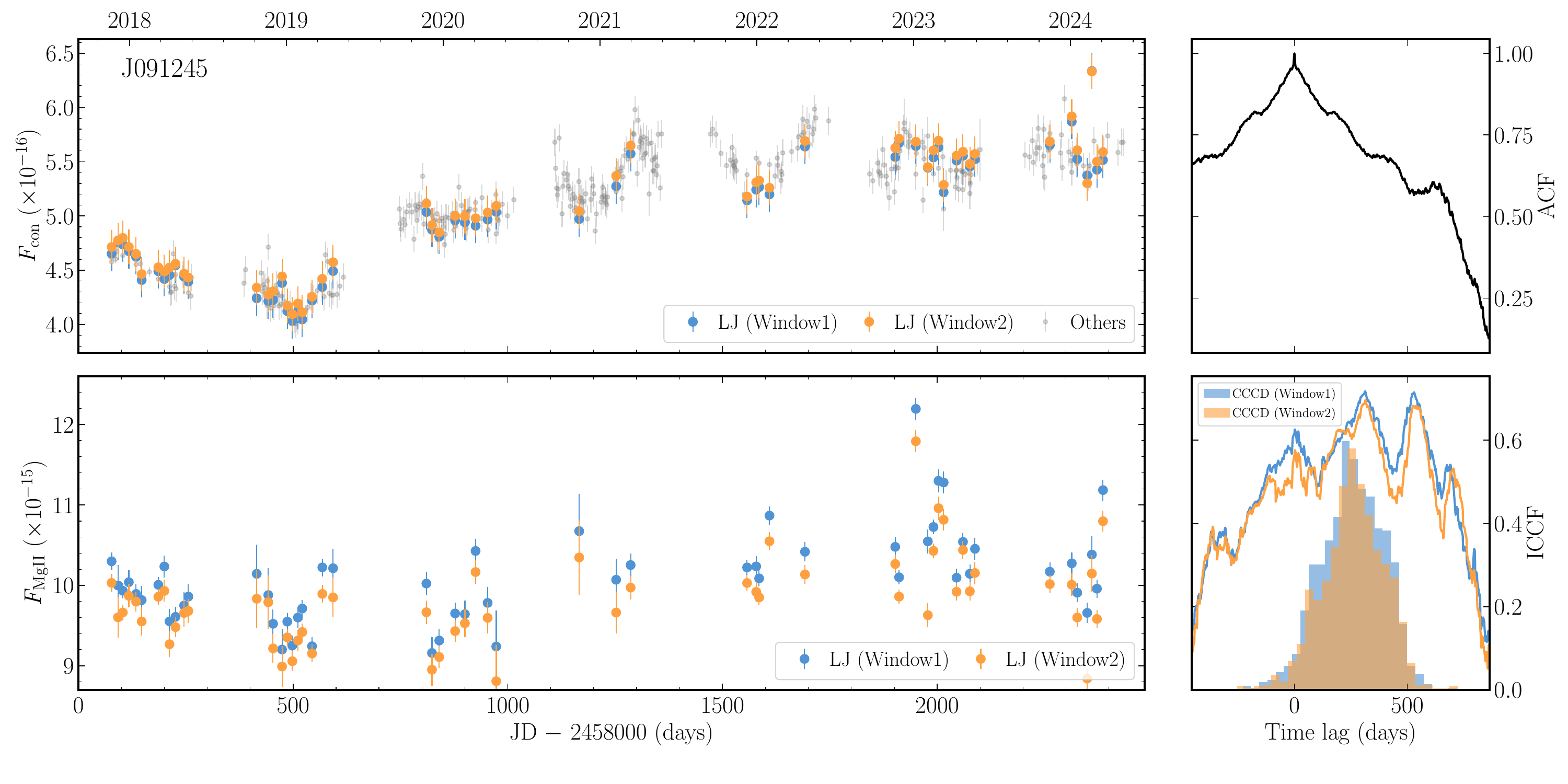}
\addtocounter{figure}{-1}
\caption{(Continued.) }
\end{figure*}
\begin{figure*}
\centering
\includegraphics[width=\textwidth]{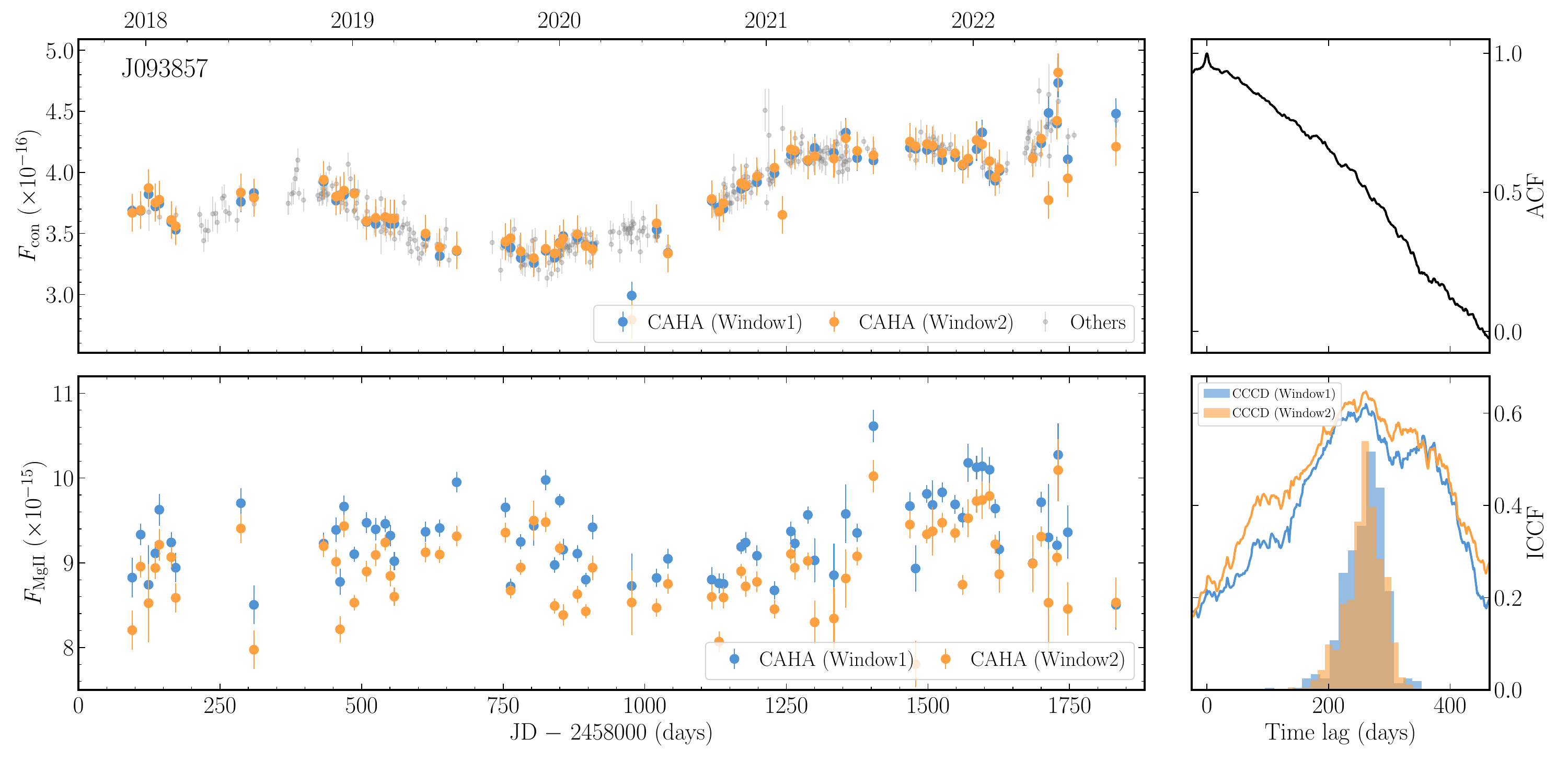}
\addtocounter{figure}{-1}
\caption{(Continued.) }
\end{figure*}
\begin{figure*}
\centering
\includegraphics[width=\textwidth]{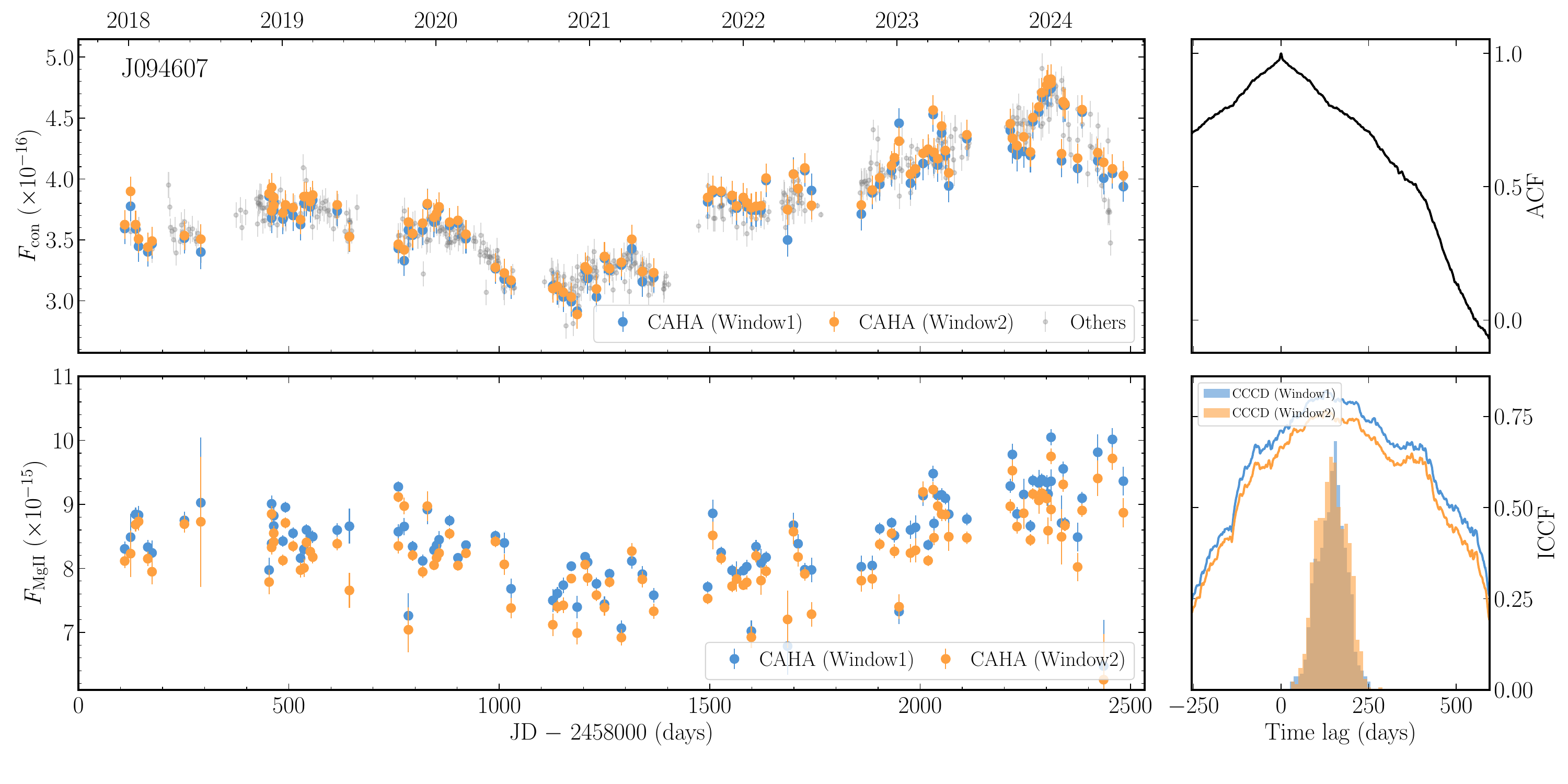}
\addtocounter{figure}{-1}
\caption{(Continued.) }
\end{figure*}
\begin{figure*}
\centering
\includegraphics[width=\textwidth]{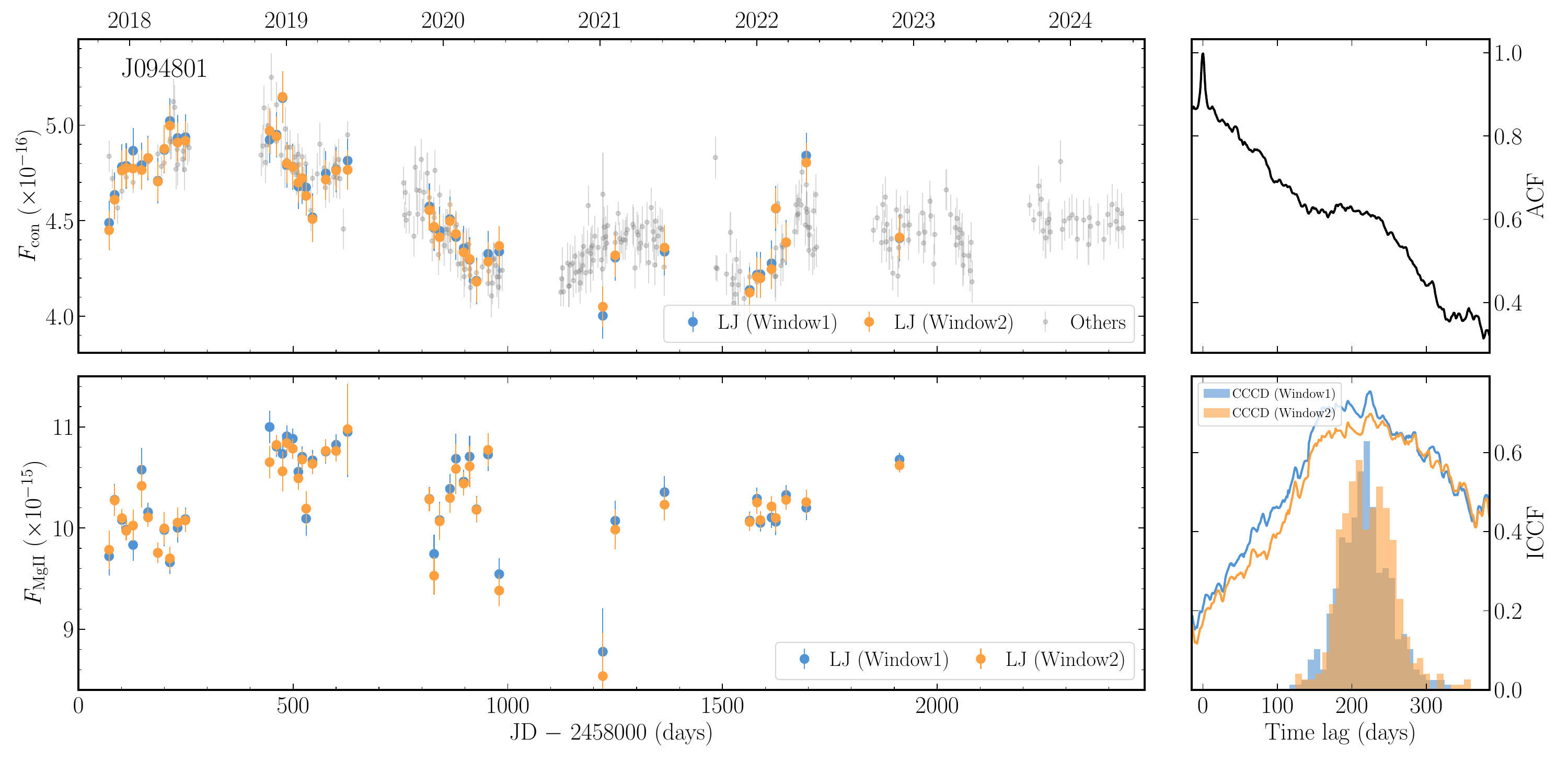}
\addtocounter{figure}{-1}
\caption{(Continued.) }
\end{figure*}
\begin{figure*}
\centering
\includegraphics[width=\textwidth]{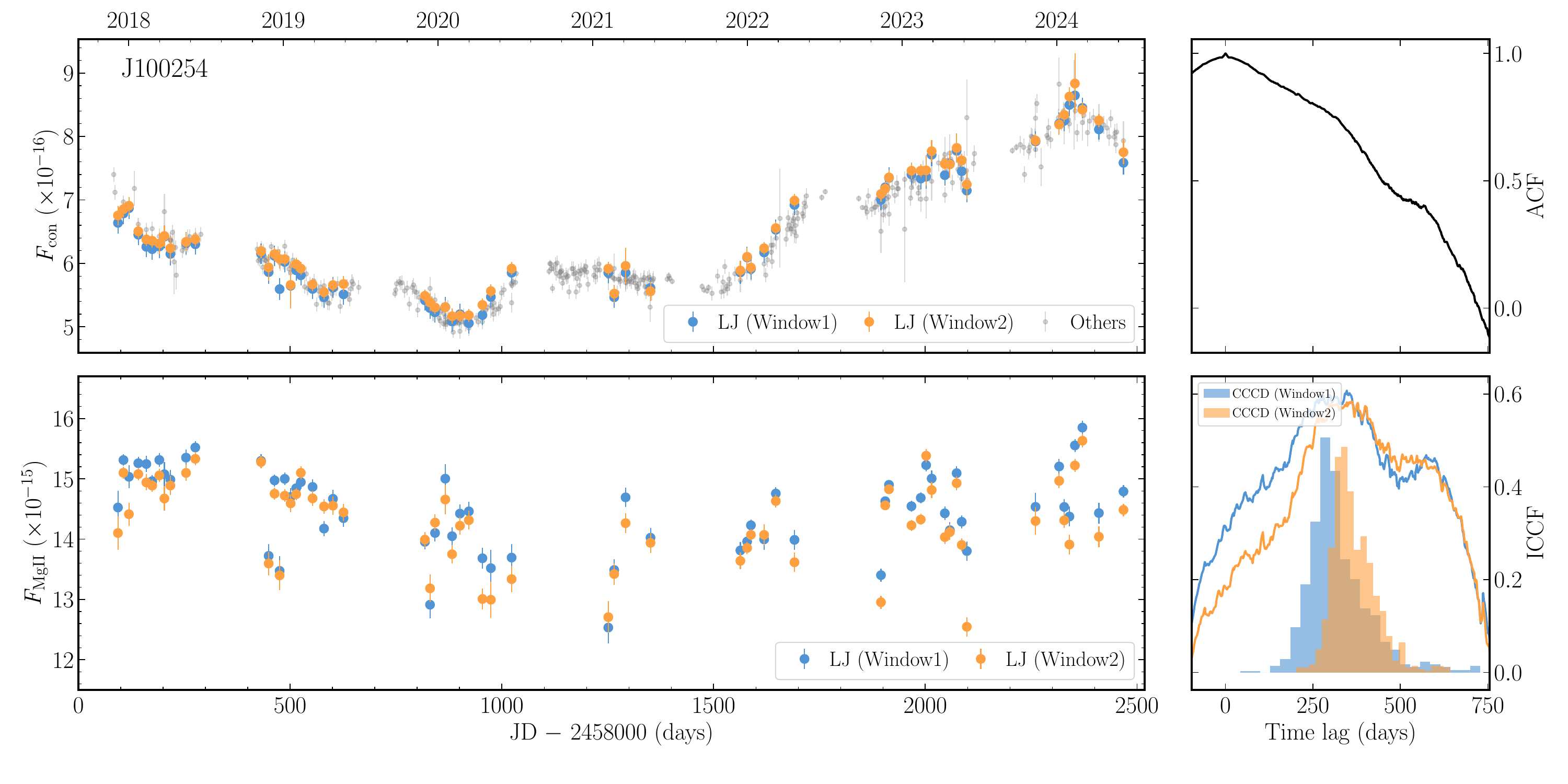}
\addtocounter{figure}{-1}
\caption{(Continued.) }
\end{figure*}
\begin{figure*}
\centering
\includegraphics[width=\textwidth]{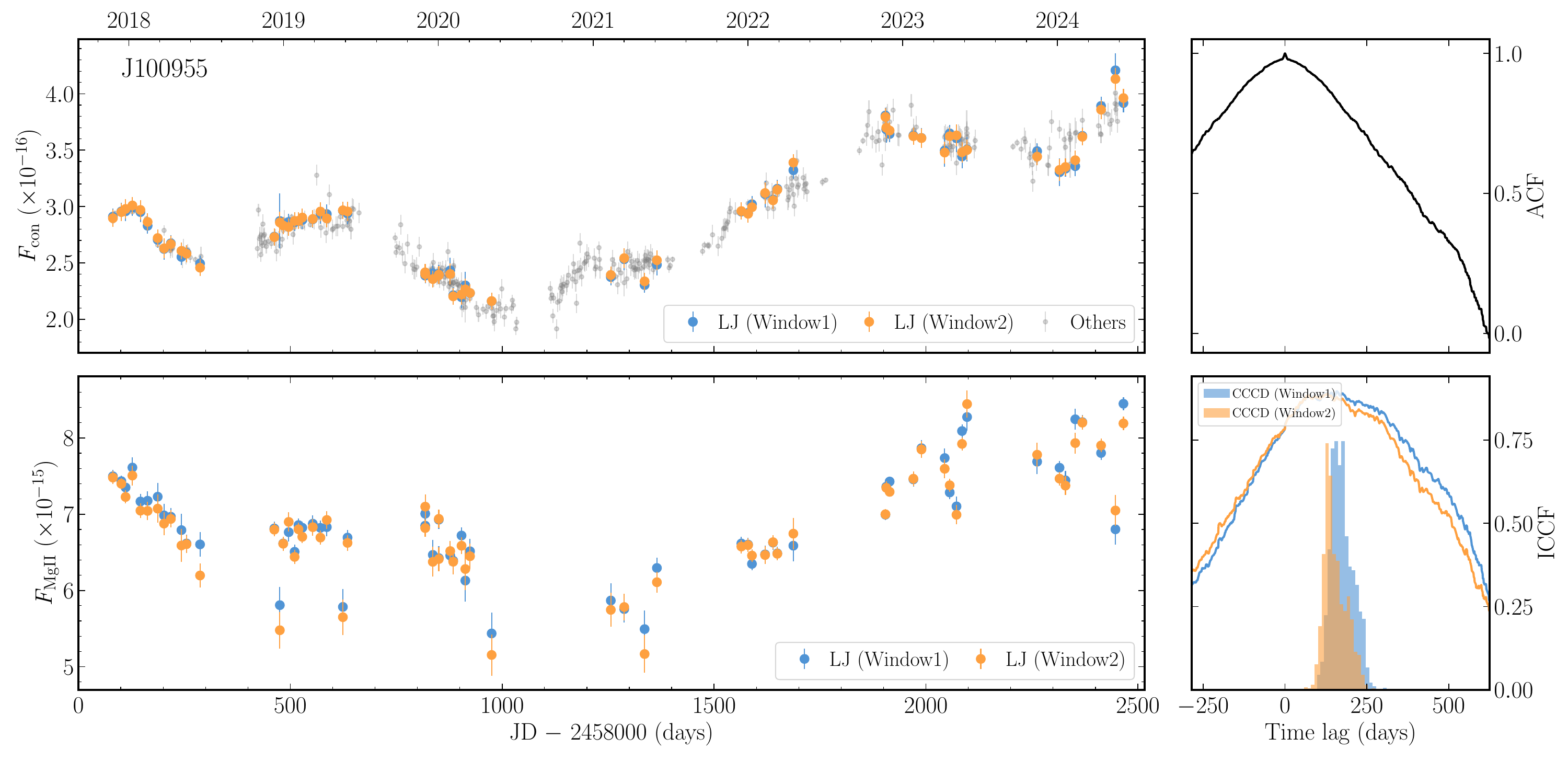}
\addtocounter{figure}{-1}
\caption{(Continued.) }
\end{figure*}

%%%%%%%%%

\begin{deluxetable}{cccc}
\tabletypesize{\footnotesize}
\tablecaption{The results of partial correlation analysis between $\Delta R_{\rm MgII}$ and $\log\dot{\mathscr{M}}$ when taking the lag $\tau_{\rm MgII}$ or the luminosity $L_{3000}$ as a confounding variable.  $\rho$ and $p$ are the partial correlation coefficient and the $p$-value, respectively.     \label{tab_partial}}
\tablehead{\colhead{~Correlation~}   & \colhead{Confounding Variable }
  &  ~~$\rho$&  ~~$p$~~
}
\startdata
\multirow{2}{*}{$\Delta R_{\rm MgII}$ vs. $\log\dot{\mathscr{M}}$} & $\tau_{\rm MgII}$ &-0.85&$9.0\times10^{-27}$\\ & 
$L_{3000}$&-0.70&$9.9\times10^{-15}$
\enddata
\end{deluxetable}


\begin{thebibliography}{99}
\bibitem[Bahk et al.(2019)]{Bahk2019} Bahk, H., Woo, J.-H., \& Park, D.\ 2019, \apj, 875, 50. doi:10.3847/1538-4357/ab100d
\bibitem[Barth et al.(2013)]{Barth2013} Barth, A.~J., Pancoast, A., Bennert, V.~N., et al.\ 2013, \apj, 769, 128. doi:10.1088/0004-637X/769/2/128
\bibitem[Barth et al.(2015)]{Barth2015} Barth, A.~J., Bennert, V.~N., Canalizo, G., et al.\ 2015, \apjs, 217, 26. doi:10.1088/0067-0049/217/2/26
\bibitem[Bentz et al.(2009)]{Bentz2009} Bentz, M.~C., Walsh, J.~L., Barth, A.~J., et al.\ 2009, \apj, 705, 199. doi:10.1088/0004-637X/705/1/199
\bibitem[Bentz et al.(2010)]{Bentz2010} Bentz, M.~C., Horne, K., Barth, A.~J., et al.\ 2010, \apjl, 720, L46. doi:10.1088/2041-8205/720/1/L46
\bibitem[Bentz et al.(2013)]{Bentz2013}
  Bentz, M.~C., Denney, K.~D., Grier, C.~J., et al.\ 2013, \apj, 767, 149. doi:10.1088/0004-637X/767/2/149
\bibitem[Blandford \& McKee(1982)]{BM1982} Blandford, R.~D. \& McKee, C.~F.\ 1982, \apj, 255, 419. doi:10.1086/159843
\bibitem[Boroson \& Green(1992)]{boroson1992} Boroson, T.~A. \& Green, R.~F.\ 1992, \apjs, 80, 109. doi:10.1086/191661
\bibitem[Boroson(2002)]{Boroson2002} Boroson, T.~A.\ 2002, \apj, 565, 78. doi:10.1086/324486
\bibitem[Cackett et al.(2015)]{Cackett2015} Cackett, E.~M., G{\"u}ltekin, K., Bentz, M.~C., et al.\ 2015, \apj, 810, 86. doi:10.1088/0004-637X/810/2/86
\bibitem[Chen et al.(2023)]{chen2023} Chen, Y.-J., Liu, J.-R., Zhai, S., et al.\ 2023, \mnras, 522, 3, 3439. doi:10.1093/mnras/stad1136
\bibitem[Czerny et al.(2019)]{Czerny2019} Czerny, B., Olejak, A., Ra{\l}owski, M., et al.\ 2019, \apj, 880, 46. doi:10.3847/1538-4357/ab2913
\bibitem[De Rosa et al.(2018)]{DeRosa2018} De Rosa, G., Fausnaugh, M.~M., Grier, C.~J., et al.\ 2018, \apj, 866, 133. doi:10.3847/1538-4357/aadd11
\bibitem[Dong et al.(2011)]{dong2011} Dong, X.-B., Wang, J.-G., Ho, L.~C., et al.\ 2011, \apj, 736, 2, 86. doi:10.1088/0004-637X/736/2/86
\bibitem[Du et al.(2014)]{Du2014}
  Du, P., Hu, C., Lu, K.-X., et al.\ 2014, \apj, 782, 45. doi:10.1088/0004-637X/782/1/45
\bibitem[Du et al.(2015)]{Du2015} Du, P., Hu, C., Lu, K.-X., et al.\ 2015, \apj, 806, 22. doi:10.1088/0004-637X/806/1/22
\bibitem[Du et al.(2016a)]{Du2016} Du, P., Lu, K.-X., Zhang, Z.-X., et al.\ 2016a, \apj, 825, 126. doi:10.3847/0004-637X/825/2/126
\bibitem[Du et al.(2016b)]{du2016FP} Du, P., Wang, J.-M., Hu, C., et al.\ 2016b, \apjl,  818, 1, L14. doi:10.3847/2041-8205/818/1/L14
\bibitem[Du et al.(2016c)]{Du2016VI} Du, P., Lu, K.-X., Hu, C., et al.\ 2016c, \apj, 820, 1, 27. doi:10.3847/0004-637X/820/1/27
\bibitem[Du et al.(2018a)]{Du2018} Du, P., Zhang, Z.-X., Wang, K., et al.\ 2018a, \apj, 856, 1, 6. doi:10.3847/1538-4357/aaae6b
\bibitem[Du et al.(2018b)]{Du2018mahaI} Du, P., Brotherton, M.~S., Wang, K., et al.\ 2018b, \apj, 869, 142. doi:10.3847/1538-4357/aaed2c
\bibitem[Du \& Wang(2019)]{DW2019} Du, P. \& Wang, J.-M.\ 2019, \apj, 886, 42. doi:10.3847/1538-4357/ab4908
\bibitem[Du et al.(2023)]{Du2023} Du, P., Zhai, S., \& Wang, J.-M.\ 2023, \apj, 942, 112. doi:10.3847/1538-4357/aca52a
\bibitem[Du(2024)]{Du2024} Du, P.\ 2024, Zenodo, v0.8.1. doi:10.5281/zenodo.12702951
\bibitem[Edelson et al.(2002)]{Edelson2002} Edelson, R., Turner, T.~J., Pounds, K., et al.\ 2002, \apj, 568, 610. doi:10.1086/323779
\bibitem[Fausnaugh et al.(2016)]{Fausnaugh2016} Fausnaugh, M.~M., Denney, K.~D., Barth, A.~J., et al.\ 2016, \apj, 821, 56. doi:10.3847/0004-637X/821/1/56
\bibitem[Ferrarese \& Ford(2005)]{FF2005} Ferrarese, L. \& Ford, H.\ 2005, \ssr, 116, 523. doi:10.1007/s11214-005-3947-6
\bibitem[Gaskell \& Peterson(1987)]{Gaskell1987} Gaskell, C.~M. \& Peterson, B.~M.\ 1987, \apjs, 65, 1. doi:10.1086/191216
\bibitem[Goad et al.(1999)]{Goad1999} Goad, M.~R., Koratkar, A.~P., Axon, D.~J., et al.\ 1999, \apjl, 512, L95. doi:10.1086/311884
\bibitem[GRAVITY Collaboration et al.(2018)]{GRAVITY2018} GRAVITY Collaboration, Sturm, E., Dexter, J., et al.\ 2018, \nat, 563, 657. doi:10.1038/s41586-018-0731-9
\bibitem[GRAVITY Collaboration et al.(2020)]{GRAVITY2020} GRAVITY Collaboration, Amorim, A., Baub{\"o}ck, M., et al.\ 2020, \aap, 643, A154. doi:10.1051/0004-6361/202039067
\bibitem[GRAVITY Collaboration et al.(2021)]{GRAVITY2021} GRAVITY Collaboration, Amorim, A., Baub{\"o}ck, M., et al.\ 2021, \aap,  648, A117. doi:10.1051/0004-6361/202040061
\bibitem[Greene \& Ho(2005)]{Greene2005} Greene, J.~E. \& Ho, L.~C.\ 2005, \apj, 630, 122. doi:10.1086/431897
\bibitem[Grier et al.(2012)]{Grier2012} Grier, C.~J., Peterson, B.~M., Pogge, R.~W., et al.\ 2012, \apj, 755, 60. doi:10.1088/0004-637X/755/1/60
\bibitem[Grier et al.(2013)]{Grier2013} Grier, C.~J., Martini, P., Watson, L.~C., et al.\ 2013, \apj, 773, 90. doi:10.1088/0004-637X/773/2/90
\bibitem[Guo et al.(2020)]{Guo2020} Guo, H., Shen, Y., He, Z., et al.\ 2020, \apj, 888, 58. doi:10.3847/1538-4357/ab5db0
\bibitem[Ho et al.(2012)]{Ho2012} Ho, L.~C., Goldoni, P., Dong, X.-B., et al.\ 2012, \apj, 754, 1, 11. doi:10.1088/0004-637X/754/1/11
\bibitem[Ho \& Kim(2014)]{Ho2014} Ho, L.~C. \& Kim, M.\ 2014, \apj,  789, 1, 17. doi:10.1088/0004-637X/789/1/17
\bibitem[Homayouni et al.(2020)]{Homayouni2020} Homayouni, Y., Trump, J.~R., Grier, C.~J., et al.\ 2020, \apj, 901, 55. doi:10.3847/1538-4357/ababa9
\bibitem[Hryniewicz et al.(2014)]{Hryniewicz2014} Hryniewicz, K., Czerny, B., Pych, W., et al.\ 2014, \aap, 562, A34. doi:10.1051/0004-6361/201322487
\bibitem[Hu et al.(2008)]{Hu2008} Hu, C., Wang, J.-M., Ho, L.~C., et al.\ 2008, \apj, 687, 1, 78. doi:10.1086/591838
\bibitem[Hu et al.(2015)]{Hu2015} Hu, C., Du, P., Lu, K.-X., et al.\ 2015, \apj,  804, 2, 138. doi:10.1088/0004-637X/804/2/138
\bibitem[Hu et al.(2021)]{Hu2021} Hu, C., Li, S.-S., Yang, S., et al.\ 2021, \apjs, 253, 20. doi:10.3847/1538-4365/abd774
\bibitem[Hu et al.(2025)]{Hu2025} Hu, C., Yao, Z.-H., Chen, Y.-J., et al.\ 2025, \apjs, 278, 2, 61. doi:10.3847/1538-4365/add40b
\bibitem[Jiang et al.(2024)]{jiang2024} Jiang, D., Onoue, M., Jiang, L., et al.\ 2024, \apj,  975, 2, 214. doi:10.3847/1538-4357/ad7d09
\bibitem[Kaspi et al.(2000)]{Kaspi2000} Kaspi, S., Smith, P.~S., Netzer, H., et al.\ 2000, \apj, 533, 631. doi:10.1086/308704
\bibitem[Kelly(2007)]{Kelly2007} Kelly, B.~C.\ 2007, \apj, 665, 1489. doi:10.1086/519947
\bibitem[Khadka et al.(2022)]{Khadka2022} Khadka, N., Zaja{\v{c}}ek, M., Panda, S., et al.\ 2022, \mnras,  515, 3, 3729. doi:10.1093/mnras/stac1940
\bibitem[Kokubo et al.(2014)]{Kokubo2014} Kokubo, M., Morokuma, T., Minezaki, T., et al.\ 2014, \apj, 783, 46. doi:10.1088/0004-637X/783/1/46
\bibitem[Korista \& Goad(2000)]{KG00} Korista, K.~T. \& Goad, M.~R.\ 2000, \apj, 536, 284. doi:10.1086/308930
\bibitem[Korista \& Goad(2004)]{KG04} Korista, K.~T. \& Goad, M.~R.\ 2004, \apj, 606, 2, 749. doi:10.1086/383193
\bibitem[Kormendy \& Ho(2013)]{KH2013} Kormendy, J. \& Ho, L.~C.\ 2013, \araa, 51, 511.
\bibitem[Kova{\v{c}}evi{\'c}-Doj{\v{c}}inovi{\'c} \& Popovi{\'c}(2015)]{Kovacevic2015} Kova{\v{c}}evi{\'c}-Doj{\v{c}}inovi{\'c}, J. \& Popovi{\'c}, L. {\v{C}}.\ 2015, \apjs,  221, 2, 35. doi:10.1088/0067-0049/221/2/35
\bibitem[Le et al.(2020)]{Le2020} Le, H.~A.~N., Woo, J.-H., \& Xue, Y.\ 2020, \apj, 901, 35. doi:10.3847/1538-4357/abada0
\bibitem[Li et al.(2014)]{Li2014}
  Li, Y.-R., Wang, J.-M., Hu, C., et al.\ 2014, \apjl, 786, L6. doi:10.1088/2041-8205/786/1/L6
\bibitem[Li et al.(2016)]{Li2016}
  Li, Y.-R., Wang, J.-M., \& Bai, J.-M.\ 2016, \apj, 831, 206. doi:10.3847/0004-637X/831/2/206
\bibitem[Li et al.(2018)]{Li2018} Li, Y.-R., Songsheng, Y.-Y., Qiu, J., et al.\ 2018, \apj,  869, 2, 137. doi:10.3847/1538-4357/aaee6b
\bibitem[Li et al.(2022)]{Li2022} Li, Y.-R., Wang, J.-M., Songsheng, Y.-Y., et al.\ 2022, \apj,  927, 1, 58. doi:10.3847/1538-4357/ac4bcb
\bibitem[Li(2024a)]{li2024pycali} Li, Y.-R.\ 2024a, Zenodo, v0.2.3. doi:10.5281/zenodo.10700132
\bibitem[Li(2024b)]{li2024mica} Li, Y.-R.\ 2024b,  Zenodo, v2.1.3. doi:10.5281/zenodo.11082109
\bibitem[Li et al.(2025)]{Li2025} Li, Y.-R., Shangguan, J., Wang, J.-M., et al.\ 2025, \apj, 988, 1, 42. doi:10.3847/1538-4357/addf40
\bibitem[Lira et al.(2018)]{Lira2018} Lira, P., Kaspi, S., Netzer, H., et al.\ 2018, \apj, 865, 56. doi:10.3847/1538-4357/aada45
\bibitem[Maoz et al.(1990)]{Maoz1990}
  Maoz, D., Netzer, H., Leibowitz, E., et al.\ 1990, \apj, 351, 75. doi:10.1086/168445
\bibitem[Mart{\'\i}nez-Aldama et al.(2020)]{MA2020} Mart{\'\i}nez-Aldama, M.~L., Zaja{\v{c}}ek, M., Czerny, B., et al.\ 2020, \apj, 903, 86. doi:10.3847/1538-4357/abb6f8
\bibitem[Marziani et al.(2001)]{marziani2001} Marziani, P., Sulentic, J.~W., Zwitter, T., et al.\ 2001, \apj,  558, 2, 553. doi:10.1086/322286
\bibitem[Marziani et al.(2013)]{Marziani2013} Marziani, P., Sulentic, J.~W., Plauchu-Frayn, I., et al.\ 2013, \aap,  555, A89. doi:10.1051/0004-6361/201321374
\bibitem[Masci et al.(2019)]{Masci2019} Masci, F.~J., Laher, R.~R., Rusholme, B., et al.\ 2019, \pasp,  131, 995, 018003. doi:10.1088/1538-3873/aae8ac
\bibitem[McGill et al.(2008)]{McGill2008} McGill, K.~L., Woo, J.-H., Treu, T., et al.\ 2008, \apj, 673, 703. doi:10.1086/524349
\bibitem[McLure \& Jarvis(2002)]{MJ2002} McLure, R.~J. \& Jarvis, M.~J.\ 2002, \mnras, 337, 109. doi:10.1046/j.1365-8711.2002.05871.x
\bibitem[Mej{\'\i}a-Restrepo et al.(2016)]{Mejia-Restrepo2016} Mej{\'\i}a-Restrepo, J.~E., Trakhtenbrot, B., Lira, P., et al.\ 2016, \mnras, 460, 1, 187. doi:10.1093/mnras/stw568
\bibitem[Metzroth et al.(2006)]{MOP2006} Metzroth, K.~G., Onken, C.~A., \& Peterson, B.~M.\ 2006, \apj, 647, 901. doi:10.1086/505525
\bibitem[Morton(1991)]{Morton1991} Morton, D.~C.\ 1991, \apjs, 77, 119. doi:10.1086/191601
\bibitem[Onken \& Peterson(2002)]{OP2002} Onken, C.~A. \& Peterson, B.~M.\ 2002, \apj, 572, 746. doi:10.1086/340351
\bibitem[Onken et al.(2004)]{Onken2004} Onken, C.~A., Ferrarese, L., Merritt, D., et al.\ 2004, \apj,  615, 2, 645. doi:10.1086/424655
\bibitem[Pan et al.(2025)]{pan2025} Pan, Z., Jiang, L., Guo, W.-J., et al.\ 2025, \apj, 987, 1, 48. doi:10.3847/1538-4357/add7dd
\bibitem[Pancoast et al.(2011)]{pancoast2011} Pancoast, A., Brewer, B.~J., \& Treu, T.\ 2011, \apj, 730, 2, 139. doi:10.1088/0004-637X/730/2/139
\bibitem[Pancoast et al.(2014)]{pancoast2014} Pancoast, A., Brewer, B.~J., Treu, T., et al.\ 2014, \mnras,  445, 3, 3073. doi:10.1093/mnras/stu1419
\bibitem[Panda et al.(2019)]{Panda2019MS} Panda, S., Marziani, P., \& Czerny, B.\ 2019, \apj,  882, 2, 79. doi:10.3847/1538-4357/ab3292
\bibitem[Peterson(1993)]{Peterson1993} Peterson, B.~M.\ 1993, \pasp, 105, 247. doi:10.1086/133140
\bibitem[Peterson et al.(1998)]{Peterson1998} Peterson, B.~M., Wanders, I., Horne, K., et al.\ 1998, \pasp, 110, 660. doi:10.1086/316177
\bibitem[Peterson \& Wandel(1999)]{PW1999} Peterson, B.~M. \& Wandel, A.\ 1999, \apjl, 521, L95. doi:10.1086/312190
\bibitem[Peterson \& Wandel(2000)]{PW2000} Peterson, B.~M. \& Wandel, A.\ 2000, \apjl, 540, L13. doi:10.1086/312862
\bibitem[Peterson et al.(2004)]{Peterson2004} Peterson, B.~M., Ferrarese, L., Gilbert, K.~M., et al.\ 2004, \apj, 613, 682. doi:10.1086/423269
\bibitem[Planck Collaboration et al.(2020)]{Planck2020}
  Planck Collaboration, Aghanim, N., Akrami, Y., et al.\ 2020, \aap, 641, A6. doi:10.1051/0004-6361/201833910
\bibitem[Popovi{\'c} et al.(2019)]{popovic2019} Popovi{\'c}, L. {\v{C}}., Kova{\v{c}}evi{\'c}-Doj{\v{c}}inovi{\'c}, J., \& Mar{\v{c}}eta-Mandi{\'c}, S.\ 2019, \mnras, 484, 3, 3180. doi:10.1093/mnras/stz157
\bibitem[Press et al.(1992)]{Press1992} Press, W.~H., Teukolsky, S.~A., Vetterling, W.~T., et al.\ 1992, , Numerical recipes in FORTRAN. The art of scientific computing. 
\bibitem[Prince et al.(2022)]{Prince2022} Prince, R., Zaja{\v{c}}ek, M., Czerny, B., et al.\ 2022, \aap, 667, A42. doi:10.1051/0004-6361/202243194
\bibitem[Prince et al.(2023)]{Prince2023} Prince, R., Zaja{\v{c}}ek, M., Panda, S., et al.\ 2023, \aap, 678, A189. doi:10.1051/0004-6361/202346738
\bibitem[Rakshit et al.(2020)]{Rakshit2020} Rakshit, S., Stalin, C.~S., \& Kotilainen, J.\ 2020, \apjs, 249, 1, 17. doi:10.3847/1538-4365/ab99c5
\bibitem[Richards et al.(2006)]{Richards2006} Richards, G.~T., Lacy, M., Storrie-Lombardi, L.~J., et al.\ 2006, \apjs, 166, 470. doi:10.1086/506525
\bibitem[Rodr{\'\i}guez-Pascual et al.(1997)]{Rodriguez1997} Rodr{\'\i}guez-Pascual, P.~M., Alloin, D., Clavel, J., et al.\ 1997, \apjs, 110, 9. doi:10.1086/312996
\bibitem[Salviander et al.(2007)]{Salviander2007} Salviander, S., Shields, G.~A., Gebhardt, K., et al.\ 2007, \apj, 662, 131. doi:10.1086/513086
\bibitem[Schlafly \& Finkbeiner(2011)]{schlafly2011} Schlafly, E.~F. \& Finkbeiner, D.~P.\ 2011, \apj,  737, 2, 103. doi:10.1088/0004-637X/737/2/103
\bibitem[Shen et al.(2008)]{Shen2008} Shen, Y., Greene, J.~E., Strauss, M.~A., et al.\ 2008, \apj, 680, 169. doi:10.1086/587475
\bibitem[Shen et al.(2011)]{Shen2011} Shen, Y., Richards, G.~T., Strauss, M.~A., et al.\ 2011, \apjs, 194, 45. doi:10.1088/0067-0049/194/2/45
\bibitem[Shen \& Liu(2012)]{Shen2012} Shen, Y. \& Liu, X.\ 2012, \apj, 753, 2, 125. doi:10.1088/0004-637X/753/2/125
\bibitem[Shen \& Ho(2014)]{shen2014} Shen, Y. \& Ho, L.~C.\ 2014, \nat,  513, 7517, 210. doi:10.1038/nature13712
\bibitem[Shen et al.(2016)]{Shen2016} Shen, Y., Horne, K., Grier, C.~J., et al.\ 2016, \apj, 818, 30. doi:10.3847/0004-637X/818/1/30
\bibitem[Shen et al.(2019)]{Shen2019} Shen, Y., Hall, P.~B., Horne, K., et al.\ 2019, \apjs, 241, 2, 34. doi:10.3847/1538-4365/ab074f
\bibitem[Shen et al.(2024)]{Shen2024} Shen, Y., Grier, C.~J., Horne, K., et al.\ 2024, \apjs, 272, 26. doi:10.3847/1538-4365/ad3936
\bibitem[Shin et al.(2019)]{Shin2019} Shin, J., Nagao, T., Woo, J.-H., et al.\ 2019, \apj, 874, 1, 22. doi:10.3847/1538-4357/ab05da
\bibitem[Shin et al.(2021)]{shin2021} Shin, J., Woo, J.-H., Nagao, T., et al.\ 2021, \apj, 917, 2, 107. doi:10.3847/1538-4357/ac0adf
\bibitem[Sulentic et al.(2000)]{sulentic2000} Sulentic, J.~W., Zwitter, T., Marziani, P., et al.\ 2000, \apjl,  536, 1, L5. doi:10.1086/312717
\bibitem[Sun et al.(2015)]{Sun2015} Sun, M., Trump, J.~R., Shen, Y., et al.\ 2015, \apj, 811, 42. doi:10.1088/0004-637X/811/1/42
\bibitem[Trevese et al.(2007)]{Trevese2007} Trevese, D., Paris, D., Stirpe, G.~M., et al.\ 2007, \aap, 470, 491. doi:10.1051/0004-6361:20077237
\bibitem[Tsuzuki et al.(2006)]{Tsuzuki2006} Tsuzuki, Y., Kawara, K., Yoshii, Y., et al.\ 2006, \apj, 650, 57. doi:10.1086/506376
\bibitem[U et al.(2022)]{U2022} U, V., Barth, A.~J., Vogler, H.~A., et al.\ 2022, \apj, 925, 52. doi:10.3847/1538-4357/ac3d26
\bibitem[Valdes et al.(2004)]{Valdes2004} Valdes, F., Gupta, R., Rose, J.~A., et al.\ 2004, \apjs, 152, 2, 251. doi:10.1086/386343
\bibitem[Verner et al.(1999)]{Verner1999} Verner, E.~M., Verner, D.~A., Korista, K.~T., et al.\ 1999, \apjs, 120, 101. doi:10.1086/313171
\bibitem[Vestergaard \& Wilkes(2001)]{VW2001} Vestergaard, M. \& Wilkes, B.~J.\ 2001, \apjs, 134, 1. doi:10.1086/320357
\bibitem[Vestergaard \& Peterson(2006)]{Vestergaard2006} Vestergaard, M. \& Peterson, B.~M.\ 2006, \apj, 641, 689. doi:10.1086/500572
\bibitem[Villafa{\~n}a et al.(2023)]{villafana2023} Villafa{\~n}a, L., Williams, P.~R., Treu, T., et al.\ 2023, \apj, 948, 2, 95. doi:10.3847/1538-4357/accc84
\bibitem[Wang et al.(2009)]{Wang2009} Wang, J.-G., Dong, X.-B., Wang, T.-G., et al.\ 2009, \apj, 707, 1334. doi:10.1088/0004-637X/707/2/1334
\bibitem[Wang et al.(2014)]{Wang2014} Wang, J.-M., Du, P., Hu, C., et al.\ 2014, \apj, 793, 108. doi:10.1088/0004-637X/793/2/108
\bibitem[Wang et al.(2020)]{Wang2020} Wang, J.-M., Songsheng, Y.-Y., Li, Y.-R., et al.\ 2020, Nature Astronomy, 4, 517. doi:10.1038/s41550-019-0979-5
\bibitem[Wills et al.(1980)]{Wills1980} Wills, B.~J., Netzer, H., Uomoto, A.~K., et al.\ 1980, \apj, 237, 319. doi:10.1086/157871
\bibitem[Woo(2008)]{Woo2008} Woo, J.-H.\ 2008, \aj, 135, 1849. doi:10.1088/0004-6256/135/5/1849
\bibitem[Woo et al.(2015)]{Woo2015} Woo, J.-H., Yoon, Y., Park, S., et al.\ 2015, \apj, 801, 38. doi:10.1088/0004-637X/801/1/38
\bibitem[Woo et al.(2018)]{Woo2018} Woo, J.-H., Le, H.~A.~N., Karouzos, M., et al.\ 2018, \apj, 859, 138. doi:10.3847/1538-4357/aabf3e
\bibitem[Woo et al.(2024)]{Woo2024} Woo, J.-H., Wang, S., Rakshit, S., et al.\ 2024, \apj, 962, 67. doi:10.3847/1538-4357/ad132f
\bibitem[Yang et al.(2020)]{Yang2020} Yang, Q., Shen, Y., Chen, Y.-C., et al.\ 2020, \mnras, 493, 5773. doi:10.1093/mnras/staa645
\bibitem[Yang et al.(2023)]{YangJY2023} Yang, J., Wang, F., Fan, X., et al.\ 2023, \apjl, 951, L5. doi:10.3847/2041-8213/acc9c8
\bibitem[Yang et al.(2024)]{yang2024} Yang, S., Du, P., \& Wang, J.-M.\ 2024, \apjs, 274, 2, 24. doi:10.3847/1538-4365/ad61e3
\bibitem[Yu et al.(2021)]{Yu2021} Yu, Z., Martini, P., Penton, A., et al.\ 2021, \mnras, 507, 3771. doi:10.1093/mnras/stab2244
\bibitem[Yu et al.(2023)]{Yu2023} Yu, Z., Martini, P., Penton, A., et al.\ 2023, \mnras, 522, 4132. doi:10.1093/mnras/stad1224
\bibitem[Zaja{\v{c}}ek et al.(2020)]{Zajacek2020} Zaja{\v{c}}ek, M., Czerny, B., Martinez-Aldama, M.~L., et al.\ 2020, \apj, 896, 146. doi:10.3847/1538-4357/ab94ae
\bibitem[Zaja{\v{c}}ek et al.(2021)]{Zajacek2021} Zaja{\v{c}}ek, M., Czerny, B., Martinez-Aldama, M.~L., et al.\ 2021, \apj, 912, 10. doi:10.3847/1538-4357/abe9b2
\bibitem[Zhu et al.(2017)]{Zhu2017} Zhu, D., Sun, M., \& Wang, T.\ 2017, \apj, 843, 30. doi:10.3847/1538-4357/aa76e7
\end{thebibliography}
\end{document}